\documentclass[a4paper,fleqn,usenatbib]{mnras}

\usepackage[T1]{fontenc}
\usepackage{ae,aecompl}

\usepackage{graphicx}	
\usepackage{amsmath}	
\usepackage{amssymb}	

\newcommand{\kms}{km\,s$^{-1}$}
\newcommand{\rsun}{$R_\odot$}
\newcommand{\msun}{$M_\odot$}
\newcommand{\bell}{$B_\ell$}
\newcommand{\nell}{$N_\ell$}
\newcommand{\rstar}{$R_\star$}
\newcommand{\xmm}{{\sc XMM}-Newton}

\title[Magnetic field and magnetosphere of Plaskett's star]{The magnetic field and magnetosphere of Plaskett's star: A fundamental shift in our understanding of the system\thanks{Based on spectropolarimetric observations obtained at the Canada-France-Hawaii Telescope (CFHT) which is operated by the National Research Council of Canada, the Institut National des Sciences de l'Univers (INSU) of the Centre National de la Recherche Scientifique of France and the University of Hawaii, as well as on observations obtained using the Narval spectropolarimeter at the Observatoire du Pic du Midi (France), which is operated by the INSU.}}

\author[Grunhut et al.]{J.H. Grunhut$^{1}$\thanks{Email: jason.grunhut@gmail.com}, G.A. Wade$^2$, C.P. Folsom$^{2,3}$, C. Neiner$^4$, O. Kochukhov$^5$, E. Alecian$^6$,  
\newauthor{M. Shultz$^7$, V. Petit$^7$, and the MiMeS \& BinaMIcS collaborations}\\
\\
$^{1}$European Southern Observatory, Karl-Schwarzschild-Strasse 2, 85748 Garching bei München, Germany\\
$^{2}$Dept. of Physics \& Space Science, Royal Military College of Canada, PO Box 17000, Station Forces, Kingston, ON, Canada, K7K7B4\\
$^{3}$Tartu Observatory, University of Tartu, Observatooriumi 1, T\~{o}ravere, 61602, Estonia\\
$^{4}$LESIA, Paris Observatory, PSL University, CNRS, Sorbonne University, Universit\'e de Paris, 5 place Jules Janssen, 92195 Meudon, France\\
$^5$Department of Physics and Astronomy, Uppsala University, Box 516, Uppsala 75120, Sweden\\
$^6$Universit\'e Grenoble Alpes, CNRS, IPAG, 38000 Grenoble, France\\
$^7$Department of Physics and Astronomy, Bartol Research Institute, University of Delaware, Newark, DE 19716, USA\\
}

\date{Accepted XXX. Received YYY; in original form ZZZ}

\pubyear{2020}

\begin{document}
\label{firstpage}
\pagerange{\pageref{firstpage}--\pageref{lastpage}}
\maketitle

\begin{abstract}
Plaskett's ``star" appears to be one of a small number of short-period binary systems known to contain a hot, massive, magnetic star. Building on the 2013 discovery investigation, we combine an extensive spectropolarimetric (Stokes $V$) dataset with archival photometry and spectropolarimetry to establish the essential characteristics of the magnetic field and magnetosphere of the rapidly rotating, broad-line component of the system. We apply Least-Squares Deconvolution (LSD) to infer the longitudinal magnetic field from each Stokes $V$ spectrum. Using the timeseries of longitudinal field measurements, in combination with CoRoT photometry and equivalent width measurements of magnetospheric spectral lines, we infer the rotation period of the magnetic star to be equal to $1.21551^{+0.00028}_{-0.00034}$\,d. 
Modeling the Stokes $V$ LSD profiles with Zeeman Doppler Imaging, we produce the first {reliable} magnetic map of an O-type star.  We find a magnetic field that is predominantly dipolar, but with an important quadrupolar component, and weak higher order components.  The dipolar component has an obliquity near $90$\degr\ and a polar strength of about 850 G, while the average field strength over the entire surface is 520 G.
We update the calculations of the theoretical magnetospheric parameters, and in agreement with their predictions we identify clear variability signatures of the H$\alpha$, H$\beta$, and He~{\sc ii}\,$\lambda 4686$ lines confirming the presence of a dense centrifugal magnetosphere surrounding the star. Finally, we report a lack of detection of radial velocity (RV) variations of the observed Stokes $V$ profiles, suggesting that historical reports of the large RV variations of the broad-line star's spectral lines may be spurious. { This discovery may motivate a fundamental revision of the historical model of the Plaskett's star as a near-equal mass O+O binary system.}
\end{abstract}


\section{Introduction}\label{intro}
\defcitealias{linder08}{L08}

Massive stars ($M\ga8$\,\msun) impart a disproportionate amount of energy and momentum into their surroundings during their short lives through their intense ultraviolet (UV) radiation fields, their strong radiatively-driven winds, and in their deaths as supernovae. Despite their rarity, they significantly impact the energetics, structure, chemical enrichment, and evolution of their host galaxies \citep{crowther10}. It is now generally believed that binary evolution impacts the majority of massive stars during their lives \citep{sana12, sana13}. Improving our understanding of the dynamical and physical properties of massive binary stellar systems will, therefore, have a broad impact on our knowledge of the stars that dominate the evolution of the Universe.

It is now well established that approximately 10\% of all main sequence and pre-main sequence isolated OBA stars host strong, stable, and globally-ordered magnetic fields \citep[e.g.][]{alecian13a, fossati15, 2017MNRAS.465.2432G, sikora2019a}. The detected fields are characteristically different from those of cool, low-mass stars, and show no clear correlations between their magnetic and physical properties (such as mass or rotation), which suggests a very different origin than the contemporaneously driven dynamo fields of the Sun and other cool stars \citep[e.g.][]{donati09, alecian13a, grunhut15}. The broadly favoured hypothesis is that these {\it fossil fields} are the remnants of the Galactic field accumulated and possibly enhanced by a dynamo field generated in an earlier phase of evolution \citep[e.g.][]{mestel01, moss01}. MHD instabilities (e.g. \citealt{tayler73}, \citealt{spruit99}, Emeriau-Viard \& Mathis, submitted) can potentially account for the observed magnetic dichotomy of magnetic and (apparently) non-magnetic stars, as only strong fields would survive \citep[e.g.][]{auriere07}. An alternative hypothesis is that the magnetic fields of ABO stars fields are the remnants of short-lived dynamo-generated fields that occur during episodic merger events or from strong binary interactions \citep[e.g.][]{ferrario09, tout08,grunhut14, langer12, wickramasinghe14, Schneider2019}, which are expected to occur with roughly a 10\% frequency among high-mass stars \citep[e.g.][]{bogomazov09, demink13}.

{HD\,47129 (Plaskett's star) is a bright ($V=6.06$) spatially unresolved system with two distinct spectroscopic O-type components.  The spectroscopic component with narrower lines is in a very clear $P_{\rm orb}=14.396257\pm 0.00095$~d orbit \citep{linder08}, and the system is understood to be a nearly equal-mass, non-eclipsing SB2 system in an approximately circular orbit \citep[e.g.][]{1987Obs...107...68S}.}

Plaskett's star is considered to be a probable member of Mon OB2, a large association located at a heliocentric distance of $1.4-1.7$~kpc \citep{2007AJ....134.1368C} containing two or three subgroups of OB stars and stellar aggregates \citep{1979Ap&SS..66..191S,2005AJ....130..721L}. The proposed Mon OB2 membership is in { reasonable agreement with the Gaia eDR3 distance of Plaskett's star ($1.283\pm 0.122$~kpc; \citealt{2021AJ....161..147B}).} These subgroups are characterized by a relatively large range of characteristic ages, from 0.2-0.6 Myr \citep[the dynamical age of NGC 2244, the youngest component of the association;][]{1967ApJ...147..965M} to 2 Myr \citep[the main sequence turnoff age of NGC 2244;][]{2002AJ....123..892P} to 4 Myr \citep[the dynamical age of the H~{\sc i} shell of the Rosette;][]{2007AJ....134.1368C} to 20-25 Myr \citep{1976ApJ...210...65T} for the most evolved stars in the association.

According to the analysis of \citet{linder08}, the system is composed of an O8III/I spectroscopic component (referred to in this paper as `the narrow-line star') 
of mass $(45.4\pm 2.4)\sin^3 i~M_\odot$, and an O7.5V/III spectroscopic component (referred to here as `the broad-line star') 
of mass $(47.3\pm 0.3)\sin^3 i~M_\odot$. Despite their very similar masses, the narrow-line component appears to be $\sim$2$\times$ brighter than the broad-line component in the optical \citep{linder08}. The system inclination $i=71\pm 9\degr$ was estimated by \citet{1978PASP...90..163R} using linear polarimetry. \citet{1978PASP...90..163R} and \citet{bagnuolo92} noted that an inclination significantly larger than $\sim 70\degr$ would result in eclipses (which are not observed), while an inclination significantly smaller would imply improbably large masses of the components. According to this analysis, with a total mass of approximately $112~M_\odot$, HD~47129 is one of the most massive known O-type binaries.

\citet{linder08} identified serious inconsistencies between the characteristics of the components inferred from their spectroscopic and dynamical analyses. The inferred spectral types are too late for the dynamical masses. The optical brightness ratio inferred from equivalent widths (EWs) of a selection of He~{\sc i} and He~{\sc ii} lines implies that the narrow-line star is twice as bright as the broad-line star. The inferred absolute magnitudes of the components are consistent with main sequence or subgiant evolutionary states, in contradiction to the inferred giant luminosity classes. The radii implied by the dynamically-inferred masses and measured surface gravities of \citet{linder08} (approximately $22~R_\odot$ for both components) disagree with those inferred from the Stefan-Boltzmann law ($14~R_\odot$ and $10.5~R_\odot$ for the narrow-line and broad-line components, respectively) using the luminosities and temperatures of \citet{linder08}. While \citet{linder08} note that some of this tension could be relieved by increasing the distance from the assumed 1.5~kpc, this would result in { significant conflict with the precise Gaia eDR3 parallax.} 

As implied above, the two stars exhibit substantially different line widths, implying very different rotational velocities. \citet{linder08} measured projected rotational velocities from several spectral lines, deriving $v\sin i$ ranging from 60-75~\kms\ for the narrow-line star and from 230-310~\kms\ for the broad-line star. These very different rotational velocities are unexpected given the obvious circularization of the orbit.

The system is clearly chemically peculiar. According to \citet{linder08} \citep[see also][]{2017A&A...607A..82M}, the narrow-line star is strongly N enhanced (16 times solar) and C depleted (3\% of solar), while the broad-line star is N depleted (20\% of solar) and He overabundant (1.5 times solar). The peculiar chemistry, in combination with the mass/luminosity mismatch and rapid rotation of the secondary, has led investigators \citep{bagnuolo92, linder08} to speculate that Plaskett's star is a post-Roche-lobe overflow (RLOF) system.

The system was first identified as an X-ray emitter by the ROSAT All-Sky Survey \citep{1996A&AS..118..481B}. \citet{2006MNRAS.370.1623L} reported analysis of \xmm\ observations of Plaskett's star, revealing the system to be a hard, luminous and variable X-ray emitter ($kT_{\rm max} \simeq 1.4$~keV, $\log L_{\rm x}/L_{\rm bol} = -6.0$). \citet{2017MNRAS.465.2160K} furthermore detected radio emission at 1 and 3 cm. These observations are qualitatively consistent with the historical interpretation \citep[e.g.][]{wiggs92} of the system as a colliding-wind binary (CWB).

\citet{grunhut13} reported the discovery of a magnetic field associated with the broad-line component of the Plaskett system. With only a limited number of observations, no robust conclusion about the characteristics of the magnetic field or the wind-field interaction could be drawn. However, it was proposed that the rotationally-flattened wind of the broad-line star \citep{wiggs92,linder08} was a consequence of magnetic confinement of the broad-line star's wind. As this system is the only known short-period, O-type binary with a rapidly rotating magnetic component, it may provide a unique guide towards our understanding of the formation and evolution of magnetism in massive stars, and of the potential role magnetism may play in binary evolution of these systems.

The investigations summarized above demonstrate the complexity of the Plaskett system, combining binarity, magnetism, interacting winds, and non-standard stellar evolution. This article aims to clarify the rotational, magnetic, and magnetospheric properties of the broad-line component, elaborating on the preliminary findings reported by \citet{grunhut13}. In Sect.~\ref{obs_sec} we present the new and archival observations used in our analysis, including a discussion of the extraction of magnetic, spectroscopic and photometric measurements. In Sect.~\ref{period_sec} we perform a period analysis of the polarimetric, spectroscopic, and photometric data to search for rotationally-modulated variability associated with the broad-line star. Sect.~\ref{mag_sect} presents the magnetic modelling and the inferred field properties based on the polarimetric observations. In Sect.~\ref{magnetosphere_sect} we present a detailed analysis of the observed emission variations associated with the magnetically-confined wind of the broad-line component. Lastly, we summarise our findings and conclusions in Sect.~\ref{discussion_sect}. 

\section{Observations}\label{obs_sec}
\subsection{Spectropolarimetric data}
Between 2012 February 4 and 2013 March 4 a total of 63 high-spectral resolving power ($R=\frac{\lambda}{\Delta\lambda}\sim65\,000$), spectropolarimetric observations were acquired with the Echelle SpectroPolarimetric Device for the Observations of Stars (ESPaDOnS), mounted on the Canada-France-Hawaii Telescope (CFHT), and its twin instrument Narval, mounted on the T{\' e}lescope Bernard Lyot (TBL). (For a detailed discussion of the formation and analysis of Stokes $V$ in an SB2 spectrum see \citealt{2019MNRAS.489.5669P}.) Observations obtained before 2012 April were previously described by \citet{grunhut13}. All observations obtained prior to 2013 were acquired within the context of the CFHT and TBL Magnetism in Massive Stars (MiMeS) Large Programmes over the course of 21 different nights \citep{wade16}. The remaining spectra obtained in 2013 were acquired within the context of the Binarity and Magnetic Interactions in various classes of Stars (BinaMIcS) CFHT Large Programme \citep{neiner13, alecian15}.

Each spectropolarimetric sequence consisted of four individual sub-exposures with exposure times ranging from 410 to 900 seconds that were taken with different configurations of the polarimetric retarders to acquire circularly polarized Stokes $V$ spectra \citep[e.g.][]{donati97}. The individual sub-exposures were processed using the automated reduction package {\sc libre-esprit} to produce unpolarized Stokes $I$ and circularly polarized Stokes $V$ spectra in the wavelength range of $3700 - 10\,500$\,\AA, following the double-ratio procedure described by \citet{donati97}. This process ensures that all spurious polarization signatures are removed to first order. Diagnostic null spectra were also determined by combining the four sub-exposures in such a way that the polarization cancels out \citep{wade16}. This allows us to verify that no spurious polarization signals are present in the processed data. { After verifying that no significant nightly variability was present, all spectra obtained during a given night were co-added and normalised to the continuum.} The peak signal-to-noise ratio (S/N) of the { co-added} polarimetric sequences ranges from 600-2000 per 1.8\,\kms\ spectral pixel. A log of these observations is provided in Table~\ref{obs_log}. 

\subsubsection{Polarimetric measurements}
To increase the S/N of the polarimetric spectra, we applied the Least-Squares-Deconvolution (LSD) procedure of \citet{donati97}. Following the analysis of \citet{grunhut13}, we adopted a line mask that uses a small subset of lines (11) based on an LTE synthetic spectrum of an O8 giant. The lines (He\,{\sc i} $\lambda$4026, 4471, 4713, 5015, He\,{\sc ii} $\lambda$4200, 4541, C\,{\sc iv} $\lambda$5801, 5811, N\,{\sc iii} $\lambda$4511, 4515, and O\,{\sc iii} $\lambda$5592) are mostly in absorption in our spectra, although there is some evidence of emission contamination in some of the lines. This mask is similar to that employed by \citet{donati06a} for HD\,191612, which proved to yield the most significant Zeeman detections in that and in several other magnetic O stars. 

\begin{table}
\centering
\caption{Journal of polarimetric observations listing the date, the heliocentric Julian date at mid-exposure (2\,450\,000+), the instrument employed (E=ESPaDOnS, N=Narval), the number of spectroscopic exposures obtained per night, the exposure time per sub-exposure, the orbital phase according to the ephemeris of \citet{linder08}, the peak S/N per 1.8\,\kms\ pixel in the observed spectrum, the evaluation of the detection level of the Stokes $V$ Zeeman signature within the line profile according to the criteria of \citet{donati97} (DD = definite detection, MD = marginal detection, ND = no detection), and the adopted velocity bin for the given detection level. The observations obtained during the first eight nights were previously discussed by \citet{grunhut13}.}\label{obs_log}
\begin{tabular}{|c@{}|c@{}|c@{}|c@{}|c@{}|c@{}|c@{}|c@{}|r@{}}
\hline
\ & Mid. && N & $t_{\rm exp}$ & Orb. & Pk & Det. & { Bin}\\
Date & HJD & { Inst} & Exp. & (s) & Phase & S/N & Flag & (km/s) \\
\hline
12-02-04	&	5961.8588 & E	&	8	&	600	&	0.295	&	2080	&	DD	& 30.6\\
12-02-09	&	5966.8777 & E	&	16	&	600	&	0.644	&	2462	&	MD	& 14.4\\
12-02-10	&	5967.7682 & E	&	8	&	600	&	0.706	&	1637	&	DD	& 46.8\\
12-02-12	&	5969.7755 & E	&	16	&	600	&	0.845	&	3000	&	DD	& 46.8\\
12-03-13	&	6000.4172 & N	&	4	&	1200	&	0.974	&	2648	&	MD	& 10.8\\
12-03-14	&	6001.3654 & N	&	4	&	1200	&	0.039	&	2518	&	ND	& 43.2\\
12-03-23	&	6010.3588 & N	&	4	&	1200	&	0.664	&	2459	&	ND	& 41.4\\
12-03-25	&	6012.3443 & N	&	4	&	1200	&	0.802	&	2410	&	MD	& 41.4\\
12-09-25	&	6196.0913 & E	&	8	&	900	&	0.566	&	1999	&	DD	& 43.2\\
12-09-27	&	6198.0942 & E	&	8	&	900	&	0.705	&	2131	&	ND	& 10.8\\
12-09-28	&	6199.0824 & E	&	8	&	900	&	0.773	&	1121	&	DD	& 41.4\\
12-10-01	&	6202.1013 & E	&	8	&	900	&	0.983	&	2338	&	DD	& 39.6\\
12-11-29	&	6261.1259 & E	&	8	&	900	&	0.083	&	2314	&	MD	& 1.8\\
12-11-30	&	6262.0630 & E	&	12	&	900	&	0.148	&	2396	&	DD	& 46.8\\
12-12-02	&	6264.1276 & E	&	8	&	900	&	0.292	&	1273	&	ND	& 1.8\\
12-12-10	&	6272.1204 & E	&	8	&	900	&	0.847	&	2268	&	ND	& 1.8\\
12-12-21	&	6282.9363 & E	&	8	&	900	&	0.598	&	2679	&	ND	& 43.2\\
12-12-22	&	6284.1162 & E	&	8	&	900	&	0.680	&	1647	&	MD	& 41.4\\
12-12-26	&	6288.0984 & E	&	6$^*$	&	900	&	0.957	&	2503	&	MD	& 1.8\\
12-12-27	&	6289.1048 & E	&	8	&	900	&	0.027	&	2301	&	DD	& 1.8\\
12-12-28	&	6289.9823 & E	&	8	&	900	&	0.087	&	1837	&	DD	& 41.4\\
13-02-20	&	6343.8556 & E	&	4	&	900	&	0.830	&	2149	&	ND	& 14.4\\
13-02-21	&	6344.8483 & E	&	1$^*$	&	900	&	0.774	&	-	&	-	& - \\
13-02-28	&	6351.8751 & E	&	16	&	410	&	0.387	&	1968	&	DD	& 46.8\\
13-03-01	&	6352.9188 & E	&	16	&	410	&	0.459	&	998	&	MD	& 45.0\\
13-03-02	&	6353.9112 & E	&	16	&	410	&	0.528	&	1072	&	MD	& 41.4\\
13-03-03	&	6354.7522 & E	&	16	&	410	&	0.587	&	601	&	MD	& 1.8\\
13-03-04	&	6355.8976 & E	&	16	&	410	&	0.666	&	1218	&	MD	& 1.8\\
\hline
\multicolumn{8}{l}{$^*$Incomplete polarimetric sequence}\\
\end{tabular}
\end{table}

For weak Stokes $V$ signals, the detection probability can sometimes be enhanced by optimising the velocity pixel size used for extracting an LSD profile \citep{2017MNRAS.465.2432G}. 
Adopting a single width of the velocity bin for a set of observations of a single star is a strategy that is most commonly adopted, but due to the large width of the line profile, the low amplitude of the Zeeman signatures, and the relatively high noise level, this strategy is not optimal for the analysis of this system.
Therefore, in order to improve our ability to detect weak Zeeman signatures, we adapted the velocity bin size for each observation.  This was done by extracting LSD profiles using different velocity pixel sized, ranging from 1.8\,\kms\ to 46.8\,\kms. 

The likelihood that a Zeeman signature was detected in the Stokes $V$ profile of an LSD spectrum was computed by measuring the $\chi^2$ of Stokes $V$ with respect to zero within the confines of the line profile, interpreted using the false alarm probability (FAP) criteria of \citet{donati92, donati97} for each LSD profile. A Zeeman signature is considered a definite detection (DD) if excess signal is detected within the line profile with a FAP $<10^{-5}$. A signal is considered a marginal detection (MD) if the FAP is greater than $10^{-5}$ but less than $10^{-3}$, while a higher FAP is considered a non-detection (ND). The most significant detection for each observation varied with velocity bin size; the results are reported in  Table~\ref{obs_log}, where we obtain detections of magnetic signatures in 20 of the profiles (10 DD, 10 MD). 
The Stokes $V$ profiles exhibit both positive and negative polarities, as well as crossover morphologies.  We note that interestingly, the velocity span of the Stokes $V$ profiles appears to be constant, i.e.\ roughly -450 to 450 \kms. 

\begin{figure}
\centering
\hspace{-0.15in}\includegraphics[width=3.45in]{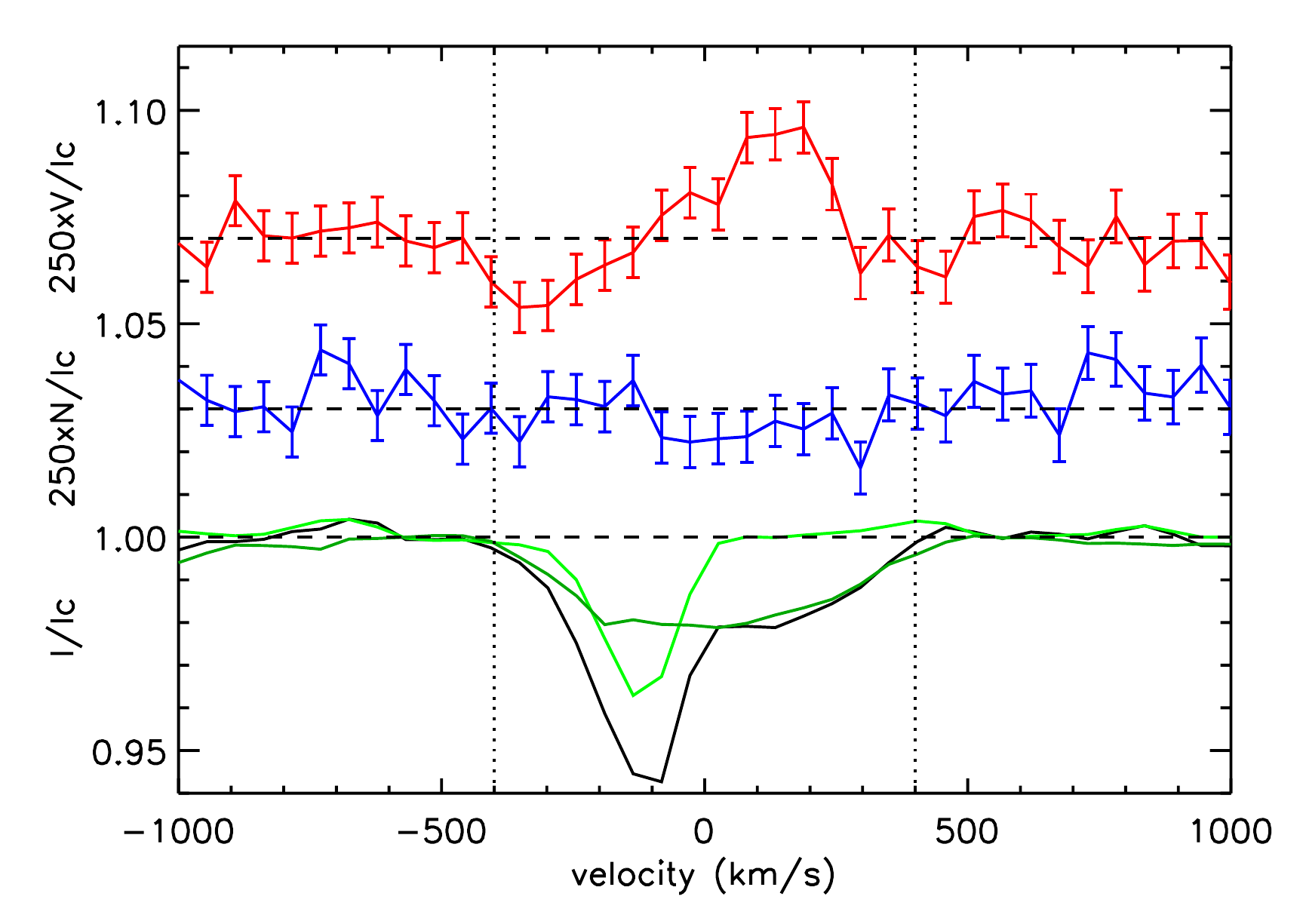}
\caption{LSD Stokes $V$ (top), diagnostic null (middle) and Stokes $I$ profile of Plaskett's star obtained on the night of 2012-10-01. The profiles have been rebinned to a velocity bin of 54\,\kms\ and the $V$ and $N$ profiles have been expanded by the indicated factor and shifted for display purposes. The Stokes $I$ profile (observed in black, disentangled profiles assuming the stationary solution of the narrow-line and broad-line components in green) shows the clear blend of the two components. Vertical dotted lines are included to illustrate the full width of the broad-line profile. A clear Zeeman signature is detected in Stokes $V$ with a position and width that is consistent with the broad-line component. No excess signal is found in the corresponding diagnostic null profile.}\label{lsd_examp_fig}
\end{figure}

All extracted LSD profiles were scaled to a wavelength of 500\,nm, a line depth of 0.1 times the continuum, and a Land{\' e} factor of 1.2. An example of an observation with a clear detection of a Zeeman signature in the LSD profile is provided in Fig.~\ref{lsd_examp_fig}.

\begin{figure}
\centering
\hspace{-0.15in}\includegraphics[width=3.45in]{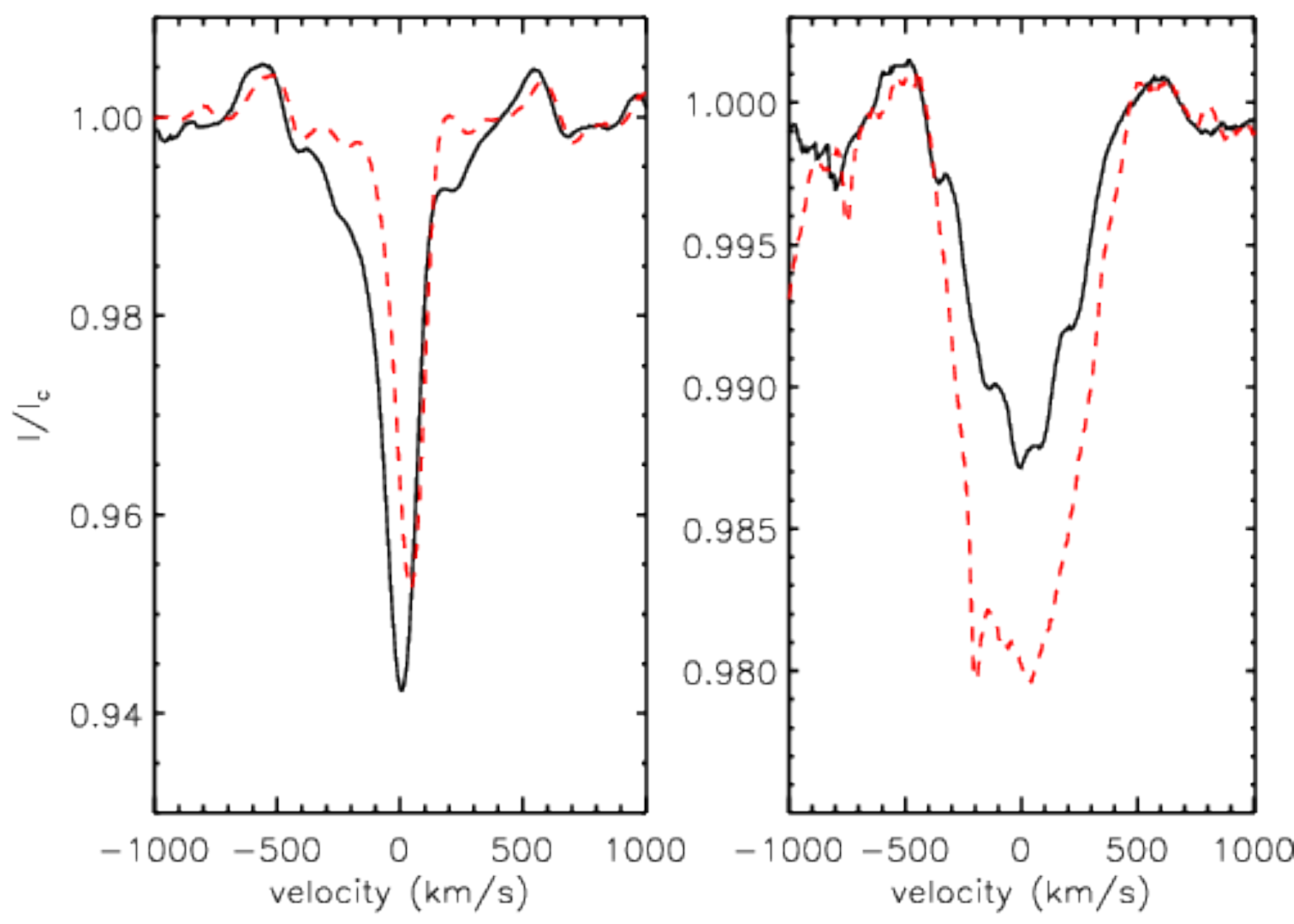}
\caption{Comparison of the two disentangling solutions, with the narrow-line component on the left and the broad-line component on the right. The solid black lines illustrate the variable-RV solution, while the dashed red lines illustrate the static-RV solution.}\label{disentangle_fig}
\end{figure}

{ \subsubsection{Disentangling of the LSD profiles}

The Stokes $I$ profile of the broad-line magnetic star is contaminated by the presence of the narrow-line component (see Fig.~\ref{lsd_examp_fig}). In order to infer the longitudinal magnetic field \bell\ of the broad-line component without contribution from the intensity profile of the narrow-line star, we followed the method discussed by \citet{grunhut13}. This disentangling approach works in the wavelength domain (or velocity for LSD profiles) and proceeds iteratively, first by generating a line model for one component by subtracting a model for the other component from the observations, shifting the residuals to a common rest frame, and summing to produce an averaged line profile (weighted by uncertainties).  Then an updated line model for the other component is generated using the same procedure, subtracting the new line model for the first star from the observations.  This is iterated until the two line profiles no longer change significantly. See \citet{gonzalez06} for more details.

As previously discussed by \citet{grunhut13}, this method neglects any intrinsic equivalent width (EW) variability. Nonetheless, as argued by \citet{grunhut13}, we expect that variable wind emission formed above the photosphere, where the magnetic field is weaker, is the primary source of non-RV related variations, but does not contribute significantly to the Stokes $V$ profile and corresponding \bell\ measurements. The validity of these approximations can be diagnosed from the scatter of the \bell\ measurements (Sect.~\ref{bell_sec} and Fig.~\ref{all_plot}.)	

As will be further discussed in Sect~\ref{sect-ZDI2}, our attempts to map the magnetic field of the star using Zeeman-Doppler Imaging (ZDI) led us to consider models in which the radial velocity of the broad-line star was constant. As a consequence we have computed two disentangling solutions for the broad-line star: one assuming the RV variations adopted by \citet{linder08}, and a second in which the RV is held stationary. In both solutions the narrow-line star varies in RV according to \citet{linder08}.

The chief impact of adopting the stationary RV solution for the broad-line star is to broaden and deepen the broad-line component. The disentangled profiles for both models are shown in Fig.~\ref{disentangle_fig}.}

\subsubsection{Longitudinal magnetic field}

The longitudinal magnetic field (\bell) was computed from the disentangled LSD profiles of the broad-line star extracted with a uniform velocity spacing of 5.4\,\kms\ for consistency, using the first-order moment method discussed by \citet{rees79} and Eq.~1 of \citet{wade00}. We performed the measurements using both disentangling solutions. 
The wavelength $\lambda$ and Land{\' e} factor $g$ used were the scaling values of the LSD profiles, 500\,nm and 1.2, respectively. The integration range was chosen to be between -450 and 450\,\kms. 
A similar quantity was computed from the diagnostic null profile (\nell) using the same integration range. This latter value should be consistent with zero in the absence of systematic errors.  The \bell\ measurements for individual observations are presented in Table \ref{online_bl_tab}. The measurements are presented for only the stationary solution, since the measurements for the two solutions differ only marginally.

\subsubsection{Equivalent width measurements}\label{eqw_sect}
{ From the Stokes $I$ spectra of the individual polarimetric subexposures,} we measured the EW variations of a number of spectral lines. The measurements followed the procedure previously discussed by \citet{wade12a} and briefly summarised here: each spectral region is locally re-normalized by dividing the spectrum by a line of the form $y = mx + b$, fit to the continuum regions around the line of interest prior to any measurement. The integration was carried out over a sufficient velocity width to capture the full width of the broad-line profile as well as the velocity-shifted narrow-line profile. The EW measurements therefore include contributions from the narrow-line profile. The 1$\sigma$ uncertainties were calculated by propagating the individual pixel uncertainties in quadrature. The EW measurements are reported in Table~\ref{online_ew_tab}.

\subsection{Photometric data}
We utilised 72 {\it Hipparcos} \citep{perryman97} photometric measurements acquired on 54 different nights between 1990 March 11 and 1993 March 12.

A second photometric dataset used in this analysis was acquired with the {\it CoRoT} \citep[Convection, Rotation and planetary Transits; ][]{baglin06} satellite between 2008 October 8 and 2008 November 12. This dataset was already discussed in detail by \citet{mahy11} and the details are not repeated here. The raw photometry for Plaskett's star contains 92\,696 datapoints, but we removed all measurements with a non-zero flag (corresponding to potentially corrupted data), which provided 83\,359 datapoints subsequently used in our analysis.

{A third photometric dataset was acquired by {\it TESS} \citep[Transiting Exoplanets Survey Satellite; ][]{2014SPIE.9143E..20R,2015JATIS...1a4003R}. The star was observed with TESS camera \#1 in Sector 6. The observations spanned 21.7 d, between 2018 Dec.\ 15 and 2019 Jan.\ 6  (BJD 2458468 and 2458490), and consisted of 14829 data points. These data are currently the subject of a detailed analysis by Stacey et al. (in prep.); here we provide a discussion of their period content limited specifically to constraining the rotation of the magnetic star.}

\section{Period Analysis}\label{period_sec}
A detailed characterisation of the short-term (i.e. during $\sim 1$~month) photometric variability of Plaskett's star was presented by \citet{mahy11}, while a similar spectroscopic analysis was presented by \citet{palate14}. \citet{mahy11} identified the orbital period of the system and a number of other frequencies that they considered to be possibly due to non-radial pulsations (we find one of these is the rotation period of the broad-line star). \citet{palate14} found three frequencies consistent with the photometric analysis, but could not confidently ascribe an origin to them.  With the spectropolarimetric dataset acquired here, we carried out a new variability analysis, with the specific goal to search for periodicity that is attributable to rotational modulation by the magnetic broad-line component. 

\subsection{Spectropolarimetric data}\label{bell_sec}
Following the procedure adopted for the analysis of other magnetic O-type stars \citep[e.g.][]{grunhut12c, wade12a,2017MNRAS.465.2432G}, we analysed the \bell\ measurements and the EW measurements of prominent magnetospheric emission lines, such as H$\alpha$, H$\beta$, H$\gamma$, and He\,{\sc ii} $\lambda$4686, to search for variability related to rotational modulation. The analysis was carried out with the \citet{sc96} technique using the code developed by \citet{townsend10}. This technique reports the analysis of variance (ANOVA) statistic $\Theta$. To improve the temporal sampling we used the individual (unbinned) polarimetric spectra for the magnetic measurements (63 spectra in total) and the spectra corresponding to individual sub-exposures for the EW measurements (255 spectra in total). The log of observations and a summary of these measurements is provided in Tables~\ref{online_bl_tab} and \ref{online_ew_tab}, which can be found in Appendix~\ref{online_tab_sec}.

\citet{linder08} report projected rotational velocities inferred from various lines in the disentangled spectrum of the broad-line star ranging from 245-310\,\kms. {While our LSD profiles extracted from the disentangled spectra assuming the Linder RV solution yield compatible results, we obtain a somewhat larger $v\sin i=360\pm40$\,\kms\ from the LSD profiles corresponding to the disentangling assuming a stationary broad-line star}. Considering this range of $v\sin i$ and adopting the equatorial radius of this star implied by the luminosity and temperature of \citet{linder08} ($10.5~R_\odot$), the rotation period must be $\la 2.2$\,d assuming rigid rotation. On the other hand, using the radius computed from the mass and surface gravity of \citet{linder08} ($R_\star = 22$\,\rsun) a period of $\la$4.5\,d is implied. We therefore conservatively searched for periodicity in the range of 0.1-20\,d, with the upper bound chosen to probe periods similar to the orbital period ($\sim$14.4\,d). 

\begin{figure}
\includegraphics[width=3.2in]{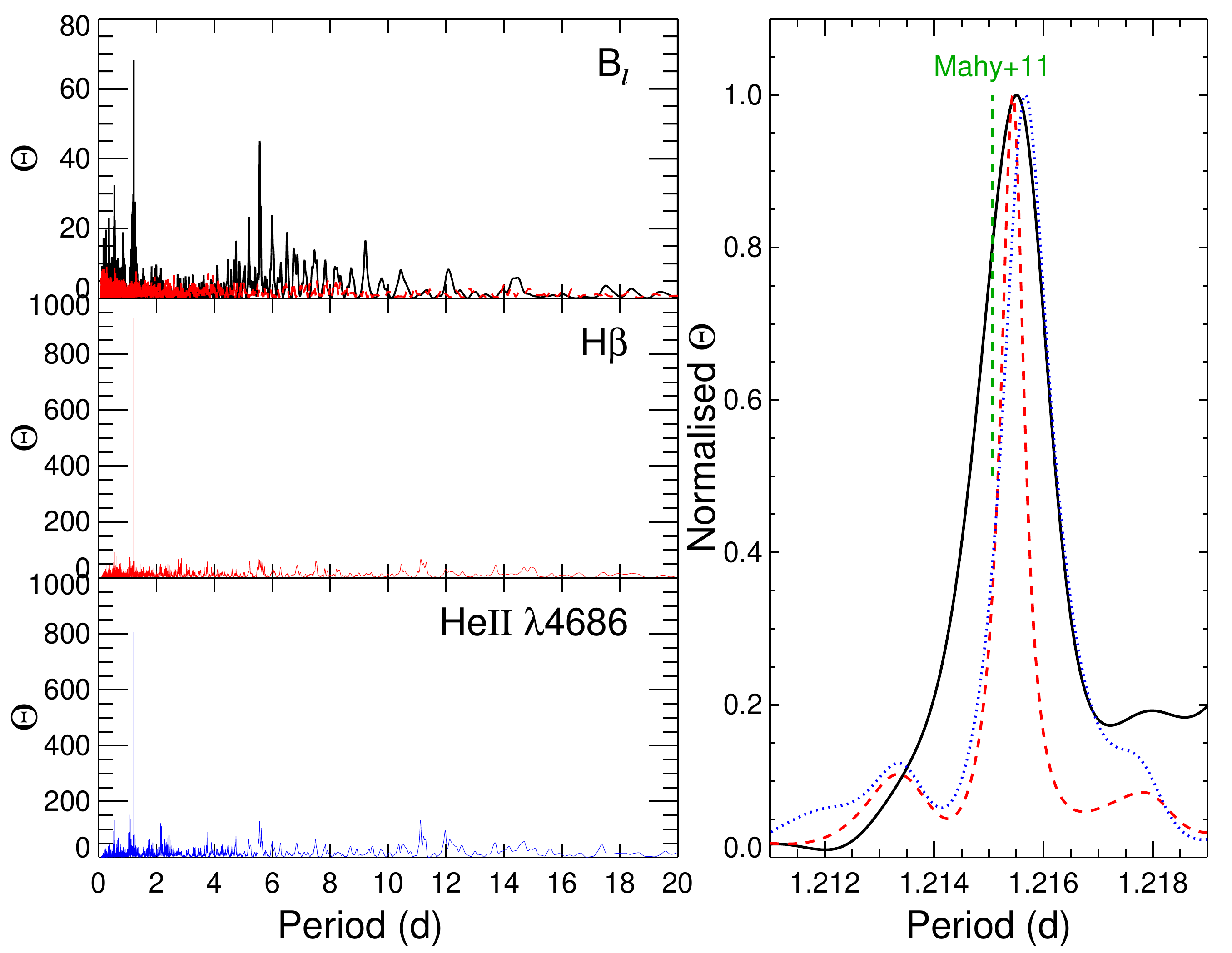}
\caption{Periodograms obtained from the \bell\  measurements (top panel, solid black), the H$\beta$ (middle panel, red) and the He\,{\sc ii} $\lambda$4686 (bottom panel, dotted blue) EW measurements. Significant power is present in all periodograms at $\sim$1.215\,d. The right panel provides a narrow view of the periodograms about 1.215\,d, normalised to their maximum power. The peak period identified by \citet{mahy11} from the {\it CoRoT} lightcurve is also indicated for comparison. Periodograms obtained from the EW measurements included contributions from the first harmonic. The panel for \bell\ (top) also presents the periodogram after prewhitening with the 1.215\,d period, as discussed in the text (dashed red).}
\label{periodogram}
\end{figure}

The periodogram obtained from the \bell\ measurements (see top panel of left frame of Fig.~\ref{periodogram}) shows several strong peaks. The strongest peak in the periodogram occurs at $1.21551^{+0.00028}_{-0.00034}$\,d, which we note is consistent with one of the periods identified by \citet{mahy11} from the {\it CoRoT} photometry. The uncertainty is derived from $\chi^2$ statistics corresponding to sinusoidal fits to the data. When phased with this period, the \bell\ measurements show coherent variations that are well-fit by a sinusoid (reduced $\chi^2$, $\chi_r^2=1.37$). 
The next strongest peak in the periodogram occurs around 5.565\,d, which appears to be an alias of the the 1.215\,d period, since no significant peaks remain in the 5.565\,d region after computing the periodogram from data prewhitened by subtracting a sinusoidal fit to the phased $B_\ell$ curve with a period of 1.215\,d (see Fig.~\ref{periodogram} for comparison)\footnote{While we performed this analysis for both sets of \bell\ measurements (i.e. for both disentangling solutions), we report results only for the stationary solution since the variation of \bell\ is unaffected.}.

\begin{figure}
\includegraphics[width=3.2in]{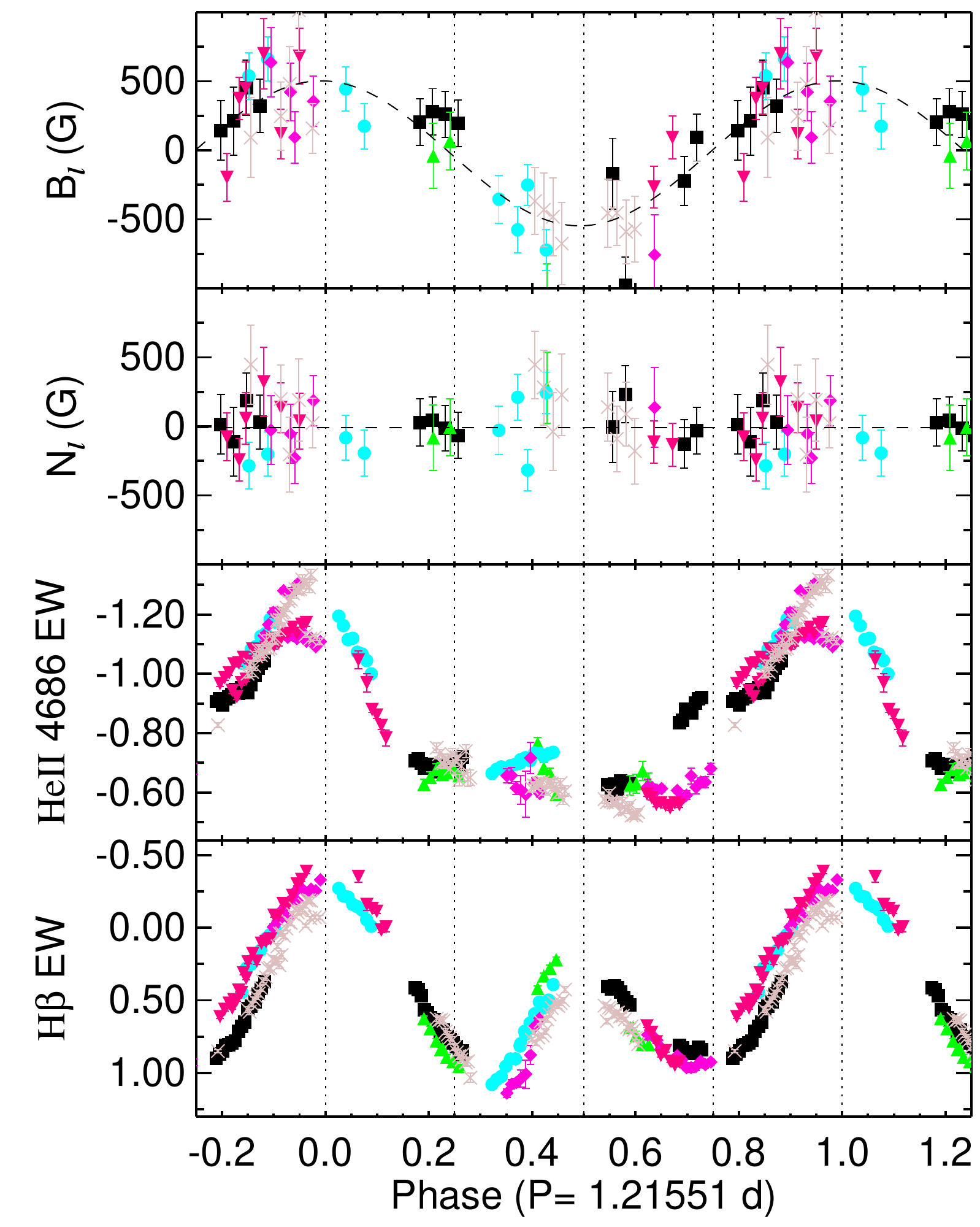}
\caption{Observational data phased using the rotational ephemeris of Eq.~\ref{ephemeris}. The top panel shows the \bell\ measurements from the LSD profiles and the sinusoidal fit to the data. The next panel shows the same measurements performed on the null LSD profiles with the corresponding best-fitting linear fit. Only points with $\sigma<300$\,G are shown for display purposes. As evident from the data and the sinusoidal fits, the \bell\ measurements show clear sinusoidal variations that are not observed in the null measurements. The He\,{\sc ii} $\lambda$4686 EW measurements are presented in the next panel, while the H$\beta$ EW variations are presented in the bottom panel. To illustrate the cycle-to-cycle variations of the EW measurements we plot different epochs of observations with different colours and symbols.}
\label{all_plot}
\end{figure}

Unlike the periodogram of the \bell\ measurements, the periodograms of several prominent emission lines (e.g. H$\alpha$, H$\beta$, H$\gamma$, He\,{\sc ii}\,$\lambda$4686) are more complex.
We also computed the periodograms of the EWs of four emission lines: H$\alpha$, H$\beta$, H$\gamma$, and He\,{\sc ii}\,$\lambda$4686. In each case the periodogram is dominated by a strong peak at 0.607\,d and a weaker peak at $\sim 1,215$\,d. Since the shorter period is the first harmonic of the longer period, this suggests a double-wave variation of the EWs according to the period determined from the \bell\ measurements, similar to the emission variation observed in the magnetic O9.5\,IV star HD\,57682 \citep{grunhut12c}. Indeed, when phased with the 1.215\,d period (see Fig.\ \ref{all_plot}), the measurements show more coherent phasing than with the 0.607\,d period.

Since the EW variations yield a most-prominent period that appears to be a harmonic of the single-wave \bell\ period, we proceeded to obtain periodograms of the H$\alpha$, H$\beta$ and He\,{\sc ii} measurements using the multiharmonic fitting capability of the \citet{sc96} technique, by including contributions from the first harmonic only. Periodograms from the \bell\ measurements, H$\beta$ and He\,{\sc ii} EW measurements are displayed Fig.~\ref{periodogram}. The H$\beta$ periodogram shows a single significant peak at $1.21543\pm0.00004$\,d. The periodogram of the He\,{\sc ii} measurements shows only minor differences compared to the hydrogen lines, with the strongest peak occurring at $1.21582\pm0.00006$\,d and another peak at $\sim$2.4\,d (i.e. about twice longer). Therefore, we conclude that all spectropolarimetric and spectroscopic variable quantities provide a consistent period solution.

Adopting the \bell\ period and the phase of maximum positive \bell\ as HJD$_0$ we obtain the following ephemeris:
\begin{equation}\label{ephemeris}
HJD =  2455961.000^{+48}_{-34} + 1.21551^{+28}_{-34}\cdot E,
\end{equation}
where the 1$\sigma$ uncertainties are listed for the last digits only. We note that the periodic spectroscopic variations are roughly consistent with the repeating episodic nature of the shell lines of this star, as reported by \citet{struve58}.

In Fig.~\ref{all_plot} we show the \bell, \nell, He\,{\sc ii} $\lambda$4686, and H$\beta$ EW measurements phased according to Eq.~\ref{ephemeris}. The \bell\ measurements show a clear sinusoidal variation with a central value of $-2 \pm 27$\,G and a semi-amplitude of $513 \pm 41$\,G. The maximum positive \bell\  occurs at phase 0 and maximum negative field occurs at phase 0.5. We find no evidence for any variability in the \nell\ measurements and these measurements are fully consistent with scatter due only to random noise about 0\,G.  

The He\,{\sc ii} $\lambda$4686 EW measurements show a strong emission peak (minimum EW) around phase 0, likely with a weaker secondary peak at about phase 0.45 (at least for one epoch, as indicated by different symbols/colours). The H$\beta$ EW curve reaches maximum emission at about phase 0, and a secondary emission peak around phase $\sim$0.50. Minimum emission appears to be reached around phases $\sim$0.30 and $\sim$0.75. The EW variations corresponding to different rotation cycles show significant systematic differences; however, the \bell\ measurements show no such effect. The EW variations, their origin, and the cycle-to-cycle variations are further discussed in Sect.~\ref{magnetosphere_sect}.

\subsection{Photometric data}\label{phot_data_sect}

\begin{figure}
\includegraphics[width=3.2in]{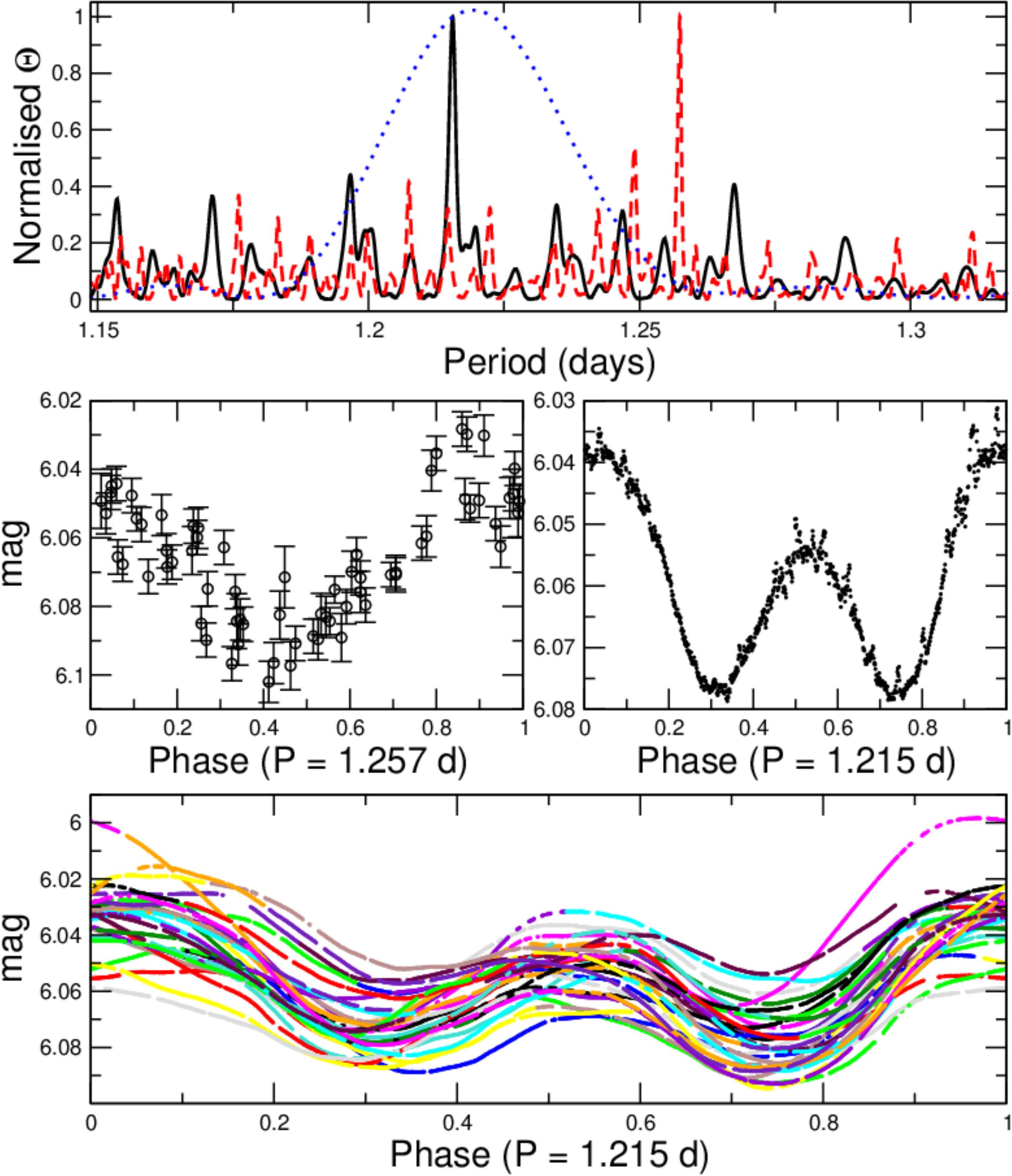}
\caption{Top: Periodogram obtained from the \bell\ measurements (solid-black), the {\it Hipparcos} photometry (dashed red) and the {\it CoRoT} lightcurve (dotted blue). Middle: Phased lightcurves of the {\it Hipparcos} (left) and {\it CoRoT} (right) photometry, phase-folded with the maximum-power period from their individual periodograms, while adopting $T_0$ according to Eq~\ref{ephemeris}. The {\it CoRoT} data have been binned in phase to emphasise the overall periodicity. Bottom: Phase-folded lightcurves of the {\it CoRoT} photometry for each individual rotation cycle. The data have been binned for display purposes.} \label{hip_bl_comp_fig}
\end{figure}

In the photometric analysis carried out by \citet{mahy11} a fundamental frequency of $f=0.823$\,d$^{-1}$ was found that corresponds to a period of $\sim$1.215\,d, which is fully consistent with the periods detected in the longitudinal field and EW measurements (See Fig.~\ref{hip_bl_comp_fig}). We carried out our own analysis of the {\it CoRoT} data, with a focus on finding rotational variability.  Using the \citet{sc96} technique we confirm the presence of significant power at the $\sim$1.215\,d period and its first harmonic ($P\sim0.60$\,d). An analysis of the {\it TESS} photometry (discussed in detail by Stacey et al. (in prep.) and not shown here) yields qualitatively similar results and dominant periods that are compatible with those obtained from {\it CoRoT}.

Because of the short duration ($\sim$30\,d) of the {\it CoRoT} dataset, the peaks in the associated periodogram are significantly broader than that obtained from the spectropolarimetric measurements (see the upper panel of Fig.~\ref{hip_bl_comp_fig}), implying that we cannot use this dataset to determine a higher precision period. We therefore proceeded to analyse the {\it Hipparcos} photometry, which was obtained about 15 years before the {\sc CoRoT} data. These data were acquired over a baseline of three years, but with substantially coarser sampling. 

In Fig.~\ref{hip_bl_comp_fig} we compare the periodograms obtained from the {\it Hipparcos} and {\sc CoRoT} datasets
with the periodogram obtained from the \bell\  measurements in a small range of periods about the suggested rotational period. While the {\it Hipparcos} periodogram contains several peaks, the most significant power occurs at a period of $1.2574\pm0.0001$\,d. In the bottom panels of Fig.~\ref{hip_bl_comp_fig} we also plot the photometric data phased to their corresponding maximum-power periods. We note that the {\it Hipparcos} data do not phase well with either the {\it CoRoT} period or the spectropolarimetric period. The {\it CoRoT} and {\it TESS} data demonstrate very similar double-wave nature to their variations, similar to that observed in the EW variations.  

Our interpretation is that the origin of the variation is from rotationally modulated, magnetically confined wind plasma \citep[e.g.][]{townsend08}. On the other hand, the {\it Hipparcos} data are better described by a single-wave variation, but with a peak-to-peak amplitude similar to the {\it CoRoT} and {\em TESS} data. While the double-wave nature of the {\it CoRoT} and {\it TESS} photometry persisted throughout the entire timeseries, the individual cycles' lightcurves exhibit substantial differences in their details, even over the relatively short duration of the observations (see bottom panel of Fig.~\ref{hip_bl_comp_fig}). Possible reasons for this discrepancy will be addressed in Sect.~\ref{period_change_disc}. As already noted by \citet{mahy11} 
there are additional significant short-term and long-term contributions to the photometric variability of the system that cause the phased light curve to change qualitatively in amplitude and character from cycle to cycle. Some of these variations are likely due to known periodic behaviour \citep{mahy11}, but this system has also been reported to show epochs of irregular photometric behaviour \citep{morrison78b}.

Nevertheless, the spectroscopic, magnetic, and {\it CoRoT/TESS} measurements are all consistent with the dominant 1.215\,d period. Therefore, in the context of the oblique rotator model and a magnetically confined wind \citep{stibbs50, babel97a} we interpret this period as the rotation period of the broad-line star. Additional spectroscopic and photometric variations are obvious from cycle to cycle, which is probably indicative of contributions from other mechanisms. This is further discussed in Sect.~\ref{discussion_sect}. No cycle-to-cycle variations are found in the magnetic measurements. This is a natural result of the fact that there are essentially no other competing contributions to $B_\ell$, {since the $B_\ell$ values are based on largely photospheric lines}. In other words, it seems likely that the rotational modulation of \bell\ is distinct from the photometric and spectroscopic variability in that it is the fundamental origin of the variability, rather than a phenomenon that results from it.

\section{Magnetic field of the broad-line star}\label{mag_sect}

The stable, sinusoidal variation of the longitudinal field suggests that the magnetic field of the broad-line star has an important dipole topology, with a surface polar field strength of approximately 1.5\,kG. Assuming that the field can be described by a centred oblique dipole rotator model \cite[ORM; ][]{stibbs50}, which is typical of most magnetic massive stars, the observed, symmetric \bell\ variations about 0\,G imply that either $i$ or $\beta$ (or both) is (are) close to 90\degr.

The rapid rotation implied by the model derived in accordance with the larger radius, inferred from $\log g$, surpasses the theoretical breakup velocity. Hence in the following we prefer the smaller radius.

Assuming rigid rotation, the radius ($10.5~R_\odot$) computed from  the luminosity and temperature reported by \citet{linder08}, $v\sin i$ (245-310\,\kms; again from the Linder solution), and the inferred rotation period ($P_{\rm rot}=1.21551^{+0.00028}_{-0.00034}$\,d), the implied inclination of the rotation axis is $i=45\pm 10$\degr\ using the formula:
\begin{equation}
\sin i = \frac{P (v\sin i)}{50.6 R_\star},
\end{equation}
\noindent with $P$ expressed in days, $v\sin i$ in \kms, and the stellar equatorial radius $R_\star$ in solar units.

\subsection{ZDI: modelling approach and assumptions}

Given that the broad-line star is unique among magnetic O-type stars as a rapid rotator and that the observations densely sample the rotation cycle of this star, we attempt to model the magnetic field of this star using Zeeman Doppler Imaging (ZDI). To increase the S/N of the Stokes $V$ profiles, we combined all profiles obtained each night into mean nightly profiles with 54\,\kms\ pixel bins. To avoid phase smearing, we first verified that the combined spectra span less than 5 percent of the rotation period, and that the individual profiles of each night exhibited no significant differences with respect to the nightly mean. Observations from 23 Mar.\ and 2 Dec.\ 2012 were omitted due to their particularly low S/N. 

In order to carry out this mapping, we used the ZDI code of \citet{folsom2018} that is based on the the work of \citet{brown91}, \citet{donati97b} and \citet{donati06b}. The magnetic inversion uses the time series of rotationally modulated Stokes $V$ profiles, and derives the simplest magnetic field geometry that can reproduce the observations, using the maximum entropy method of \citet{skilling84}. Using the method of \citet{donati06b}, the field topology is described as a spherical harmonic decomposition including radial, tangential poloidal, and tangential toroidal harmonic modes with angular degree $\ell=1-15$, and azimuthal orders $m=0 - \ell$. This routine attempts to find a model with maximum entropy and a $\chi^2$ less than or equal to a target value.  In practice it proceeds iteratively, with early iterations weighted towards minimizing $\chi^2$ to a target value, and once that is achieved later iterations weighted towards a solution that maximizes the negative entropy while not exceeding that target $\chi^2$. Local Stokes $V$ line profiles are calculated in the weak field approximation, using the derivative of the Stokes $I$ profile.  Local Stokes $I$ profiles can be Voigt profiles but in this work they were approximated by a Gaussian.  The weak field approximation is reasonable given the large competing broadening and the relatively weak magnetic field (i.e. below or near 1 kG).  A Gaussian local Stokes $I$ profile is reasonable since the local line profiles are dominated by turbulent broadening. The local Stokes $V$ profiles are then integrated across the stellar disk, including a rotational Doppler shift and a linear limb-darkening law.  Finally a Gaussian instrumental profile is applied to produce the disc-integrated Stokes $V$ profile.  A more detailed description is provided by \citet{folsom2018} in their Appendix B, and an earlier description of this line model is { given} by \citet{Wade2014}.

The adopted velocity binning is quite large for a typical application of ZDI. However, given the very large turbulent broadening of O stars, we do not expect much structure in the $V$ profiles smaller than this, so the binning should not affect the resolution of the magnetic map too much.  The ZDI code of \citet{folsom2018} is not adapted to such a large pixel binning, in particular if the binning is larger than the local line profile width and instrumental resolution, numerical problems may arise.  We modified the code to calculate line profiles on a finer velocity grid than the observation, with 10 sub-pixels evenly distributed in the $\pm 27$ \kms\ around each 54 \kms\ LSD pixel.  The sub-pixels are then summed together, after disk integration and convolution with the instrumental profile, to produce the final pixel. This models the binning of the LSD profile and avoids potential numerical artefacts.  In testing, we find this addition to the ZDI code has a relatively small impact on the resulting magnetic map if it is appropriately regularized. However, it does reduce the ability of the code to fit noise or over-fit pixels inside the line, and thus limits the smallest $\chi^2$ that can be reached.  

For input parameters of the local line model we used the scaling wavelength and Land\'e factor of the LSD profile (500\,nm and 1.2, respectively), and a linear limb-darkening coefficient of 0.31 \citep{claret04}.  
The width of the local Gaussian profile is controlled by turbulent broadening, since this is much larger than other local line broadening processes for most lines; but the amount of turbulent broadening is uncertain. As macroturbulent broadening is much less than $v\sin i$, it cannot be constrained reliably by the observations. This issue is exacerbated considering the imperfections in the disentangling and LSD processes.  \citet{Sundqvist2013a} investigated turbulent broadening in a sample of very slowly rotating magnetic O stars, and found Gaussian distributions to be a reasonable approximation with velocities between 20 and 60\,\kms.  We adopted this as the range of values and performed the analysis with a width of 20\,\kms, then repeated the analysis with a 60\,\kms\ width, but found that had a very small impact on our results.

The rotation period was determined in Sect.\ \ref{period_sec}, and we used the ephemeris of Eq.~\ref{ephemeris}. An instrumental profile with $R=65000$ was used, although this has little impact given the large turbulent broadening.
The $v\sin i$ and line depth were set by fitting the disk integrated model profile to the disentangled Stokes $I$ profiles. 

For the variable-RV disentangled profile, with a Gaussian { turbulent} width of 20\,\kms\ we find a best fit $v\sin i$ of 331\,\kms, while for 60\,\kms\ of turbulent broadening the best fit $v\sin i$ is 321\,\kms.  The static-RV profile is somewhat wider, yielding a best fit $v\sin i$ of 357 \kms\ for a Gaussian width of 20 \kms.  For a Gaussian width of 60 \kms\ the best fit $v\sin i$ is 352 \kms. These no longer agree with the the $v\sin i = 310$ \kms\ of \citet{linder08}.

There are apparent emission features at the edges of the variable-RV disentangled Stokes $I$ line profile. If they are treated as distortions to the continuum and we attempt to remove them through re-normalization, the fit $v\sin i$ increases to 368 \kms.  This is likely a worse treatment of the Stokes $I$ profiles, as emission is clearly present in the full spectrum, but one should be aware that there may be 30 or 40 km/s of systematic uncertainty.  The 331 and 321 \kms\ values of $v\sin i$ are acceptably close to the $310 \pm 20$ \kms\ of \citet{linder08}.  

As there is no evidence of contribution in the Stokes $V$ profile from the narrow-line star (see Sect.~\ref{sec_narrow_line_mag}), the treatment of the Stokes $V$ profiles does not rely on spectral disentangling.  The disentangling is used only to constrain $v\sin i$ and the line depth of the broad-line component.  The $V$ profile is modeled as arising from the broad-line star alone.

The results of the various ZDI analyses are summarized in Table~\ref{zditable}.

\begin{table*}
    \centering
    \caption{Summary of ZDI models tested.  The models marked with * are plotted in Figs.~\ref{fig-ZDIfits} and \ref{fig-ZDImaps}. \label{zditable}} 
    \begin{tabular}{lccccccccccc}
    \hline
    Assumptions & $v_{\rm turb.}$ & $v\sin i$ & $i$ & $\chi^2_r$ & entropy & $\langle B\rangle$ & pol. & dip. & $B_{\rm dip}$ & $\beta$ & \\
                & (\kms)  & (\kms)    & (\degr) &    &    &  (G) & (\%tot.) & (\%pol.) & (G) & (\degr) & Success?\\
    \hline
    Linder RVs * & 20 & 331 & $56 \pm 6$ & 1.5 & -20872 & 954 & 84 & 55 & 1214 & 57 & N\\
    Linder RVs & 60 & 321 & $56 \pm 6$ & 1.5 & -54810 & 1347 & 76 & 36 & 1314 & 54 & N\\
    Linder RVs, phase offset 0.09 & 20 & 331 & $61 \pm 5$ & 1.5 & -10130 & 746 & 90 & 58 & 1037 & 64 & N\\
    Linder RVs, phase offset 0.10 & 60 & 321 & $59 \pm 5$ & 1.5 & -18997 & 946 & 87 & 46 & 1149 & 66 & N\\
    Linder RVs, $P_{\rm rot} = 2.431$ d & 20 & 331 & $56 \pm 6$ & 1.5 & -45360 & 1246 & 80 & 20 & 720 & 9 & N\\
    Linder RVs, $v\sin i$ from Stokes $V$ & 20 & $387$ & $56 \pm 6$ & 1.5 & -9620 & 837 & 87 & 74 & 1314 & 64 & N\\
    constant RV * & 20 & 357 & $48 \pm 4$     & 1.0 & -3980 & 517 & 96 & 69 & 843 & 89 & Y \\
    constant RV & 60 & 352 & $47^{+5}_{-4}$ & 1.0 & -4282 & 547 & 94 & 70 & 897 & 90 & Y \\
    \hline
    \end{tabular}
\end{table*}

\subsection{ZDI: A first magnetic map adopting the Linder et al. ephemeris}
\label{sect-ZDI1}

We first carried out ZDI assuming the RV variations of the broad-line star consistent with the orbital ephemeris of \citet{linder08} (i.e. using the variable-RV Stokes $I$ profile).

The inclination of the rotation axis of the broad-line component is estimated at the start of Sect.~\ref{mag_sect}. However, this calculation relies on the radius of the star, which is very uncertain. The $T_{\rm eff}$ and luminosity of \citet{linder08} provide a very different value from that implied by the $\log g$ and mass.  Thus we derived independent constraints on the inclination of the broad-line star using ZDI.  We calculated grids of ZDI models, varying the inclination from $10^\circ$ to $90^\circ$ in $1^\circ$ increments. 
First we fit all models to a target reduced $\chi^2$ of 1.5, and searched for the converged model with the maximum entropy\footnote{ For the ZDI analysis, the reduced $\chi^2$ is calculated with the number of degrees of freedom as the number of observed data points, neglecting the number of free parameters. This approximation is used since, in a regularized fitting problem, the true number of degrees of freedom is ambiguous.}.  This produced a best inclination for the variable-RV Stokes $I$ profile of 56$^\circ$ (also 56$^\circ$ for a turbulent broadening of 60 \kms). 

In order to derive statistical uncertainties, we again ran the grid of ZDI models with different inclinations, but fitting to a target entropy rather than a target $\chi^2$.  The ZDI code has the option of, rather than looking for a solution that maximizes entropy for $\chi^2$ less than or equal to a target, using a modified fitting routine to minimize $\chi^2$ for entropy greater than or equal to a target.  This has the effect of producing a best fit model for a given degree of complexity specified by the target entropy.  In practice, fitting to target $\chi^2$ or entropy produces identical results, provided the target $\chi^2$ (or entropy) are consistent with the final $\chi^2$ (or entropy) from the other fitting method.  Running a grid of ZDI models with the target entropy constraint produces a variation in $\chi^2$ for models with a constant degree of complexity.  Then from the statistics in the variation of $\chi^2$ around the minimum \citep[e.g.][]{Lampton1976}, one can place a formal confidence level on an interval of the parameter of interest, in this case inclination.  A reasonable target entropy value is generally not known a priori, so we use the maximum entropy found in the previous grid fit as the target (specifically -20870, from the previous target $\chi^2_r$ of 1.5)  From this process we find an inclination of $56 \pm 6^\circ$ (for $1\sigma$ uncertainties, ${^{+19}_{-17}}^{\circ}$ at $3\sigma$).  
{ The variation of $\chi^2_r$ with inclination angle is shown in Fig.~\ref{i_fitting_fig}. }
Repeating this procedure with a local line width of 60\,\kms\ produces an inclination of $56 \pm 6^\circ$ (${^{+20}_{-18}}^{\circ}$ at $3\sigma$). 
This inclination is somewhat different from the $i=45\pm10^\circ$ we find from our rotation period and the $10.5~R_\odot$ radius (from $T_{\rm eff}$ and luminosity), but consistent at $1\sigma$.  However, this is inconsistent with  $i=20\pm 5^\circ$ produced by the $22~R_\odot$ radius (from mass and $\log g$), thus the magnetic map supports only the smaller radius.

No ZDI model is able to fit the data to a $\chi^2_r$ below 1.4, and models with $\chi^2_r$ below 1.5 appear to be badly over-fitting the data. These results are independent of the detailed choice of line broadening parameters. (We carried out the modeling for both $(v\sin i=331~{\rm km/s},v_{\rm mac}=20~{\rm km/s})$ and $(v\sin i=321~{\rm km/s},v_{\rm mac}=60~{\rm km/s})$). This large $\chi^2_r$ appears to be due to clear discrepancies between the model and observations at some phases.  Indeed, inspecting the fit to the observations, we note serious discrepancies of the best-fit model relative to particular observations, all of which are clustered near orbital phases at which the broad-line star is expected to show the largest RV shifts. The large $\chi^2_r$ and failure to fit the data at some orbital phases implies that the magnetic map will likely be affected by systematic errors.

\begin{figure*}
\centering
\includegraphics[width=\linewidth]{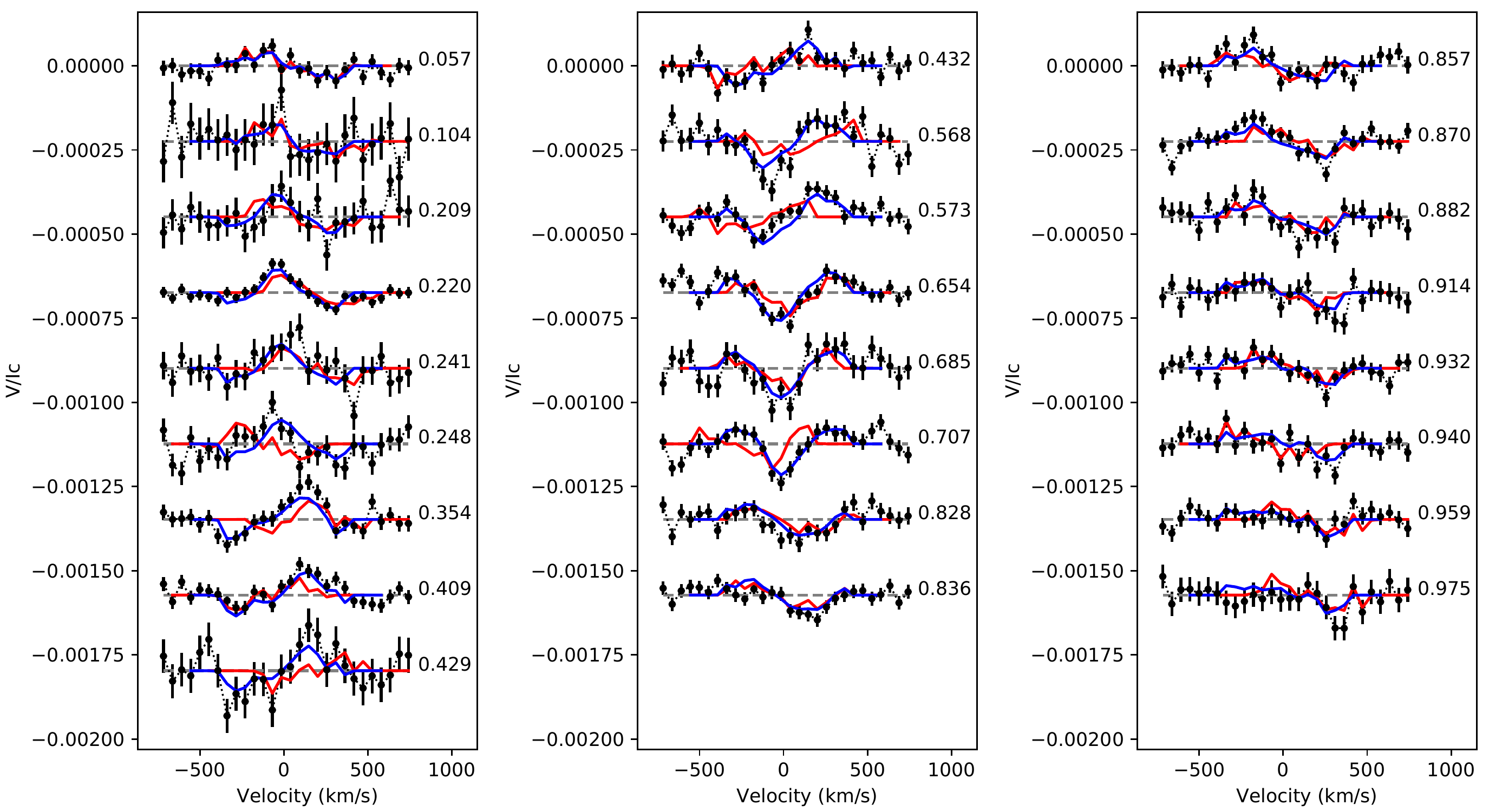}
\caption{Observed LSD $V$ profiles (black) fit with ZDI models of the broad-line component, shifted vertically for clarity.  Rotation phases are indicated on the right. Models using the radial velocity variations of \citet{linder08} 
are shown (red), but cannot reproduce the observations at several phases (in paricular, but not limited to, 0.248 and 0.707).
The model assuming a fixed velocity of the broad-line star
(blue) is also shown, and the associated profiles provide a qualitatively better fit to the observations. 
}\label{fig-ZDIfits}
\end{figure*}

\begin{figure*}
\centering
\includegraphics[width=0.47\linewidth]{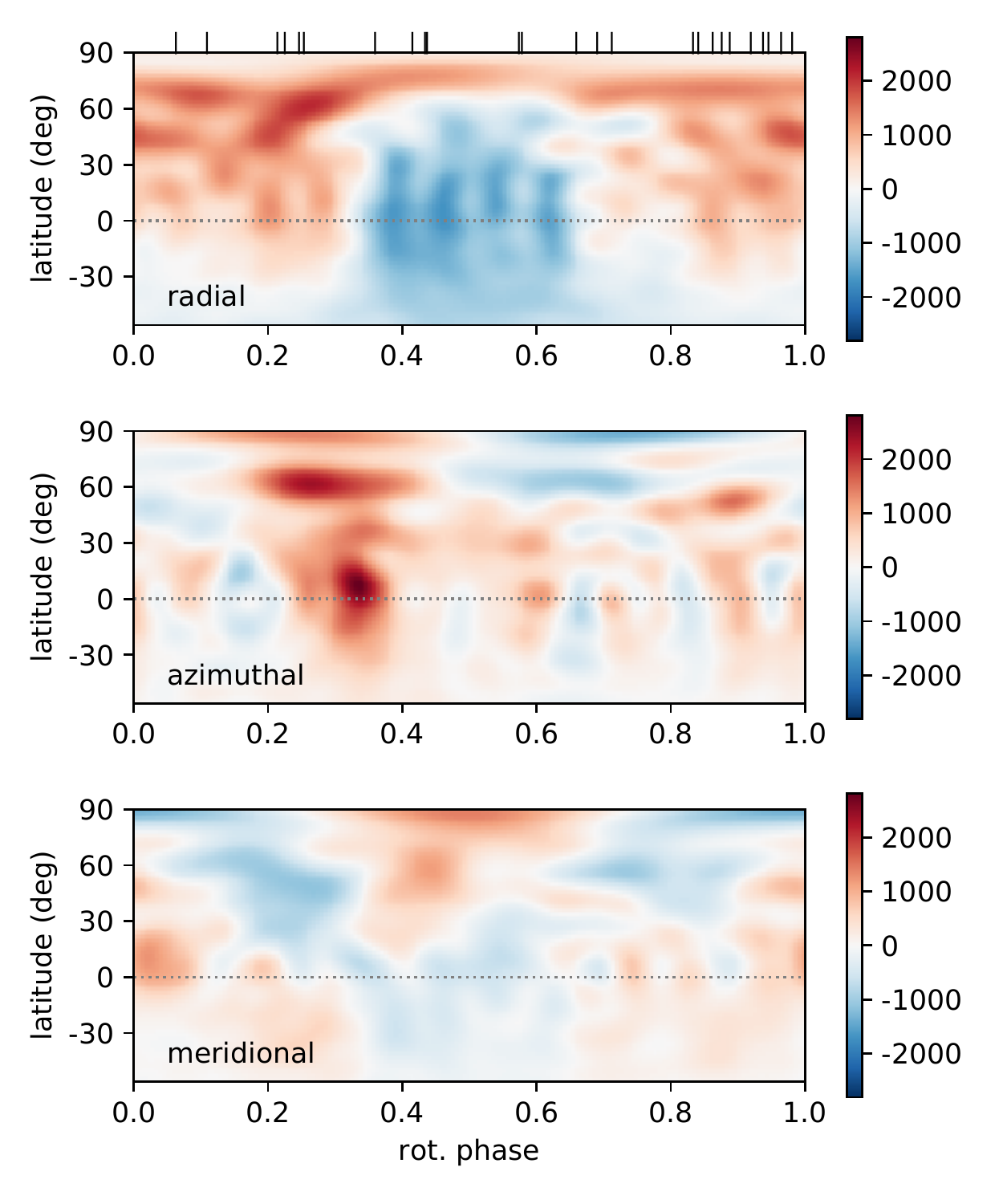}\hspace{0.5cm}
\includegraphics[width=0.47\linewidth]{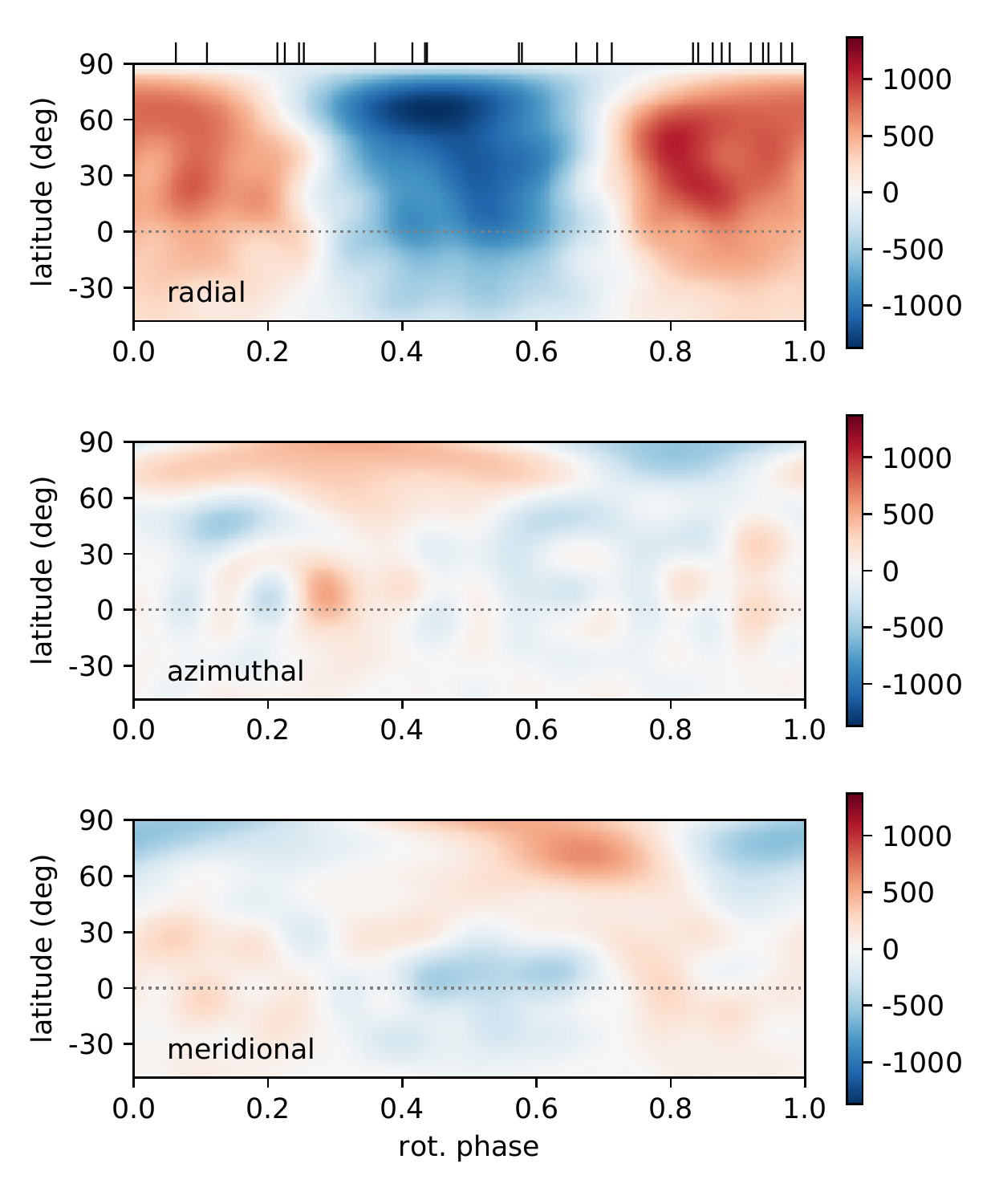}
\caption{Magnetic maps for the broad-line star with the \citet{linder08} orbital velocities (left), and a constant velocity (right). Panels show the radial, azumuthal, and meridional components of the vector magnetic field, and the colour scales show field strength in G.  Only the visible portions of the star for the best incinaltions are plotted.  Phases of observations are indicated by ticks along the top.  The higher level of complexity and small scale structure in the map including orbital velocities (left panel), together with the worse fit it provides, suggest the map is likely incorrect.}
\label{fig-ZDImaps}
\end{figure*}

We report some properties of the derived magnetic map here, but given the poor quality fit they are not likely representative of the real magnetic properties of the star. 
The fits to the observed LSD profiles are presented in Fig.~\ref{fig-ZDIfits}, and the resulting magnetic map is shown in Fig.~\ref{fig-ZDImaps} (left panel).  We describe the magnetic geometry using ratios of $B^2$ (more specifically $\oint \mathbf{B}\cdot\mathbf{B} \rm{d}\Omega$) as an approximation of the magnetic energy density, evaluated from components of the spherical harmonic description of the field (See Fig.~\ref{energy_plot_fig}).  The magnetic field reconstructed using the orbital velocities is largely poloidal (84\% magnetic energy) but with a significant toroidal component (16\% energy). The poloidal magnetic field has 55\% energy in the dipole ($\ell = 1$) mode, and 22\% in the quadrupole ($\ell = 2$), with significant energy in higher $\ell$ modes (particularly $\ell \leq 5$).  This is reflected by a large amount of small scale structure in the magnetic map, superimposed on a mostly dipolar field.  The small scale structure of the map reaches nearly 3 kG, while the dipole has a strength of only 1.2 kG.  This suggests that, even at a $\chi^2_r$ of 1.5, this model may be over-fitting portions of the line.  This magnetic geometry is inconsistent with that reconstructed using Stokes $V$ profiles for other magnetic O, B, and A-type stars, which usually have dominantly dipolar magnetic fields, sometimes with important quadrupolar components, but generally not with very strong small structures \citep[e.g.][]{2020pase.conf...89K}.  Small scale departures from a pure dipole are often found if observed at high enough S/N, but they do not approach the strength of the dipole itself.  The global dipolar component in this map may be approximately correct, but the smaller structure appears to be driven by over-fitting portions of the line profile, while other portions of line profiles remain unfit or badly fit.

We performed a number of experiments to better understand the origins of the poor quality of the fit to the Stokes $V$ observations and the distorted magnetic map.

First we considered the possibility of a phase offset of the broad-line star's RV variation due to the long time span between our observations and those employed by \citet{linder08}, notwithstanding that there is no evidence of this from the narrow-line star's RV variation (see \citealt{grunhut13}).  
To this aim we ran a grid of ZDI models with the ephemeris of \citet{linder08} but assuming a phase offset (relative to the ephemeris of \citealt{linder08}), varying from 0 to 1 in steps of 0.01, of the broad-line star's RV variation.
The simplest (maximum entropy) map for a reduced $\chi^2_r$ of 1.5 is achieved for an offset value of 0.09 cycles for a turbulent broadening of 20 \kms\ (or 0.10 for a turbulent broadening of 60 \kms)\footnote{ Line profiles for this grid of models were calculated for a range of -800 to 800 \kms\ in the heliocentric frame, to use a consistent set of observed pixels, and reached a $\chi^2_r$ of 1.4.  In the rest of the analysis we use a range of -600 to 600 km/s about the shifting line center (i.e. relative to the star), to reduce the number of extraneous continuum pixels used.  The $\chi^2_r$ calculated with this narrower range with less continuum is approximately 1.5, and this $\chi^2_r$ should be used for comparison with other values in this analysis}.  
In addition to the fact that these phase offsets are difficult to reconcile with the narrow-line star's RV curve, none of the phase-shifted models are able to substantially improve the agreement between the model and the observations, and all yield maps that share the same qualitative distortions as described above.

Secondly, we considered the possibility that the poor fit may be due to adopting an erroneous rotation period. Specifically, we investigated whether our rotation period could be the first harmonic of the real value. Thus we computed new maps assuming a rotational period of 2.431 d ($=2\times 1.2155$~d).  We note that there is no strong peak near this period in the periodogram for $B_l$ or H$\beta$, which makes such a period unlikely unless the two halves of the magnetic/EW curves are identical.  We ran this test only for a turbulent broadening of 20 \kms.   Models with this period can achieve a $\chi^2_r$ of 1.5, but again appear to badly over fit the data at this $\chi^2_r$, with more than two times the entropy and a quadrupole-dominated magnetic map.  More reasonable target $\chi^2_r$ are between 1.55 and 1.6, and the map is still quadrupole-dominated.  Fitting to the same entropy as the 1.215~d period gives $\chi^2_r = 1.558$.   We conclude that adopting the longer period does not resolve any of the issues identified with the 1.215~d maps, and yields a magnetic field structure that is even less plausible.

As a final experiment, we considered the possibility that $v\sin i$ from the disentangled Stokes $I$ profiles might be underestimated.  Stokes $V$ profiles typically provide a much weaker constraint on $v\sin i$ than Stokes $I$, and in some cases only an upper limit, but they do contain some information about this parameter.
We ran a grid of ZDI models with $v\sin i$ ranging from 300 to 450 \kms, in 1 \kms\ steps.  This test was done assuming a turbulent broadening of 20 \kms.  Fitting to a target $\chi^2_r$ of 1.5 all models converged, and the map with the best entropy (i.e. the simplest model) corresponded to $v\sin i = 387$ \kms. Fitting to this entropy (-9620) as a target, and using the variation in the achieved $\chi^2$, we found $v\sin i = 387 \pm 8$ \kms ($^{+37}_{-20}$ at $3\sigma$).\footnote{We note that we can achieve a somewhat lower $\chi^2_r$ if we adopt this higher $v\sin i$, at the risk of over-fitting the data.  Thus we reran the grid with a target $\chi^2_r$ of 1.4, and the corresponding entropy (-21400), and we found $v\sin i = 385 \pm 7$ \kms ($>369$ \kms\ at $3\sigma$). A $\chi^2_r = 1.3$ can be reached, although it appears to substantially over-fit the data, and only models with $v\sin i$ above 360 \kms\ can reach this $\chi^2_r$.} 


We conclude that, while a larger $v\sin i$ improves the fit at some phases, there are still large discrepancies present at other phases, particularly where the RV shift of the broad-line star is expected to be large.  In some profiles with large velocity shifts, a much higher $v\sin i$ would be needed to fully span the pixels with apparent signal in Stokes $V$.  This would be in clear contradiction to the observed Stokes $I$ line widths, and thus is not an adequate solution to this discrepancy.

\subsection{ZDI: A second magnetic map adopting a constant RV}
\label{sect-ZDI2}

In attempting to perform ZDI with the \citet{linder08} orbital velocities, we failed to achieve a satisfactory fit to the data, and consequently derived a map that was strongly distorted and unreliable.  Inspecting the fit to the observations, we observed that the majority of discrepancies appear at phases corresponding to large orbital RVs predicted by the ephemeris of \citet{linder08}.  We carried out a number of experiments to attempt to resolve the issues, to no avail. 

In this section we consider a simple alternative that is motivated by our observation that the largest discrepancies occur at phases of high predicted orbital RV shift. Specifically, we consider that the velocity of the broad-line star is constant, or at least that the RV variation of this star is significantly smaller than reported by \citet{linder08} and all previous studies. This, of course, would have important implications for our understanding of the architecture of the Plaskett system; this will be discussed later.

Utilizing the static-RV disentangled Stokes $I$ profile, we repeated the ZDI analysis. The spherical harmonics were restricted to $\ell=1-10$, as the extra degrees of freedom proved to be unnecessary. 

\begin{figure}
\centering
\includegraphics[width=3.15in]{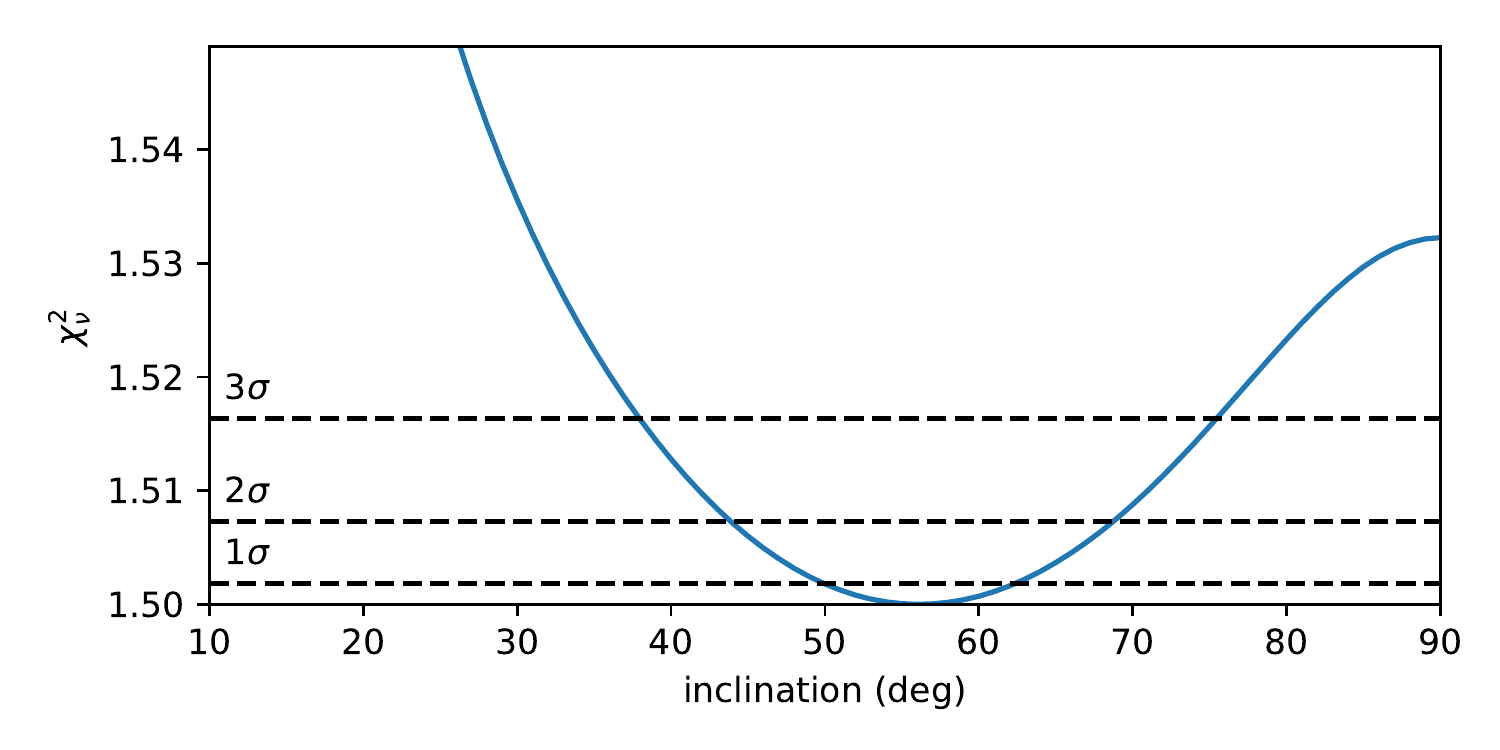}
\includegraphics[width=3.15in]{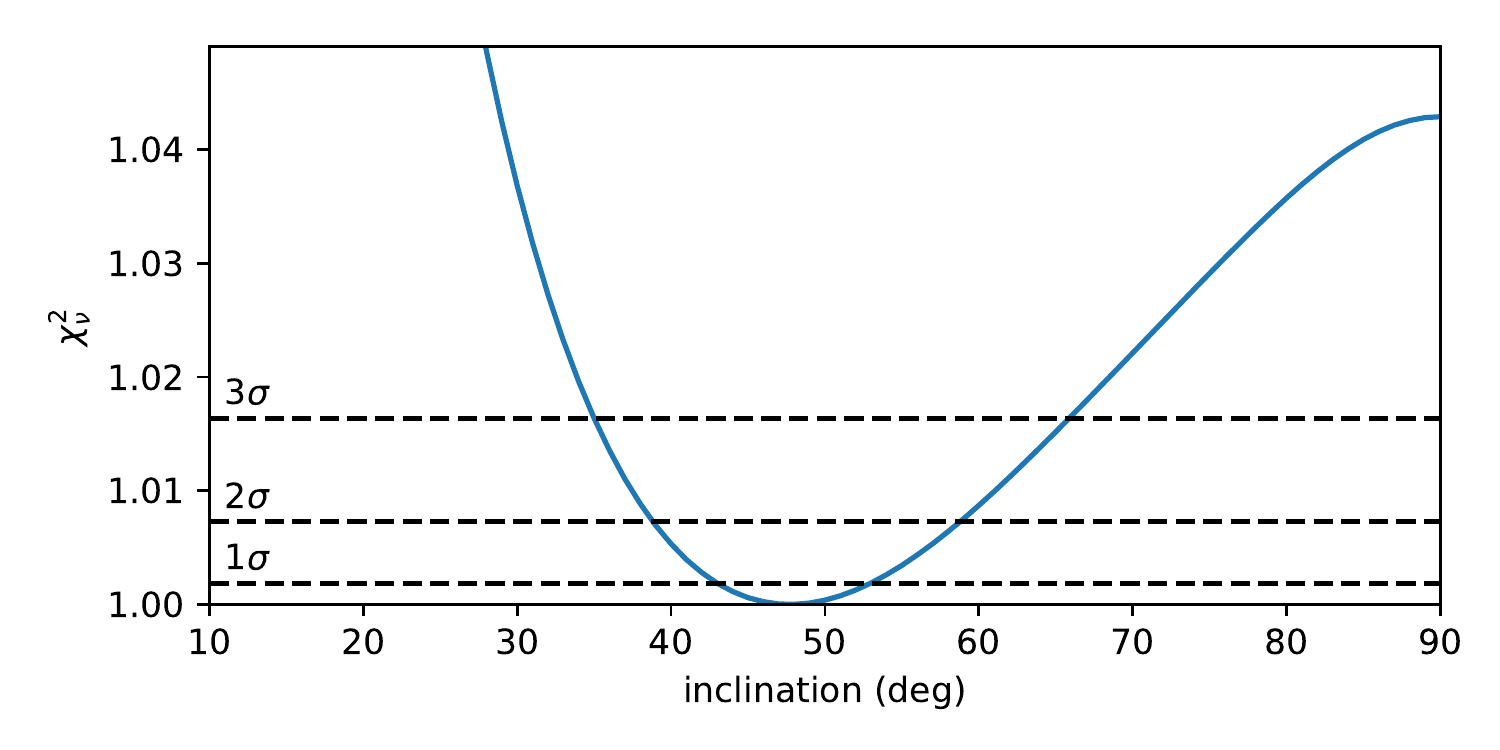}
\caption{Reduced $\chi^2$ versus inclination angle $i$. Top panel - models derived adopting the RV variation of \citet{linder08} for the broad-line star. Bottom panel - models for which the broad-line star was assumed to be stationary.\label{i_fitting_fig}}
\end{figure}

The inclination of the rotation axis of the broad-line star was one again derived using the grid search as above. This time, all models could be fit with a target reduced $\chi^2_r$ of 1.0, and the model with maximum entropy produced a best inclination of 48$^\circ$.  
We again calculated a second grid fitting to target maximum entropy, using the value from the best model in the first grid (-3072).   From this process (see Fig.~\ref{i_fitting_fig}) we find an inclination of $48 \pm 4^\circ$ (at $1\sigma$, ${^{+17}_{-12}}^{\circ}$ at $3\sigma$). 
If we use a turbulent broadening of 60\,\kms\ instead, we find a best fit value of ${47^{+5}_{-4}}^\circ$ (${^{+19}_{-12}}^{\circ}$ at $3\sigma$). 
This inclination is in good agreement with the $i=45\pm10^\circ$ we find based on our rotation period and the $10.5~R_\odot$ radius.  However it clearly disagrees with the $i=20\pm 5^\circ$ inferred from the alternate $22~R_\odot$ radius, which argues that the smaller radius is more likely correct. 

\begin{figure*}
\includegraphics[width=6.8in]{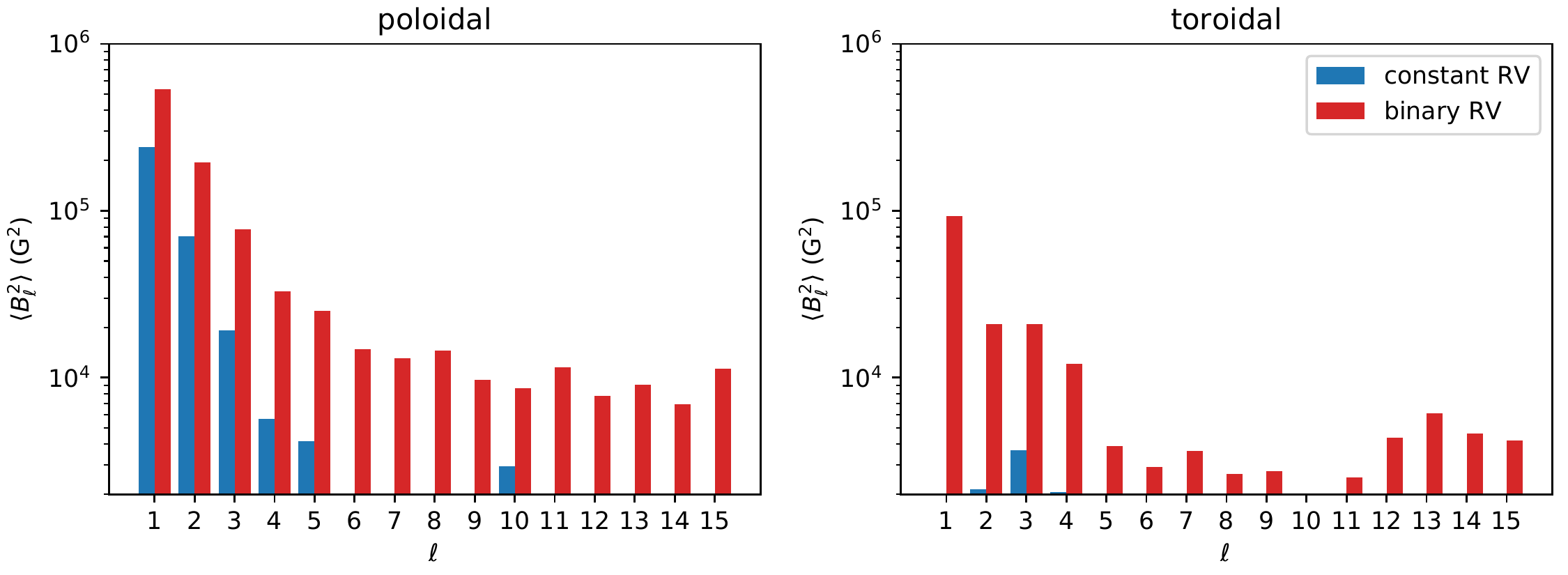}
\caption{
Comparison of the magnetic energies by angular degree $\ell$ for magnetic maps with (red) and without (blue) the orbital velocity for the broad-line star.  The poloidal (left) and toroidal (right) components of the field are shown.  Bars indicate energy in spherical harmonics of degree $\ell$.  {Here the magnetic energy is approximated by $\langle B^{2} \rangle = \oint \mathbf{B} \cdot \mathbf{B}\, \mathrm{d}\Omega / 4\pi$.}
}\label{energy_plot_fig}
\end{figure*}

The final magnetic map is presented in Fig.~\ref{fig-ZDImaps} (right panels), and the corresponding fits to the LSD Stokes $V$ profiles are presented in Fig.~\ref{fig-ZDIfits}.  
{An illustration of the strengths of the spherical harmonic coefficients of this map is shown in Fig.~\ref{energy_plot_fig}.}
This magnetic map is predominantly poloidal, corresponding to 96\% of the magnetic energy (as estimated from $\langle B^2 \rangle = \oint \mathbf{B}\cdot\mathbf{B} \rm{d}\Omega$).  The dipole ($\ell=1$) contains 69\% of the poloidal energy, while the quadrupole ($\ell=2$) contains 20\%, and the octupole ($\ell=3$) contains 6\% of the poloidal energy.  Thus the magnetic field is inferred to be largely dipolar with an important quadrupole component.  The magnetic energy is only 1.6\% symmetric about the rotation axis (as defined by $m=0$ spherical harmonics), so the field is almost entirely non-axisymmetric.  The tangential components of the magnetic field are weaker than the radial component, and in particular the tangential dipolar and quadrupolar components are weaker than would be expected from the radial component for a simple potential dipole or quadrupole.  If we use the radial dipolar component to estimate dipole quantities, the strength at the magnetic pole is 843 G, and the obliquity is 89$^\circ$ from the rotation axis.  The magnetic field has a surface averaged (unsigned) magnetic field strength $\langle B \rangle =  \oint |\mathbf{B}| \rm{d}\Omega / 4\pi $ of 517 G, or 722 G over just the fully visible hemisphere.
If we assume a turbulent broadening of 60 \kms\ instead, we get a magnetic field that is 94\% poloidal and 1.7\% axisymmetric.  The poloidal energy is 70\% dipolar, 20\% quadrupolar, and 5\% octupolar.  The radial dipole has a strength of 897 G and an obliquity of 90$^\circ$. The surface averaged strength is 547 G, and the average on the fully visible hemisphere is 760 G.  Thus the uncertainty in the turbulent broadening introduces a $\sim$6\% uncertainty in the magnetic field strength, while the geometry is largely unaffected.  

There are some important similarities and differences between the magnetic maps including an orbital velocity variation (Sect.~\ref{sect-ZDI1})  and with a fixed orbital velocity (this Section).  This is obvious from inspection of the magnetic maps (Fig.~\ref{fig-ZDImaps}), and we compare the geometry more quantitatively in Fig.~\ref{energy_plot_fig}.  Both maps require a strong radial dipole with a large obliquity in order to explain the relatively simple Stokes $V$ profiles that reverse sign.  Both maps also contain a significant quadrupole field that provides some asymmetry.  However, the poloidal component of the field in the map including binary motion contains a large amount of weaker small-scale structure superimposed on the radial dipole.  This is clear in Fig.~\ref{fig-ZDImaps}, and can be seen in Fig.~\ref{energy_plot_fig} where the poloidal energy decreases with $\ell$ but then plateaus at a non-zero value for higher $\ell$.  Typically this small scale structure is understood to be a consequence of fitting noise, which suggests we may be over-fitting portions of the line profile with this map, even at $\chi^2_r = 1.5$.
Both maps are largely poloidal, but there is clearly more toroidal energy in the binary RV map (Fig.~\ref{energy_plot_fig} right panel).  In Stokes $V$, which is sensitive to the line-of-sight component of the magnetic field, the toroidal magnetic field is only detectable near the limb of the star.  The binary RVs shift much of the observed Stokes $V$ signal towards the edge of the line at some phases, i.e.\ closer to the limb, which likely drives the increase in the toroidal field.  The energy in higher $\ell$ toroidal modes seems to largely help concentrate the azimuthal field in a few stronger spots, which would only be clearly  visible in Stokes $V$ at some phases.  

There is an important difference in the overall strength of the magnetic maps and this, unlike the geometry, is sensitive to the disentangling of the Stokes $I$ profiles.  The Stokes $V$ profiles are interpreted (and modeled) with reference to the Stokes $I$ line strength.  In disentangling, the relative strengths of the blended lines depend on the motions assumed.  When we disentangle the profiles adopting the orbital RV variation of the broad-line star, the broad-line component is weaker and the narrow-line component is stronger.  In that solution, at larger RVs, the full depth of the line near the narrow-line component is attributed to that component.  When we disentagle with a constant RV for the broad-line component, the narrow-line becomes a dip on top of a stronger line at all phases. As a consequence of the deeper line and larger $v\sin i$ in the constant RV disentangled profiles, fitting the Stokes $I$ line requires almost twice the strength relative to the binary RV disentangled profiles.  This larger line strength produces a magnetic map with a weaker magnetic field, which drives the difference in strengths apparent in Figs.~\ref{fig-ZDImaps} and \ref{energy_plot_fig}.

Overall, the ZDI map with a constant velocity provides a substantially better fit to the data than the map including the literature orbital velocities, both qualitatively and quantitatively (as indicated by $\chi^2_r$).  The magnetic map with constant velocity is also simpler, has a higher entropy, and is more consistent with the magnetic geometries of other known O and B stars.  Thus we strongly favour the magnetic map with a constant velocity, although this implies we may need to reconsider the architecture of the Plaskett system.

\section{Magnetosphere}\label{magnetosphere_sect}

\subsection{Predicted properties}

The common picture of the magnetosphere of a strongly magnetic star undergoing rapid rotation is that wind plasma is centrifugally supported and magnetically confined to accumulate into dense regions (or clouds) along gravitocentrifugal potential minima, which co-rotate with the host star \citep[e.g.][]{shore90, shore93, townsend05, uddoula08,petit13}. Magnetic confinement of the wind should occur if the local magnetic energy density is stronger than the local wind energy density, as characterised by the magnetic confinement parameter $\eta_* = B_{\rm eq}^2 R_\star^2/\dot{M}v_\infty$ \citep{uddoula02}, given the star's equatorial surface field strength ($B_{\rm eq}=B_{\rm d}/2$, for a dipole), the stellar equatorial radius $R_\star$, and the wind terminal momentum ($\dot{M}v_\infty$, for a magnetically unperturbed wind feeding rate $\dot{M}$ and the wind terminal velocity $v_\infty$). The wind is expected to be confined out to a distance where the energy density is balanced by the magnetic energy density (the Alfv{\' e}n radius, given by $R_{\rm A}/R_\star \sim 0.3 + (\eta_* + 0.25)^{1/4}$ for a dipole field in the magnetic equatorial plane; \citealt{uddoula02}). At further distances the wind dominates and the magnetic field lines are dragged with the wind and stretched to open field lines. The wind plasma is centrifugally supported beyond the Kepler, or co-rotation, radius $R_{\rm K}= 3/2\omega^{-2/3}R_{\rm p}$, where $\omega$ is the rotational frequency of the star and $R_{\rm p}$ is the polar radius; \citealt{uddoula08}). Inside the Kepler radius magnetically confined plasma is expected to fall back onto the star on the free-fall time-scale \citep{uddoula02}, while beyond this distance the plasma is centrifugally supported against infall. If $R_{\rm A}$ is beyond $R_{\rm K}$ then the centrifugally supported plasma accumulates, forming magnetospheric clouds as observed, for example, in the archetypical magnetic Bp star $\sigma$ Ori E \citep{1978ApJ...224L...5L,townsend05,2015MNRAS.451.2015O} and other rapidly rotating B stars \citep[e.g.][]{grunhut12b,2017MNRAS.465.2517W,2021MNRAS.504.3203S}.

Theoretical estimates of the magnetospheric properties of the broad-line star rely on its physical properties, in particular its wind characteristics ($\dot M$ and $v_\infty$, as would be seen in the absence of a magnetic field), its radius ($R_\star$), and its magnetic field strength ($B_{\rm d}$). Using the wind properties determined from the recipe of \citet{vink01} and $T_{\rm eff}=33$~kK, $\log L/L_\odot=5.1$, and $M_\star=56~M_\odot$, we obtain $\log\dot M=-7.1$ [$M_\odot$/yr] and $v_\infty=3500$~km/s. In combination with a $\sim 1$~kG magnetic dipole and $R_\star=10.8\pm 1.4~R_\odot$ \citep[][where the uncertainties correspond to the min/max radii from their fixed luminosity and the uncertainty in $T_{\rm eff}$]{linder08}, we obtain a minimum $\eta_\star=380$, and a minimum Alfv\'en radius $R_{\rm A}=4.4~R_\star$. Adopting $v\sin i = 360$~km/s from the ``stationary" disentangling solution, we obtain an oblateness $R_{\rm pol}/R_{\rm eq} = 0.90\pm 0.03$, $i = 46\pm 10\degr$, and Kepler radius $R_{\rm K} = 1.55\pm 0.2 R_{\rm eq}$.

As $R_{\rm K} < R_{\rm A}$, the magnetospheric parameters of the broad-line component are in agreement with the requirement for the formation of a rigidly-rotating, centrifugal magnetosphere \citep[CM; ][]{townsend05,petit13}. The broad-line component of Plaskett's star is, so far, the only O star known to host a CM, due to its combination of relatively strong magnetic field and particularly rapid rotation. However, it should be borne in mind that, due to the discrepancies between the properties inferred by \citet{linder08} via dynamics versus spectroscopy { and in light of inconsistent orbital properties suggested in this work versus previous work}, there is considerable uncertainty in the exact values of $R_{\rm A}$ and $R_{\rm K}$.

\begin{figure*}
\centering
\includegraphics[width=3.4in]{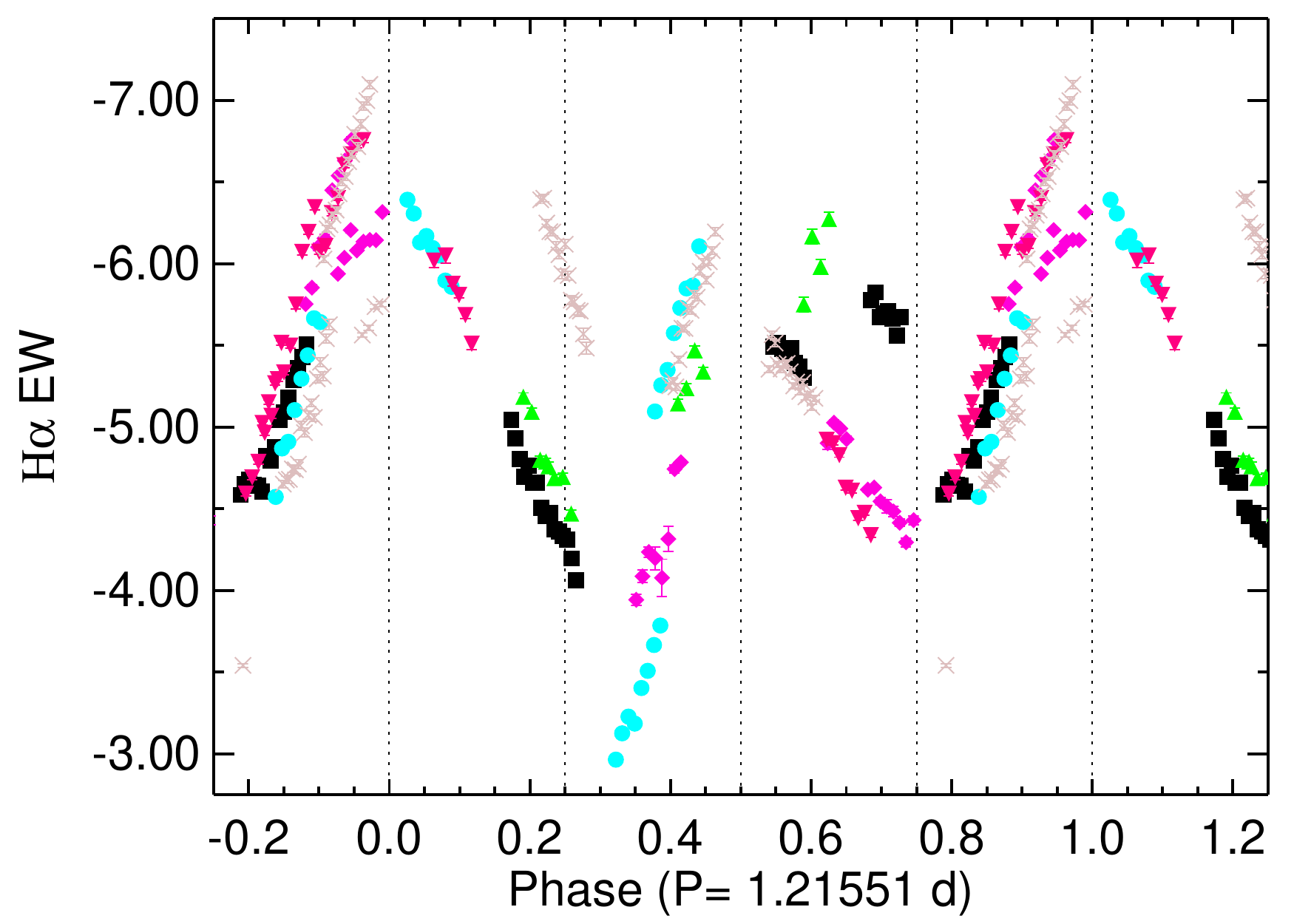}
\includegraphics[width=3.4in]{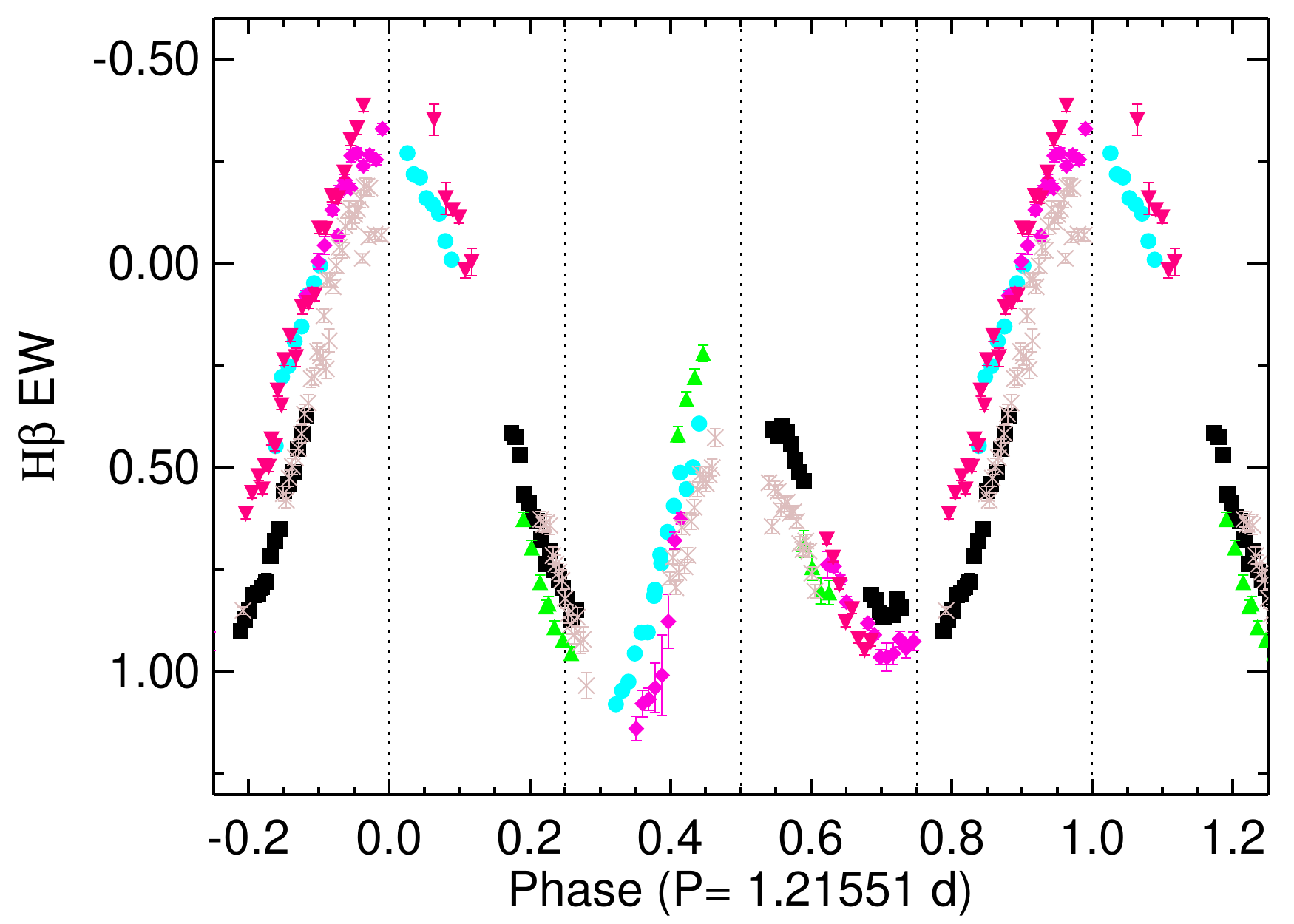}
\includegraphics[width=3.4in]{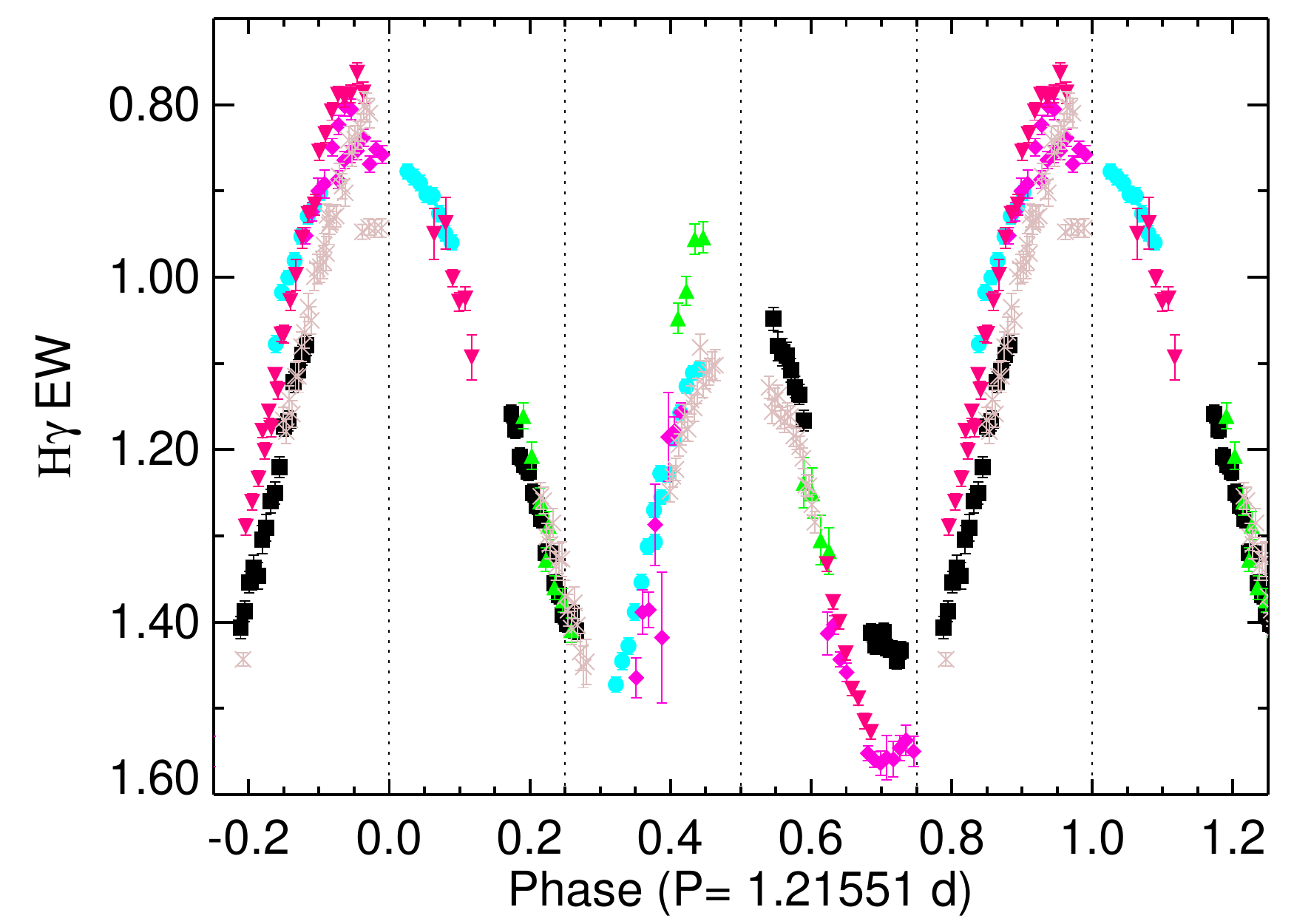}
\includegraphics[width=3.4in]{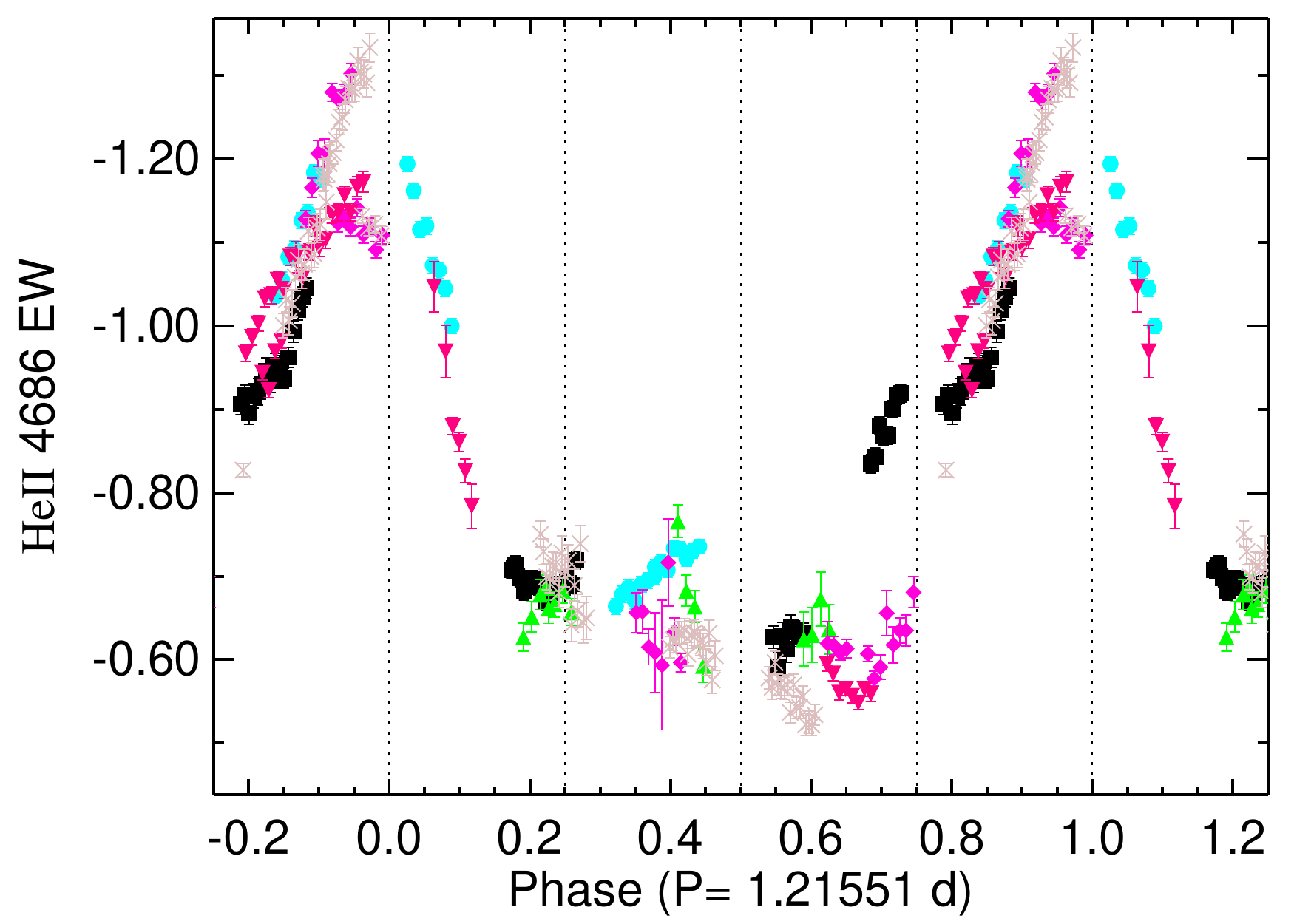}
\caption{Phased EW measurements (in \AA) from H$\alpha$, H$\beta$, H$\gamma$ and He\,{\sc ii} $\lambda$4686. The axis is inverted (EW increases downwards) to reflect the contribution from emission (emission increases upwards). Dotted lines are also included to indicate phases 0.0, 0.25, 0.5, 0.75, and 1.0. Different colours represent different epochs of observation.}\label{all_ew_fig}
\end{figure*}

\subsection{Observed properties}
In agreement with the results derived above, \citet{grunhut13} previously noted the presence of high-velocity emission, consistent with the presence of a CM surrounding the star. With the spectroscopic dataset acquired here, and with the guidance of the theoretical calculations described above, we aim to characterise the properties of the magnetosphere. 

\subsubsection{Equivalent width variations}

To characterise the observed magnetospheric properties, we first analysed the EW variations (see Sect.~\ref{eqw_sect}) of several hydrogen Balmer lines and the He\,{\sc ii} $\lambda$4686 line. The EW variations are illustrated in Fig.~\ref{all_ew_fig}. The Balmer lines all show double-peaked emission variation, with maximum emission occurring at rotation phase 0, and a secondary peak occurring at phase 0.5. Emission minima occur at phases $\sim$0.30 and $\sim$0.75. H$\alpha$ shows the most significant variation with a maximum peak-to-peak amplitude of about 4~\AA, while an average peak-to-peak variation is about 2.5-3\,\AA; however,  H$\alpha$ also shows significant systematic differences from one epoch to another. 
(Each epoch corresponds to a different observing run, and while one observing run typically spans an orbital period, there are large time gaps between runs.) 
This can also be seen in the EW curves of H$\beta$ and H$\gamma$, where most of the epoch-to-epoch changes occur around emission maximum and emission minima. Phase $\sim$0.30 shows maximal absorption for the Balmer lines. These variations are consistent with the general expectations from magnetosphere models and observations of other rapidly-rotating stars with high obliquity angles: the magnetosphere forms two higher-density regions (clouds) that produce double-peaked emission variations \citep[e.g.][]{townsend05, townsend08, oksala12,2019MNRAS.490..274S}. However, the He\,{\sc ii} EW measurements present a somewhat different behaviour. Maximum emission occurs at phase 0, followed by a rapid decrease until phase $\sim$0.20. After this the emission continues to slowly decrease until it reaches minimum emission around phase 0.75. There is no obvious secondary emission peak, although one epoch shows a brief rise in emission around phase 0.45. 

\begin{figure*}
\centering
\includegraphics[width=2.25in]{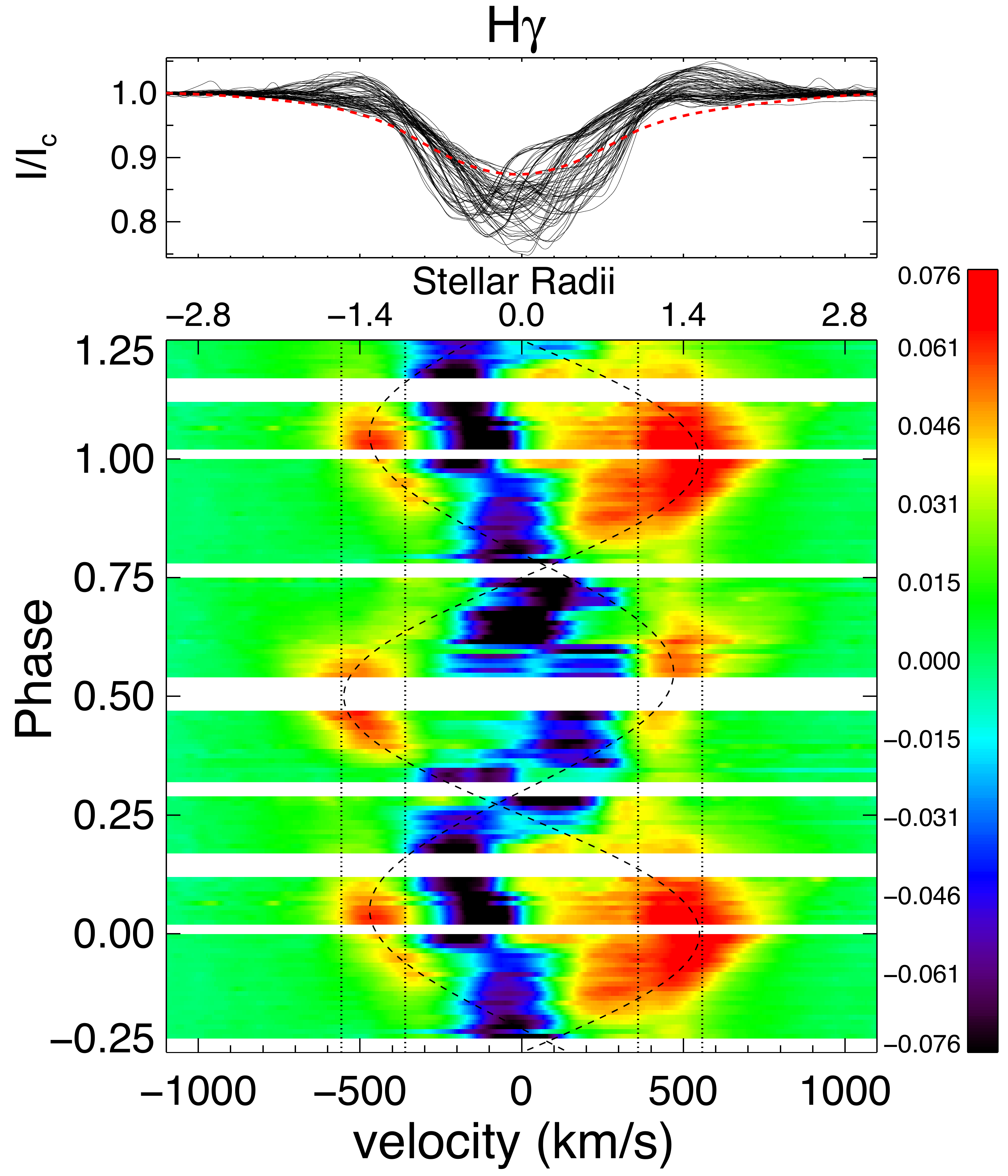}
\includegraphics[width=2.25in]{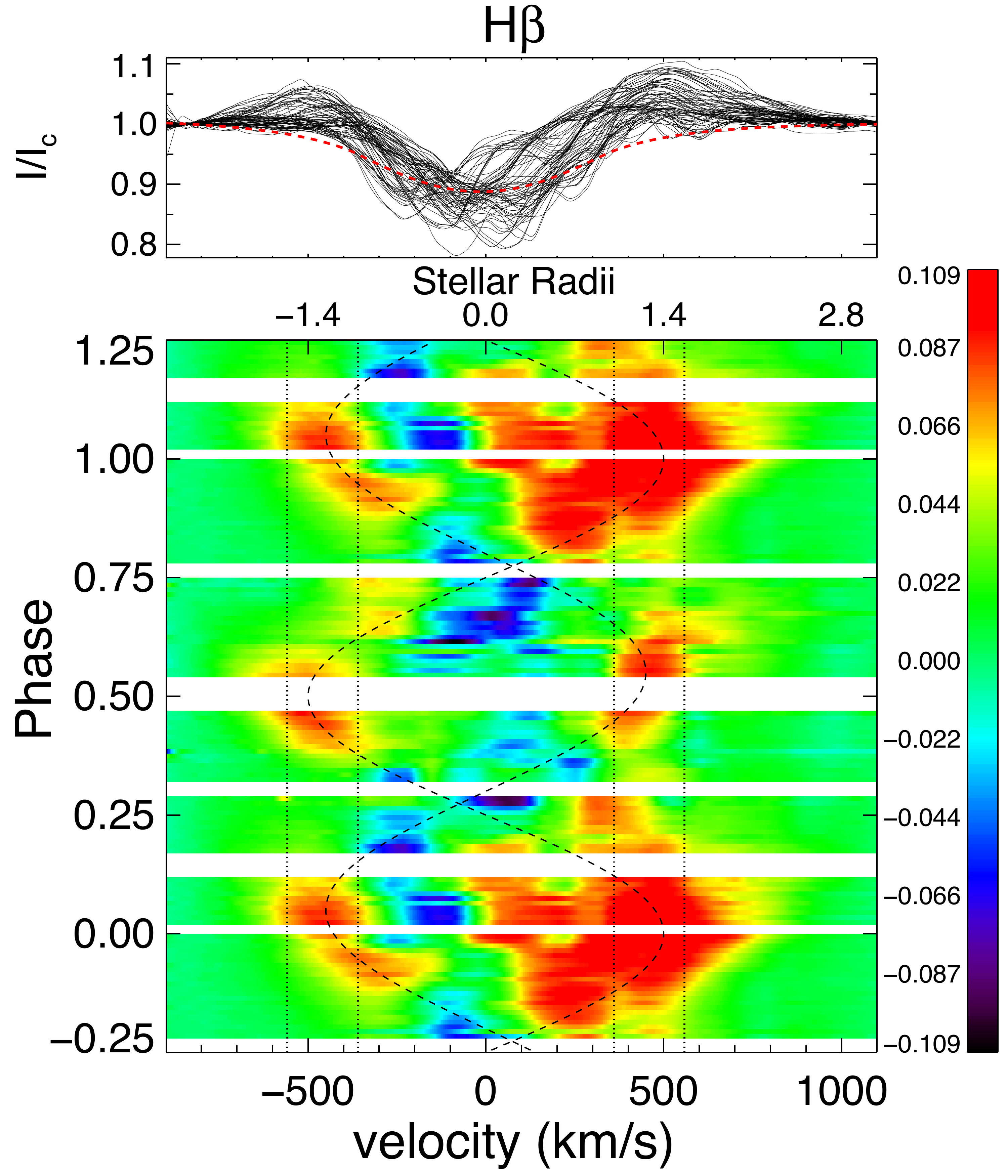}
\includegraphics[width=2.25in]{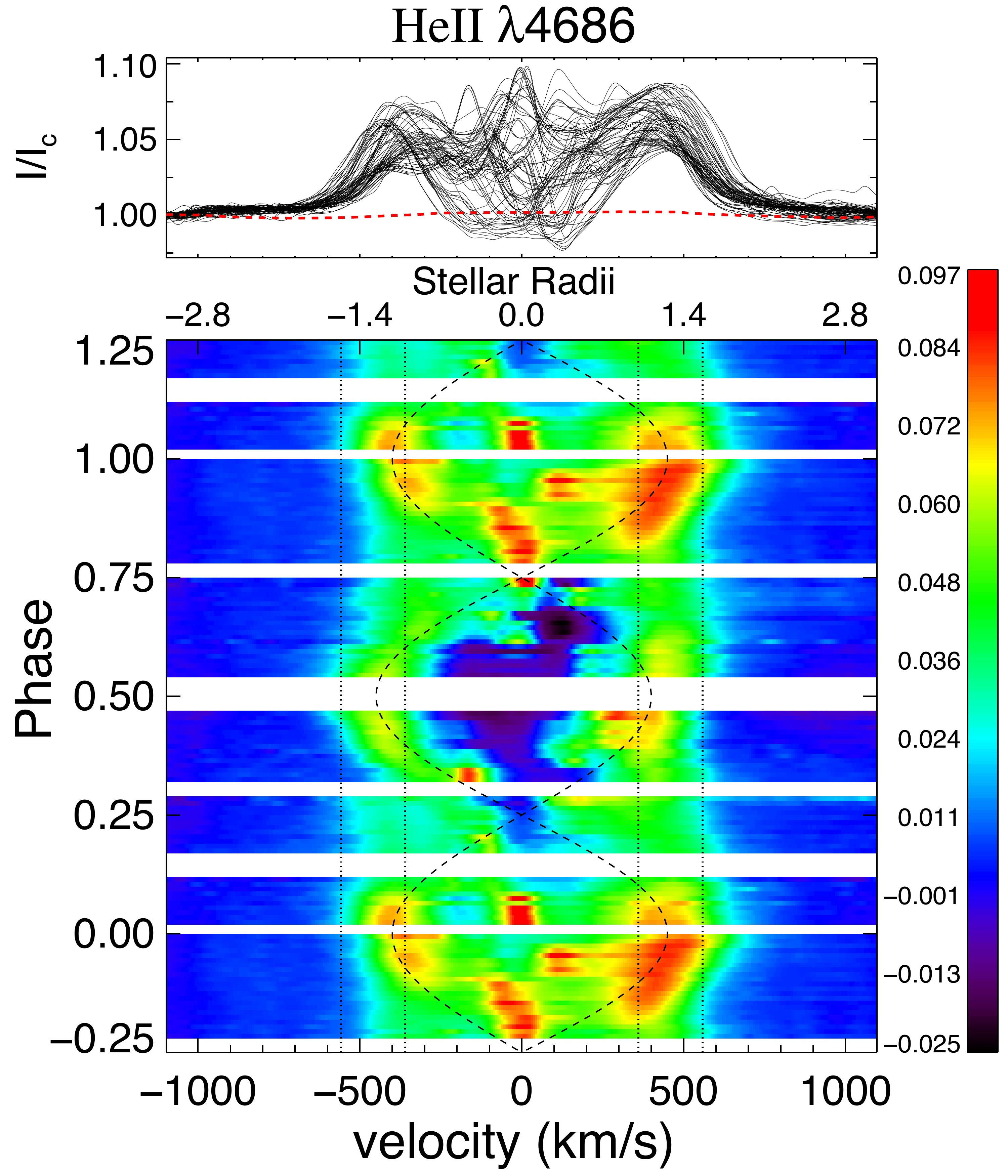}\\
\caption{Variations of the circumstellar magnetosphere phased with the rotation period. Shown are the differences between the observed profiles and a NLTE {\sc tlusty} photospheric model to highlight the full emission contribution of the magnetosphere. The vertical dotted lines indicate at lower velocity/radii reflect the projected rotational velocity (and radius) of the broad-line star, while the higher velocity lines represent our estimated Kepler radius ($\sim$1.55\,\rstar). Also shown are dashed curves to highlight the perceived motion of the emission features as discussed in the text. }\label{all_dyn_fig}
\end{figure*}

\subsubsection{Line profile variations}
We next studied the line profile variations (LPVs) of these emission lines. As shown in previous studies \citep[e.g.][]{townsend05b, bohlender11, grunhut12b, oksala12, rivinius13a, Sikora2016,2017MNRAS.465.2517W,2019MNRAS.490..274S}, emission LPVs resulting from CMs show characteristics that are distinct from other kinds of emission line variability. We investigated the LPVs by computing dynamic spectra phased with the 1.21551\,d period for H$\beta$, H$\gamma$ and He\,{\sc ii} $\lambda$4686, as shown in Fig.~\ref{all_dyn_fig} (H$\alpha$ is omitted due to the larger variation between rotation cycles). The dynamic spectra were computed by subtracting the theoretical photospheric line profile derived from a {\sc tlusty} model atmosphere \citep{lanz03} and are displayed in such a way as to show the full emission contribution from the magnetospheric plasma.  We note that the line profile of the narrow-line star was not removed or otherwise suppressed; hence it and its 14.4\,d orbital motion are evident in the figure in the form of dark traces occupying the inner $\pm 200$~km/s of the dynamic spectra. \footnote{We did attempt to remove the profile of the narrow-line star by fitting and subtracting, but the result was not really any better.}

The results show clear evidence of two nearly diametrically opposite emission features, most clearly seen in H$\beta$ and H$\gamma$, which we interpret as dense clouds in co-rotation with the star. The motion and intensity of these features are generally consistent from one epoch to another, although small differences in the line profiles are observed, as already mentioned in the analysis of the EW variations.  At phases 0 and 0.5 these clouds appear to be projected onto the sky to their fullest extent and thus show the most emission. At phase 0, the largest cloud appears at its maximum positive velocity, while the other cloud appears at its maximum negative velocity. At phase $\sim$0.5, the clouds appear at opposite velocities (and sides of the star) compared to phase 0, and so the most prominent cloud reaches its maximum negative velocity, while the other cloud reaches its maximum positive velocity. It should be noted that both clouds appear brightest at phase 0. At phase 0.5, the emission intensity of the largest cloud is considerably less than at phase 0, while there is also a slight difference for the other cloud. 
{To constrain the phases of occultation, we fit sinusoidal curves (included in Fig.~\ref{all_dyn_fig}) that indicate the expected motion of the clouds if they are in rigid corotation with the stellar surface, which appears to be the case. The curves were selected by eye such that velocity extrema correspond to the middle of the emission features, when seen in maximal emission. }
The less dense cloud crosses $v_{\rm sys}=0$\,\kms\ at phase $\sim$0.30 (travelling redward), while the denser cloud appears to cross at phase $\sim$0.75 (again travelling redward). This is consistent with the emission minima observed in the EW variations. The phase separation suggests that the two clouds are not exactly diametrically opposite, but are rather separated azimuthally by $\sim$160\degr. 
{This is likely due to the impact of the quadrupolar component of the magnetic field, and can be seen qualitatively in Fig.\ \ref{fig-ZDImaps}, top panel, where the positive magnetic region covers a wider area in phase than the negative magnetic region.}
The Balmer line dynamic spectra show absorption features about the phases where the stellar disc is occulted by the passing of the dense clouds; however, the interpretation is complicated by the fact that the spectral features of the narrow-line component could not be fully removed, despite our best attempts, and therefore it causes additional incoherent features within $\pm v\sin i$ of the broad-line star. 
We note that similar migrating absorption features were also found by \citet{palate14}. \\

The dynamic spectrum of the He\,{\sc ii} line displays emission features similar to those observed in the H lines. 
We again fit sinusoidal curves to the expected motion of the clouds if they are in rigid corotation, although the He\,{\sc ii} curves were selected to pass through the highest velocity emission features. 
According to these curves, there is a mismatch between the phases of occultation in He\,{\sc ii} relative to the Balmer lines. The orbital motion of the two He\,{\sc ii} clouds suggest they cross $v_{\rm sys}$ at phases 0.25 and 0.75. Furthermore, the high-velocity emission ($>|v|\sin i$) is generally seen at all phases, suggesting that, in addition to the dense clouds, further magnetospheric plasma is azimuthally distributed in a disc-like structure. 
This is only clearly seen in the dynamic spectrum of He\,{\sc ii}, shown in Fig.~\ref{all_dyn_fig}. 
The azimuthally distributed plasma also explains the behaviour of the He\,{\sc ii} EW variations. The EW minima in the Balmer lines are a consequence of the clouds passing behind the star. As the He\,{\sc ii} emission contains additional contribution from the disc, the occultation of the less dense cloud does not significantly reduce the total emission. Minimum emission is therefore reached at an earlier phase than in the Balmer lines as the densest cloud passes behind the star and this emission level remains relatively flat until this cloud reappears. The mismatch between the Balmer line curves and the He\,{\sc ii} curves and the differences in the emission level of the clouds when viewed at opposite quadratures are qualitatively explained by optical depth effects \citep[e.g.][]{grunhut12b}. 

Since the circumstellar plasma is bound in co-rotation, we can unambiguously map radial velocity onto the projected stellar radius, as indicated by the upper horizontal axis in Fig.~\ref{all_dyn_fig}. In doing so we can infer the spatial distances characterising the magnetosphere, with the only caveat that the emission is measured relative to a theoretical photospheric line profile and there is some outstanding uncertainty about the stellar parameters. The bulk of the emission extends to a maximum distance of $\sim$2.4\,$R_\star$ according to the H$\beta$ line and a maximum distance of $\sim$1.7\,$R_\star$ according to the He\,{\sc ii} line. The distance to the centre of the brightest part of the emission features, according to the H$\gamma$ line, is $\sim$1.45\,$R_\star$ for the strongest feature and $\sim$1.24\,$R_\star$ for the weaker one. Using the H$\beta$ line to carry out the same measurements results in slightly different values of 1.31 and 1.18\,\rstar\ for the larger and smaller feature, respectively. From the He\,{\sc ii} line, measuring from the high velocity emission peaks, we find slightly lower distances of 1.18 and 1.05\,$R_\star$. Emission appears at some phases for radii slightly less than 1\,$R_\star$ for H$\gamma$. This emission is more obvious for H$\beta$. Emission is not expected for $R<R_\star$, but the differences in the measured locations of the emission features between each line as well as the presence of emission within the projected stellar radius could result from scattering effects, a phenomenon often observed for optically thick Be star discs \citep{hummel92, hummel94}. Based on observations of other stars that host CMs, we expect to find maximum plasma build-up just beyond the Kepler radius ($R_{\rm emission~max.} \geq R_{\rm K}$ and $R_{\rm K} < 1.7\,R_\star$). 
The H$\beta$ and H$\gamma$ emission is consistent with this picture of material building up near the Kepler radius, while the He \,{\sc ii} $\lambda$4686 line suggests a somewhat different, more compact geometry. This might be a consequence of material spilling below $R_{\rm K}$, or potentially the contribution of the region of the star's dynamical magnetosphere located close to the stellar surface. 

\begin{table}
    \centering
    \caption{Distances of emission features in dynamic spectra, for three spectral lines. Columns indicate the radii corresponding to the maximum velocity at which detectable emission is observed, the velocity corresponding to maximum emission of the most prominent cloud, and the velocity corresponding to maximum emission of the less prominent cloud.} 
    \begin{tabular}{lccc}
    \hline
    line        & max.\ velocity & max.\ emission & secondary emission \\
    \hline
    H$\beta$    & $\sim$2.4\,\rstar & 1.31\,\rstar & 1.18\,\rstar\\
    H$\gamma$   & $\sim$2.2\,\rstar & 1.45\,\rstar & 1.24\,\rstar\\
    He\,{\sc ii}& $\sim$1.7\,\rstar & 1.18\,\rstar & 1.05\,\rstar\\
    \hline
    \end{tabular}
    \label{tab:my_label}
\end{table}

\section{Revisiting constraints on the magnetic field of the narrow-line star}
\label{sec_narrow_line_mag}

\citet{grunhut13} attempted to constrain the magnetic properties of the narrow-line component based on individual spectra and associated longitudinal field measurements. With the new model of the broad-line star's surface magnetic field structure, we decided to return to this issue. To this end we subtracted the corresponding Stokes $V$ signature determined from the best-fitting ZDI magnetic field model from each observed LSD profile. We then replaced each Stokes $I$ profile with the velocity-shifted disentangled Stokes $I$ profile of the narrow-line star. The resulting residual LSD profiles should represent the best description of the narrow-line star, uncontaminated by the presence of the broad-line star (in both polarized and unpolarized light). We then characterised these profiles using the same detection criteria previously discussed, and obtained non-detections for each observation.

We next constrained the upper limit of the allowed surface dipole field strength for the narrow-line star that could have remained hidden in the noise. This was accomplished using the statistical method discussed by \citet{neiner15}. This approach determines the upper limit based on the probability distribution of the FAPs obtained from a large number of synthetic Stokes $V$ profiles generated with random dipole field geometries that vary with polar field strength and contain the same noise characteristics as the observations. The Stokes $V$ profiles are constructed based on fits to the Stokes $I$ profile. The upper limits correspond to a 90 percent detection rate (i.e. a FAP $<10^{-3}$) such that the field should have been detected. The upper limits of individual profiles ranged from a lowest value of about 1200\,G (for the observation obtained on 2012-02-12) to a highest value of about 8300\,G (for the observation obtained on 2012-12-02), and are largely dependent on the S/N of the observation. Combining the individual probability distributions in the manner described by \citet{neiner15}, we find that a dipole field with a surface polar field strength of about 500\,G or higher should have been detected if the narrow-line star hosted such a field. 

{A magnetic field in the narrow-line star could not have been mistaken for the magnetic field from the broad-line star, as the Stokes $V$ signal consistently spans the width of the broad-line, implying the broad-line star has a magnetic field.  An additional detectable magnetic contribution to Stokes $V$ from the narrow-line star would have appeared as features that vary incoherently with the rotation period of the broad-line star. Such features are not observed.}

\section{Discussion and conclusions}\label{discussion_sect}

\subsection{Rotation, magnetic field, and magnetosphere of the broad-line star}

We have acquired and analyzed a new dataset of Plaskett's star consisting of 63 high resolution ($R\sim 65,000$) spectropolarimetric (Stokes $V$) observations obtained over 13 months in 2012 and 2013. The analysis of the spectra using the LSD procedure yields repeated detection of significant signal in the Stokes $V$ spectrum with width and centroid velocity coincident with the line profile of the broad-line (secondary) star. Measurements of the mean longitudinal magnetic field \bell\ from these data are found to vary with a single dominant period of $1.21551^{+28}_{-34}$\,d. 

Phasing the \bell\ measurements with this period yields a sinusoidal variation with central value of $-2 \pm 27$\,G and semi-amplitude of $513 \pm 41$\,G. The longitudinal magnetic field measurements - obtained over more than one year, corresponding to over 25 orbits of the binary and about 325 rotations of the broad-line star - phase coherently with the adopted rotational ephemeris, and exhibit scatter consistent with the independently-derived observational uncertainties. No systematic differences are observed between the longitudinal field variations measured at different epochs. These are normal, well-established characteristics of early-type stars hosting strong, dipolar, fossil magnetic fields. 

We performed direct fitting of the LSD Stokes $V$ profiles using Zeeman Doppler Imaging, and ultimately achieved a good fit to the profiles.
This yields a magnetic geometry that is predominantly a dipole (69\%) plus a modest quadrupole (20\%). However, some additional complexity in the magnetic field appears to be present, and the tangential components of the field are weaker than would be expected from the radial component of the dipolar or quadrupolar field.  While allowed by the model, very little toroidal field is found (4\%).  The global average field strength is 517 G, and the polar strength of the radial dipole is about 850~G with an obliquity of 89\degr.

One remarkable outcome of the Stokes $V$ modeling was our inability to fit the Stokes $V$ profiles while simultaneously considering the radial velocity (RV) variation of the broad-line star according to the 14.4\,d orbital period as inferred by \citet{linder08}. When the orbital RV variation was included, the theoretical best-fit Stokes $V$ profiles were frequently and obviously shifted in RV relative to their associated observations, with the profiles exhibiting the largest shifts being those corresponding to phases of RV extrema predicted by the \citet{linder08} binary model. {Even the great flexibility of a ZDI model could not account for these apparent RV shifts, resulting in a poor fit to the data and a distorted magnetic map.} We found that this discrepancy could be resolved by completely removing the 14.4\,d RV modulation from the Stokes $V$ model, i.e. by treating the broad-line star as stationary. This is an extremely puzzling result that will be analyzed and discussed in detail in a forthcoming paper.

Analysis of magnetospheric diagnostic lines (H$\alpha, \beta, \gamma$, He~{\sc ii}\, $\lambda 4686$) reveals that they also vary significantly and more-or-less coherently according to the $1.21551$\,d period derived from the magnetic measurements, implying that the bulk of the emission modulation is connected with the rotation of the broad-line star. The phased equivalent width measurements and dynamic spectra of the Balmer lines reveals the presence of two nearly diametrically-opposite emission clouds that should be located near the plane of the magnetic equator. In the He\,{\sc ii} $\lambda4686$ line, we find additional evidence for a disc-like structure, {although the two clouds in the magnetic equator are less clearly seen}. The rotational modulation of these clouds is found to enhance the absorption as the clouds travel in front of the star, to increase the emission as the clouds are maximally projected onto the sky, and to reduce the emission as the clouds pass behind the star.  

Our detailed interpretation of the origin of much of the variability observed from Plaskett's star is hampered by the large systematic uncertainty between the radius of the broad-line star implied by the dynamically-inferred mass and surface gravity ($22~R_\odot$) versus that implied by the combination of luminosity and temperature ($10.5~R_\odot$).  {However, given the apparent lack of RV variability in the broad-line star, previously determined dynamical masses are likely in error, and we tentatively conclude that the smaller radius based on temperature and luminosity is more likely correct.}  Nevertheless, it is clear that all modern datasets are strongly modulated according to the $1.21551$\,d period (and/or its first harmonic), and that this period can be naturally and self-consistently associated to the rotation of the magnetic broad-line star. { We therefore conclude that the rotation of the broad-line star coupled with its dipolar surface magnetic field is the underlying ``clock'' producing the bulk of the observed spectroscopic and photometric modulation of the Plaskett system.}

\citet{palate14} also found evidence for line profile variability in their spectroscopic dataset that was consistent with our derived $1.21551$~\,d period, although they concluded that this period leads to a large discrepancy between their radius estimates from $v \sin i$ and from surface gravity, and concluded that the rotation period of the broad-line star should be twice this period ($\sim$2.4\,d).  {However, the radius estimate from surface gravity depends on a mass estimate from the radial velocity curve.  As we have shown, the Stokes $V$ profiles does not appear to exhibit the large velocity variations reported previously, possibly calling into question the previous dynamical mass estimates as well as the radius inferred from $\log g$ and the dynamical mass.}  
Nevertheless, if we were to adopt the twice-longer period recommended by \citet{palate14} it would result in a complex, non-dipolar magnetic field structure. 
Using the ZDI code, we confirm this hypothesis as we were able to model the Stokes $V$ profiles with a dominantly quadrupolar field topology and a 2.2 d rotation period. The quality of fit of this model was similar to the 1.2\,d period model, although the dipolar component is very weak ($\sim$2 \% energy) and the entropy is much larger. While fossil fields with such complex structures are known to exist \citep[see, e.g.][]{donati06a,kochukhov11}, they are very rare. Given the uncertainties of the physical parameters of the components, and the coherence of a model of the system in which the broad-line star rotates with a $1.21551$\,d period, we consider adopting this model to be the most sound choice at present. {\em In that context, the broad-line component of Plaskett's star appears to be a rather typical magnetic early-type star, hosting a stable surface magnetic field that is approximately a tilted dipole, but with detectable departures from a pure dipole. The magnetic field directs and confines the star's wind, leading to a structured centrifugal magnetosphere that is clearly evident in optical emission lines.}

{ In this sense, the magnetic component of Plaskett's star shares many qualitative similarities with the archetypical CM-hosting star $\sigma$~Ori~E \citep{1978ApJ...224L...5L,oksala12,2015MNRAS.451.2015O}. On the other hand, a number of essential differences exist. Key among these is the significantly higher temperature of the photosphere and the wind of Plaskett, as established by the presence of developed lines of He~{\sc ii} in its spectrum. The rotation of the Plaskett broad-line star is also likely far closer to critical than $\sigma$~Ori E, and its weaker magnetic field and stronger wind result in a magnetic confinement parameter that is orders of magnitude lower. Therefore, notwithstanding the qualitative similarities, Plaskett's star clearly probes a very different quantitative region of parameter space than $\sigma$~Ori E (even considering the significant uncertainties in the physical parameters of the Plaskett system).
}

\subsection{Outstanding and unexplained characteristics of Plaskett's star}

Notwithstanding the reasonably coherent picture of a typical magnetic oblique rotator outlined above, there are a number of observational properties of Plaskett's star that our model does not naturally explain. These include:
\begin{itemize}
\item The rapid rotation of the magnetic star;
\item The epoch-to-epoch variations of mean phased light and EW curves;
\item The additional frequencies of photometric and spectroscopic variability reported in the literature;
\item The incompatibility of the period derived from {\em Hipparcos} photometry with that derived from {\em CoRoT, TESS}, and magnetic data;
\item The puzzling lack of evidence for RV variations of the observed Stokes $V$ profiles.
\end{itemize}

\subsubsection{Rotation of the secondary}

All studies of the HD\,47129 system agree that the spectrum is composed of two spectroscopic components: a narrow-lined ``primary" star (with $v\sin i\sim 75$~km/s), and a broad-lined ``secondary" star (with $v\sin i\sim 250-350$~km/s. In our study we clearly attribute the detected magnetic field to the broad-line star, and identify its rotation period to be $1.21551$\,d. Some magnetic A-type and B-type stars are known to exhibit such short rotational periods \citep{2018MNRAS.475.5144S, sikora2019a}, but they are both unobserved \citep[e.g.][]{wade15} and unexpected among the magnetic O-type stars. This is because O-type stars exhibit much stronger winds than intermediate-mass stars, resulting in magnetic braking spin-down times that are very short relative to their main sequence lifetimes \citep{uddoula08,petit13}. Indeed, the median rotation period of the known magnetic O stars is of several months, and the shortest known periods (Plaskett's Star excluded) are $\sim 7$\,d. The very short rotation period of the magnetic broad-line star - of similar duration to the shortest-known rotation periods of magnetic intermediate-mass stars - is puzzling. \citet{grunhut13} proposed that mass transfer following Roche-Lobe overflow in the binary might have been responsible for ``rejuvenating" the secondary's rotation. However, this picture may require revision given our lack of detection of RV variations of the LSD Stokes $V$ profiles.

\subsubsection{Epoch-to-epoch variations of phased light and EW curves}

In Sects.~\ref{period_sec} and \ref{magnetosphere_sect} we identified systematic differences in the phased photometric EW measurements and {\it CoRoT} photometry for data taken at the same rotation phase, but at different epochs. While some small-scale variations have been identified in similar photometric and spectroscopic measurements of other magnetic hot stars with clear evidence of magnetospheres (e.g. HD\,148937 - \citealt{wade12a}, HR\,5907 - \citealt{grunhut12b}),
the large systematic differences observed here are unprecedented. As shown by \citet{mahy11}, the {\it CoRoT} photometry presents periodic behaviour at a number of different frequencies, one of which is also the orbital period. This tells us that the photometry is modulated by several phenomena; the same could potentially be true of the spectroscopy.

In our attempt to better understand the origin of the epoch-to-epoch variations of the EW measurements, we investigated whether the deviations were a consequence of orbital variations. As evidenced by the periodograms shown in Fig.~\ref{periodogram}, there is no compelling evidence for modulation of the EW or \bell\ measurements with the orbital period. Nevertheless, we attempted to { assess the possibility of epoch-to-epoch variations with the orbital period by analyzing the EW measurements after prewhitening the data by subtracting a sinusoidal fit to the phased orbital variations. We found no significant reduction in the epoch-to-epoch variations in any of the spectral lines analyzed. }

Furthermore, a comparison of the spectra obtained at similar rotation phases and at similar orbital phases revealed just as much discrepancy as spectra obtained at similar rotation phases, but different orbital phases. Furthermore, we found no obvious correlation of the most discrepant epochs with a particular orbital phase (e.g. corresponding to conjunction and quadrature events). We do note that the data obtained at about orbital phase 0.5 (corresponding to secondary conjunction in the \citealt{linder08} solution), show maximum emission around rotational phase 0.25 (where we see the largest epoch-to-epoch differences). Similarly, measurements obtained around orbital phase 0.85 (which is close to primary conjunction), show the least emission compared to other data at the same rotation phase, suggesting there might be a trend. However, for measurements obtained at orbital phase 0.95, we see very few inconsistencies with measurements obtained at other orbital phases around rotation phase 1.0, and likewise we also find large excess emission for data taken around orbital phase 0.30 and rotation phase 0.70. We conclude that there is no obvious evidence that the epoch-to-epoch variations in the majority of EW measurements are due to orbital modulation. This does not rule out the possibility that these variations are not a consequence of colliding winds, although we suggest alternative explanations below.

One possibility is that these systematic variations are a consequence of physical changes in the structure of the magnetosphere, resulting from mass loss via centrifugal breakout or mass leakage (see \citealt{uddoula06, uddoula08, townsend13,2020MNRAS.499.5379S,2020MNRAS.499.5366O}). Unlike all other massive stars that are known to host a CM, the broad-line star is an O-star, and has a significantly higher mass-loss rate than B-stars. Therefore, mass-leakage should be enhanced relative to other CM-hosting massive stars. Using Eq.~A8 of \citet{townsend05} we indeed find that the expected breakout time for plasma trapped in the outer regions of the cloud to be on the order of a few days to about 50 days, which is in reasonable agreement with the time-scale in which we observe systematic differences. On the other hand, for plasma found nearer to the Kepler radius, the breakout time-scale is on the order of several hundreds of days, suggesting that most of the epoch-to-epoch variations may reflect on-going mass-leakage from the outer magnetosphere, rather than catastrophic breakout events that would affect the inner region.  However, it is also worth keeping in mind that the rigidly-rotating model \citep{townsend05} was developed for an ideal regime where $R_{\rm A} \gg R_{\rm K}$. Thus, the observed variability may just reflect inherent {\it dynamical} changes in the structure of the magnetosphere that likely affects the plasma distribution and its optical depth. { Here we may be seeing effects from both DM and CM components, with the former inherently stochastic, especially as might be stirred up by rotation (but not enough to prevent infall). The CM component is a likely fairly thin region fed by a strong wind, which probably does mean it has larger breakouts than would be seen in e.g. a B-type star. It may be worth noting that $B_{\rm K}\simeq 150$\,G, so just above the CM H$\alpha$ threshold \citep{2020MNRAS.499.5379S}. Emission would be much weaker in a B star with this $B_{\rm K}$, but the much stronger O star wind could result in a different presentation.}

Another possibility is that the variations are a result of systematic or episodic changes in the rotation of the broad-line star (see Sect.~\ref{period_change_disc} for further discussion related to period changes). This hypothesis is the subject of ongoing investigation that may be described in a future paper.

\subsubsection{Additional frequencies of photometric and spectroscopic variability}

\citet{mahy11} reported the detection of 43 ``main" frequencies in the CoRoT lightcurve of Plaskett's Star. Many of these represent harmonics of the $1.21551$\,d rotational period of the secondary star. In addition, two are attributable to modulation according to the 14.4\,d orbital period. Finally, there are two frequencies with considerable amplitudes corresponding to periods of 2.8\,d and 1.5\,d that were reported by \citet{mahy11}, and that are confirmed by Stacey et al. (in prep.) from their analysis of the new TESS photometry, that remain unexplained. In particular, a 2.8\,d period was previously reported by \citet{wiggs92} from analysis of the H$\alpha$ emission wings. 

The physical interpretation of these periodicities is not obvious. Neither appears to be a combination of the secondary rotational modulation and the orbital modulation. Nor do they appear to be harmonics of the orbital period. A remaining potential origin is the rotation of the narrow-line star. \citet{linder08} report the following characteristics of the narrow-line star: $\log L_\star/L_\odot=5.35$, $T_{\rm eff}=33.5$\,kK, $\log g=3.5$ (cgs), and $M_\star=54~M_\odot$. These parameters in turn imply radii of $R_\star=14~R_\odot$ (from $L$ and $T_{\rm eff}$ via the Stefan-Boltzmann law), and $R_\star=22~R_\odot$ (from $M_\star$ and $\log g$). \citet{linder08} also report a projected rotational velocity $v\sin i=60-75$~km/s for this component, which in combination with the inferred radii imply minimum rotational periods of 9.5\,d (for $R_\star=14~R_\odot$) and 14.8\,d (for $R_\star=22~R_\odot$). Hence the 2.8\,d and 1.5\,d periods cannot represent the rotation of the narrow-line star, even if we consider that they might be the first harmonics of the real rotational period.

As a consequence, at present we are unable to ascribe these two frequencies to a particular physical phenomenon in the system.

\subsubsection{The incompatible Hipparcos period}\label{period_change_disc}

As discussed in Sect~\ref{phot_data_sect}, the strongest peak in the periodogram of the {\it Hipparcos} photometry occurs at $1.2574\pm0.0001$\,d, a value that is similar to, but formally incompatible with, the adopted rotational period of the secondary. The {\it Hipparcos} data do not phase well with either the {\it CoRoT/TESS} period or the spectropolarimetric period. Moreover, the {\it CoRoT} and {\it TESS} lightcurves demonstrate very similar double-wave variations, while the {\it Hipparcos} data, when phased with its dominant period, appear to show a single-wave variation.

Period changes likely associated with magnetic field-wind coupling have already been measured for a number of stars, including HD\,37776 \citep{mikulasek08} and $\sigma$~Ori~E \citep{townsend10a}. However, in these particular cases, the rotational period was found to {\it increase} with time, which implies magnetic spin-down (or angular momentum loss), while our data suggest the opposite: the period would appear to have {\it decreased} with time.
\citet{mikulasek11b} suggests HD\,37776 and CU Vir show evidence for an unsteady period, exhibiting both an increase and decrease in their periods, while \citet{Shultz2019} found a consistent decrease in period for HD 142990.
Interpreted within the magnetically confined wind paradigm, this implies magnetic spin-up (or angular momentum gain). Assuming a constant steady change over this period of time implies not only a huge rate of change of the period of $\sim$0.67\,s\,d$^{-1}${ , but also a significant reduction in the tilt of the dipole relative to the rotation axis \citep{townsend08}}. { Since the more recent period estimates are consistent with each other, and such a large period change seems unlikely, the simplest explanation is that there is an error in the period inferred from Hipparcos data.  Perhaps the relatively sparse time sampling, lower S/N, and multiperiodic nature of the star contribute to this error, {although this will be investigated further in a future paper}. }

\subsubsection{The stable RVs of the observed Stokes $V$ profiles}

In Sect.~\ref{mag_sect} we modeled the LSD Stokes $V$ profiles of Plaskett's Star to infer the characteristics of the broad-line star's magnetic field. When we adopted the orbital solution of \citet{linder08} and applied it to our model we encountered large RV shifts of the model Stokes $V$ profiles relative to the observed profiles. The straightforward resolution was to remove the periodic variation of the broad-line star's RV from our model, i.e. to model it as a stationary star. This allowed us to model the Stokes $V$ profiles and to achieve a reasonable fit assuming a dipolar magnetic field rotating with the adopted 1.21551\,d rotational period.

To examine this phenomenon in more detail, we adopted the ephemeris of \citet{linder08} and computed a grid of ZDI models for RV amplitudes of the broad-line star ranging from zero to 200 km/s and a systemic velocity ranging from -100 to 100 km/s. The models were iterated to a uniform entropy value, and we examined the change in $\chi^2_r$. 

\begin{figure}
\centering
\includegraphics[width=9cm]{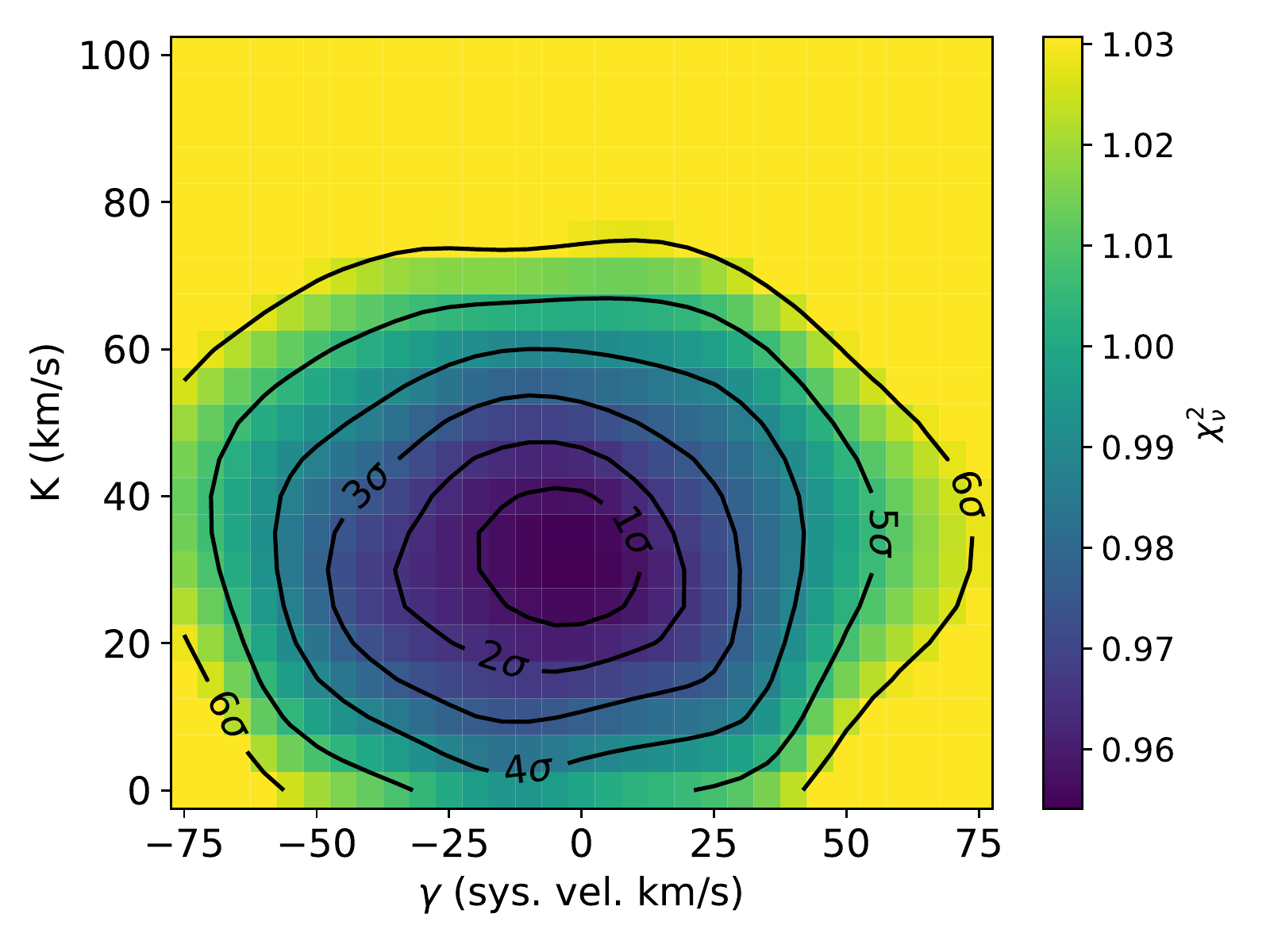}\\
\caption{Variation of $\chi^2_r$ achieved by ZDI models for a grid of systematic velocity $\gamma$ and RV semi-amplitude $K$ of the broad-line star.}\label{zdi_grid_fig}
\end{figure}

The results are illustrated in Fig.~{\ref{zdi_grid_fig}}. There are contours for different $\sigma$ levels, but given the systematic differences that exist between the observations and some models, strict interpretation of the $\chi^2_r$ contours in terms of probabilities is not straightforward.   

We find that the code can converge to a $\chi^2_r$ of 1.0 at an RV semi-amplitude $K$ of $100$~km/s, but it can't reach $\chi^2_r=1.0$ for $K=150~$km/s.  Essentially $K<70$~km/s gives good ZDI maps with a reasonable entropy.  $K$ between 70 and perhaps $125$~km/s is unlikely since those amplitudes require more complex maps with worse entropy.  And $K>150$~km/s doesn't fit the data well (it can't reach a reduced $\chi^2_r$ of unity). Recall that, while the formal best-fit RV amplitude derived from the ZDI modeling is 30~km/s, \citet{linder08} report $K=192.4\pm 6.7$~km/s. Hence our ZDI experiments indicate an RV variation amplitude of the broad-line star that is about one-sixth that previously reported.

The implications of these results are potentially profound, since they imply that circular polarization profiles associated with the secondary do not undergo the large orbital RV variations that have been consistently inferred from past studies. Given that the Stokes $V$ signatures are expected to be present only in the spectrum of the broad-line star, and are essentially unaffected by the complex circumstellar emission and variability of the system, { they provide a unique and valuable tracer of the secondary's dynamics. } The robustness and implications of this key result will be examined in further detail in a future paper.

\subsection{Conclusion}

The Plaskett's Star system continues to be an extremely interesting, complex, and poorly-understood example of massive star evolution in binary systems. We have performed the first reliable ZDI mapping of an O-type star's surface magnetic field, from which we find that the broad-line star hosts a magnetic field with characteristics that are typical of those of other hot, magnetic stars. It is the only known example of a magnetic O-type star to reside (probably) in a close binary system. Also, unlike any other known magnetic O star, it rotates rapidly, and therefore exhibits the magnetospheric properties observed in some hot magnetic B-type stars with centrifugal magnetospheres. These rotational properties are difficult to understand in the context of single-star evolution, but they may be comprehensible in terms of binary evolution. Finally, we have discovered that the Stokes $V$ profiles of the broad-line star are incompatible with the large RV variation of that star reported in numerous papers over the past century \citep[e.g.][]{plaskett22,bagnuolo92,linder08}. This may result in fundamental changes to our understanding of the composition, architecture, and history of the system, and will be the focus of a follow-up paper.

\section*{Data Availability}
The data underlying this paper are available from the Canadian Astronomy Data Centre (https://www.cadc-ccda.hia-iha.nrc-cnrc.gc.ca) and PolarBase (http://polarbase.irap.omp.eu). 

\section*{Acknowledgements}

This work has made use of data from the European Space Agency (ESA) mission
{\it Gaia} (\url{https://www.cosmos.esa.int/gaia}), processed by the {\it Gaia}
Data Processing and Analysis Consortium (DPAC,
\url{https://www.cosmos.esa.int/web/gaia/dpac/consortium}). Funding for the DPAC
has been provided by national institutions, in particular the institutions
participating in the {\it Gaia} Multilateral Agreement. GAW acknowledges support
in the form of a Discovery Grant from the Natural Science and Engineering Research
Council (NSERC) of Canada.
{ OK acknowledges support by the Swedish Research Council, the Royal Swedish Academy of Sciences, and the Swedish National Space Agency.}
{ MES acknowledges the financial support provided by the Annie Jump Cannon Fellowship, supported by the University of Delaware and endowed by the Mount Cuba Astronomical Observatory.} EA and CN acknowledge support by the "Programme National de Physique Stellaire" (PNPS) of CNRS/INSU co-funded by CEA and CNES. { We thank the referee, J.D. Landstreet, for a careful reading and thoughtful comments.}

\bibliographystyle{mnras}
\bibliography{bibtex} 

\section{Tables}\label{online_tab_sec}

\begin{table*}
\centering
\caption{Table of \bell\ measurements. Included is the HJD of mid exposure, the rotational phase according to Eq.~\ref{ephemeris}, the longitudinal field measurement (\bell), the corresponding uncertainty ($\sigma_B$) and the null field measurement (\nell) and corresponding uncertainty ($\sigma_N$).}\label{online_bl_tab}
\begin{tabular}{rrrrrrrrrrrr}
\hline
\multicolumn{1}{c}{HJD} & \multicolumn{1}{c}{Rot.} & \multicolumn{1}{c}{\bell} & \multicolumn{1}{c}{$\sigma_B$} & \multicolumn{1}{c}{\nell} & \multicolumn{1}{c}{$\sigma_N$} & \multicolumn{1}{c}{HJD} & \multicolumn{1}{c}{Rot.} & \multicolumn{1}{c}{\bell} & \multicolumn{1}{c}{$\sigma_B$} & \multicolumn{1}{c}{\nell} & \multicolumn{1}{c}{$\sigma_N$}\\
\multicolumn{1}{c}{(2450000+)} & \multicolumn{1}{c}{Phase} & \multicolumn{1}{c}{(G)} & \multicolumn{1}{c}{(G)} & \multicolumn{1}{c}{(G)} & \multicolumn{1}{c}{(G)} & \multicolumn{1}{c}{(2450000+)} & \multicolumn{1}{c}{Phase} & \multicolumn{1}{c}{(G)} & \multicolumn{1}{c}{(G)} & \multicolumn{1}{c}{(G)} & \multicolumn{1}{c}{(G)} \\
\hline
5961.8438	&	0.694	&	-223	&	177	&	-128	&	176	&	6272.1423	&	0.977	&	355	&	181	&	187	&	182	\\
5961.8735	&	0.719	&	93	&	171	&	-31	&	168	&	6282.9073	&	0.833	&	376	&	153	&	-242	&	152	\\
5966.8317	&	0.798	&	143	&	216	&	15	&	216	&	6282.9651	&	0.881	&	700	&	255	&	323	&	252	\\
5966.8614	&	0.822	&	213	&	246	&	-111	&	249	&	6284.0940	&	0.809	&	-197	&	174	&	-77	&	173	\\
5966.8916	&	0.847	&	453	&	199	&	187	&	200	&	6284.1383	&	0.846	&	450	&	189	&	58	&	187	\\
5966.9233	&	0.873	&	321	&	194	&	31	&	195	&	6288.0983	&	0.104	&	273	&	312	&	363	&	314	\\
5967.7533	&	0.556	&	-170	&	257	&	-4	&	256	&	6289.0829	&	0.914	&	120	&	180	&	137	&	181	\\
5967.7829	&	0.580	&	-979	&	206	&	233	&	207	&	6289.1265	&	0.950	&	682	&	203	&	38	&	201	\\
5969.7306	&	0.183	&	205	&	169	&	28	&	171	&	6289.9604	&	0.636	&	-266	&	152	&	-114	&	153	\\
5969.7603	&	0.207	&	279	&	168	&	45	&	167	&	6290.0040	&	0.672	&	93	&	156	&	-133	&	155	\\
5969.7904	&	0.232	&	261	&	166	&	-14	&	166	&	6343.8555	&	0.975	&	159	&	180	&	23	&	180	\\
5969.8201	&	0.256	&	196	&	168	&	-65	&	166	&	6351.8432	&	0.547	&	-456	&	247	&	140	&	246	\\
6000.4172	&	0.429	&	-1077	&	255	&	281	&	257	&	6351.8641	&	0.564	&	-454	&	235	&	-90	&	238	\\
6001.3654	&	0.209	&	-41	&	235	&	-83	&	237	&	6351.8858	&	0.582	&	-590	&	232	&	89	&	232	\\
6010.3588	&	0.608	&	-47	&	420	&	65	&	420	&	6351.9067	&	0.599	&	-572	&	236	&	-179	&	237	\\
6012.3442	&	0.241	&	67	&	207	&	-7	&	207	&	6352.8871	&	0.406	&	-368	&	242	&	446	&	244	\\
6196.0691	&	0.391	&	-251	&	151	&	-317	&	150	&	6352.9080	&	0.423	&	-431	&	268	&	282	&	271	\\
6196.1129	&	0.427	&	-722	&	151	&	244	&	150	&	6352.9292	&	0.440	&	-482	&	283	&	-38	&	281	\\
6198.0722	&	0.039	&	444	&	161	&	-83	&	162	&	6352.9501	&	0.457	&	-676	&	299	&	229	&	295	\\
6198.1159	&	0.075	&	174	&	165	&	-194	&	166	&	6353.8794	&	0.222	&	242	&	302	&	-744	&	302	\\
6199.0600	&	0.852	&	539	&	169	&	-285	&	169	&	6353.9003	&	0.239	&	703	&	333	&	-162	&	335	\\
6199.1044	&	0.889	&	661	&	159	&	-201	&	159	&	6353.9217	&	0.257	&	664	&	350	&	51	&	349	\\
6202.0790	&	0.336	&	-354	&	174	&	-29	&	176	&	6353.9426	&	0.274	&	247	&	429	&	254	&	423	\\
6202.1232	&	0.372	&	-576	&	165	&	211	&	166	&	6354.7202	&	0.914	&	248	&	247	&	199	&	248	\\
6261.1027	&	0.895	&	637	&	250	&	-27	&	249	&	6354.7411	&	0.931	&	481	&	272	&	-204	&	273	\\
6261.1488	&	0.932	&	422	&	209	&	-54	&	213	&	6354.7629	&	0.949	&	1014	&	300	&	191	&	298	\\
6262.0052	&	0.637	&	-758	&	290	&	137	&	290	&	6354.7838	&	0.966	&	523	&	308	&	-243	&	315	\\
6262.0747	&	0.694	&	-355	&	309	&	-364	&	314	&	6355.8654	&	0.856	&	92	&	288	&	448	&	286	\\
6262.1200	&	0.731	&	-247	&	347	&	-735	&	343	&	6355.8863	&	0.873	&	714	&	328	&	835	&	322	\\
6264.1052	&	0.365	&	172	&	595	&	-1608	&	599	&	6355.9085	&	0.891	&	821	&	308	&	120	&	306	\\
6264.1498	&	0.401	&	-568	&	907	&	-195	&	919	&	6355.9294	&	0.908	&	500	&	368	&	-496	&	370	\\
6272.0983	&	0.941	&	94	&	186	&	-229	&	186	\\												
\hline
\end{tabular}
\end{table*}

\begin{table*}
\tiny
\centering
\caption{Table of EW measurements. Included is the HJD of mid-exposure, the rotational phase according to Eq.~\ref{ephemeris}, the EW of H$\beta$, the corresponding uncertainty ($\sigma$), the EW of He\,{\sc ii} $\lambda$4686 and its corresponding uncertainty, and the EW of H$\alpha$ and its corresponding uncertainty.}\label{online_ew_tab}
\begin{tabular}{rrrrrrrrrrrrrrrr}
\hline
HJD & Rot. & H$\beta$ & $\sigma$ & He\,{\sc ii} & $\sigma$ & H$\alpha$ & $\sigma$ & HJD & Rot. & H$\beta$ & $\sigma$ & He\,{\sc ii} & $\sigma$ & H$\alpha$ & $\sigma$\\
(2450000+) & Phase & EW & EW & EW & EW & EW & EW & (2450000+) & Phase & EW & EW & EW & EW & EW & EW \\
\hline 
5961.8327	&	0.685	&	0.811	&	0.013	&	-0.835	&	0.011	&	-5.777	&	0.016	&	6196.0527	&	0.378	&	0.799	&	0.011	&	-0.711	&	0.009	&	-5.095	&	0.013	\\
5961.8401	&	0.691	&	0.825	&	0.013	&	-0.843	&	0.011	&	-5.824	&	0.016	&	6196.0637	&	0.387	&	0.734	&	0.011	&	-0.717	&	0.009	&	-5.256	&	0.013	\\
5961.8475	&	0.697	&	0.854	&	0.013	&	-0.881	&	0.011	&	-5.674	&	0.016	&	6196.0746	&	0.396	&	0.657	&	0.011	&	-0.707	&	0.009	&	-5.349	&	0.013	\\
5961.8550	&	0.703	&	0.867	&	0.013	&	-0.868	&	0.011	&	-5.696	&	0.016	&	6196.0855	&	0.405	&	0.593	&	0.011	&	-0.734	&	0.009	&	-5.575	&	0.013	\\
5961.8624	&	0.710	&	0.860	&	0.013	&	-0.869	&	0.011	&	-5.711	&	0.016	&	6196.0965	&	0.414	&	0.512	&	0.011	&	-0.733	&	0.009	&	-5.728	&	0.013	\\
5961.8698	&	0.716	&	0.860	&	0.013	&	-0.901	&	0.010	&	-5.665	&	0.016	&	6196.1074	&	0.423	&	0.552	&	0.011	&	-0.721	&	0.009	&	-5.848	&	0.013	\\
5961.8772	&	0.722	&	0.821	&	0.013	&	-0.917	&	0.010	&	-5.561	&	0.016	&	6196.1183	&	0.432	&	0.499	&	0.011	&	-0.731	&	0.009	&	-5.866	&	0.013	\\
5961.8847	&	0.728	&	0.842	&	0.013	&	-0.919	&	0.011	&	-5.673	&	0.016	&	6196.1292	&	0.441	&	0.392	&	0.011	&	-0.736	&	0.009	&	-6.106	&	0.013	\\
5966.8206	&	0.789	&	0.900	&	0.016	&	-0.907	&	0.013	&	-4.585	&	0.019	&	6198.0559	&	0.026	&	-0.270	&	0.011	&	-1.194	&	0.009	&	-6.391	&	0.013	\\
5966.8280	&	0.795	&	0.871	&	0.015	&	-0.917	&	0.012	&	-4.652	&	0.018	&	6198.0668	&	0.035	&	-0.219	&	0.011	&	-1.162	&	0.009	&	-6.306	&	0.013	\\
5966.8355	&	0.801	&	0.850	&	0.016	&	-0.895	&	0.013	&	-4.679	&	0.018	&	6198.0777	&	0.044	&	-0.211	&	0.012	&	-1.115	&	0.009	&	-6.130	&	0.014	\\
5966.8429	&	0.807	&	0.811	&	0.018	&	-0.916	&	0.015	&	-4.644	&	0.021	&	6198.0886	&	0.053	&	-0.160	&	0.012	&	-1.120	&	0.010	&	-6.169	&	0.014	\\
5966.8503	&	0.813	&	0.807	&	0.019	&	-0.921	&	0.016	&	-4.640	&	0.022	&	6198.0995	&	0.062	&	-0.145	&	0.012	&	-1.073	&	0.010	&	-6.095	&	0.014	\\
5966.8577	&	0.819	&	0.792	&	0.020	&	-0.931	&	0.016	&	-4.606	&	0.023	&	6198.1104	&	0.071	&	-0.122	&	0.012	&	-1.067	&	0.010	&	-6.050	&	0.014	\\
5966.8651	&	0.825	&	0.778	&	0.019	&	-0.946	&	0.016	&	-4.821	&	0.022	&	6198.1213	&	0.080	&	-0.055	&	0.012	&	-1.045	&	0.009	&	-5.896	&	0.014	\\
5966.8726	&	0.831	&	0.715	&	0.017	&	-0.935	&	0.014	&	-4.795	&	0.020	&	6198.1322	&	0.089	&	-0.010	&	0.011	&	-1.000	&	0.009	&	-5.858	&	0.013	\\
5966.8804	&	0.838	&	0.680	&	0.016	&	-0.954	&	0.013	&	-4.877	&	0.019	&	6199.0437	&	0.839	&	0.447	&	0.013	&	-1.036	&	0.010	&	-4.572	&	0.015	\\
5966.8879	&	0.844	&	0.650	&	0.015	&	-0.952	&	0.013	&	-5.042	&	0.018	&	6199.0546	&	0.848	&	0.277	&	0.012	&	-1.055	&	0.010	&	-4.869	&	0.014	\\
5966.8953	&	0.850	&	0.557	&	0.015	&	-0.937	&	0.012	&	-5.091	&	0.017	&	6199.0655	&	0.856	&	0.251	&	0.011	&	-1.083	&	0.009	&	-4.909	&	0.013	\\
5966.9027	&	0.856	&	0.540	&	0.014	&	-0.962	&	0.012	&	-5.180	&	0.017	&	6199.0764	&	0.865	&	0.190	&	0.012	&	-1.092	&	0.010	&	-5.103	&	0.014	\\
5966.9122	&	0.864	&	0.510	&	0.015	&	-0.993	&	0.013	&	-5.290	&	0.018	&	6199.0881	&	0.875	&	0.154	&	0.011	&	-1.126	&	0.009	&	-5.295	&	0.013	\\
5966.9196	&	0.870	&	0.453	&	0.015	&	-1.019	&	0.012	&	-5.360	&	0.017	&	6199.0990	&	0.884	&	0.077	&	0.012	&	-1.136	&	0.010	&	-5.438	&	0.014	\\
5966.9270	&	0.876	&	0.417	&	0.014	&	-1.035	&	0.012	&	-5.427	&	0.017	&	6199.1099	&	0.893	&	0.048	&	0.011	&	-1.184	&	0.009	&	-5.665	&	0.013	\\
5966.9345	&	0.882	&	0.373	&	0.015	&	-1.045	&	0.012	&	-5.506	&	0.017	&	6199.1208	&	0.902	&	0.005	&	0.011	&	-1.174	&	0.009	&	-5.642	&	0.013	\\
5967.7421	&	0.547	&	0.406	&	0.017	&	-0.627	&	0.014	&	-5.493	&	0.019	&	6202.0627	&	0.322	&	1.079	&	0.011	&	-0.664	&	0.009	&	-2.964	&	0.013	\\
5967.7495	&	0.553	&	0.420	&	0.021	&	-0.591	&	0.017	&	-5.513	&	0.024	&	6202.0736	&	0.331	&	1.045	&	0.012	&	-0.678	&	0.010	&	-3.125	&	0.015	\\
5967.7570	&	0.559	&	0.400	&	0.020	&	-0.628	&	0.016	&	-5.476	&	0.023	&	6202.0845	&	0.340	&	1.024	&	0.012	&	-0.686	&	0.010	&	-3.227	&	0.015	\\
5967.7644	&	0.565	&	0.414	&	0.019	&	-0.613	&	0.015	&	-5.451	&	0.022	&	6202.0954	&	0.349	&	0.954	&	0.012	&	-0.671	&	0.010	&	-3.185	&	0.014	\\
5967.7718	&	0.571	&	0.442	&	0.017	&	-0.639	&	0.014	&	-5.483	&	0.020	&	6202.1069	&	0.359	&	0.903	&	0.012	&	-0.691	&	0.009	&	-3.403	&	0.014	\\
5967.7792	&	0.577	&	0.481	&	0.016	&	-0.633	&	0.013	&	-5.392	&	0.018	&	6202.1178	&	0.368	&	0.903	&	0.011	&	-0.695	&	0.009	&	-3.508	&	0.013	\\
5967.7867	&	0.583	&	0.511	&	0.015	&	-0.628	&	0.012	&	-5.369	&	0.017	&	6202.1287	&	0.377	&	0.814	&	0.011	&	-0.699	&	0.009	&	-3.666	&	0.014	\\
5967.7941	&	0.590	&	0.533	&	0.015	&	-0.631	&	0.013	&	-5.302	&	0.018	&	6202.1396	&	0.386	&	0.713	&	0.012	&	-0.714	&	0.009	&	-3.785	&	0.014	\\
5969.7195	&	0.174	&	0.414	&	0.013	&	-0.708	&	0.010	&	-5.045	&	0.015	&	6261.0864	&	0.881	&	0.079	&	0.012	&	-1.129	&	0.010	&	-5.752	&	0.015	\\
5969.7269	&	0.180	&	0.425	&	0.013	&	-0.714	&	0.011	&	-4.931	&	0.016	&	6261.0973	&	0.890	&	0.074	&	0.014	&	-1.166	&	0.011	&	-5.853	&	0.017	\\
5969.7343	&	0.186	&	0.470	&	0.013	&	-0.698	&	0.011	&	-4.803	&	0.015	&	6261.1082	&	0.899	&	-0.005	&	0.019	&	-1.207	&	0.016	&	-6.101	&	0.023	\\
5969.7417	&	0.192	&	0.565	&	0.013	&	-0.682	&	0.010	&	-4.696	&	0.015	&	6261.1191	&	0.908	&	-0.045	&	0.021	&	-1.208	&	0.017	&	-6.155	&	0.025	\\
5969.7491	&	0.198	&	0.586	&	0.013	&	-0.696	&	0.010	&	-4.762	&	0.015	&	6261.1325	&	0.919	&	-0.131	&	0.013	&	-1.280	&	0.011	&	-6.449	&	0.016	\\
5969.7566	&	0.204	&	0.619	&	0.013	&	-0.696	&	0.011	&	-4.659	&	0.016	&	6261.1434	&	0.928	&	-0.177	&	0.014	&	-1.271	&	0.011	&	-6.537	&	0.016	\\
5969.7640	&	0.210	&	0.631	&	0.013	&	-0.689	&	0.010	&	-4.658	&	0.015	&	6261.1543	&	0.937	&	-0.203	&	0.015	&	-1.278	&	0.012	&	-6.630	&	0.017	\\
5969.7714	&	0.216	&	0.676	&	0.012	&	-0.686	&	0.010	&	-4.505	&	0.015	&	6261.1652	&	0.946	&	-0.264	&	0.016	&	-1.302	&	0.012	&	-6.757	&	0.018	\\
5969.7793	&	0.223	&	0.735	&	0.013	&	-0.670	&	0.010	&	-4.455	&	0.015	&	6261.9889	&	0.624	&	0.737	&	0.033	&	-0.620	&	0.026	&	-4.901	&	0.039	\\
5969.7867	&	0.229	&	0.703	&	0.013	&	-0.680	&	0.010	&	-4.473	&	0.015	&	6261.9998	&	0.633	&	0.742	&	0.014	&	-0.617	&	0.011	&	-5.026	&	0.017	\\
5969.7941	&	0.235	&	0.751	&	0.012	&	-0.689	&	0.010	&	-4.376	&	0.015	&	6262.0107	&	0.642	&	0.775	&	0.011	&	-0.608	&	0.009	&	-4.990	&	0.014	\\
5969.8016	&	0.241	&	0.773	&	0.013	&	-0.688	&	0.010	&	-4.362	&	0.015	&	6262.0216	&	0.650	&	0.829	&	0.014	&	-0.613	&	0.011	&	-4.926	&	0.017	\\
5969.8090	&	0.247	&	0.794	&	0.013	&	-0.686	&	0.010	&	-4.334	&	0.015	&	6262.0584	&	0.681	&	0.881	&	0.012	&	-0.607	&	0.009	&	-4.617	&	0.014	\\
5969.8164	&	0.253	&	0.820	&	0.012	&	-0.698	&	0.010	&	-4.311	&	0.015	&	6262.0693	&	0.690	&	0.909	&	0.012	&	-0.577	&	0.009	&	-4.629	&	0.014	\\
5969.8238	&	0.259	&	0.873	&	0.012	&	-0.690	&	0.010	&	-4.195	&	0.015	&	6262.0802	&	0.699	&	0.964	&	0.018	&	-0.591	&	0.015	&	-4.545	&	0.023	\\
5969.8313	&	0.266	&	0.848	&	0.013	&	-0.719	&	0.011	&	-4.060	&	0.016	&	6262.0911	&	0.708	&	0.964	&	0.034	&	-0.656	&	0.027	&	-4.513	&	0.041	\\
6000.3953	&	0.411	&	0.418	&	0.019	&	-0.766	&	0.019	&	-5.144	&	0.027	&	6262.1026	&	0.717	&	0.955	&	0.027	&	-0.618	&	0.022	&	-4.483	&	0.033	\\
6000.4099	&	0.423	&	0.332	&	0.018	&	-0.682	&	0.019	&	-5.238	&	0.025	&	6262.1135	&	0.726	&	0.919	&	0.019	&	-0.635	&	0.015	&	-4.414	&	0.023	\\
6000.4244	&	0.435	&	0.278	&	0.019	&	-0.664	&	0.019	&	-5.469	&	0.026	&	6262.1244	&	0.735	&	0.942	&	0.023	&	-0.635	&	0.018	&	-4.294	&	0.028	\\
6000.4390	&	0.447	&	0.220	&	0.019	&	-0.593	&	0.020	&	-5.339	&	0.026	&	6262.1374	&	0.746	&	0.925	&	0.023	&	-0.681	&	0.019	&	-4.430	&	0.028	\\
6001.3436	&	0.191	&	0.625	&	0.017	&	-0.627	&	0.017	&	-5.185	&	0.024	&	6264.0889	&	0.351	&	1.138	&	0.030	&	-0.656	&	0.024	&	-3.942	&	0.035	\\
6001.3581	&	0.203	&	0.695	&	0.017	&	-0.651	&	0.018	&	-5.093	&	0.025	&	6264.0998	&	0.360	&	1.077	&	0.033	&	-0.658	&	0.026	&	-4.086	&	0.039	\\
6001.3727	&	0.215	&	0.780	&	0.017	&	-0.678	&	0.018	&	-4.799	&	0.025	&	6264.1107	&	0.369	&	1.066	&	0.027	&	-0.615	&	0.021	&	-4.235	&	0.031	\\
6001.3872	&	0.227	&	0.832	&	0.018	&	-0.661	&	0.018	&	-4.760	&	0.025	&	6264.1216	&	0.378	&	1.039	&	0.060	&	-0.609	&	0.048	&	-4.197	&	0.070	\\
6010.3370	&	0.590	&	0.687	&	0.032	&	-0.625	&	0.032	&	-5.750	&	0.045	&	6264.1335	&	0.388	&	1.008	&	0.098	&	-0.593	&	0.078	&	-4.077	&	0.114	\\
6010.3516	&	0.602	&	0.743	&	0.032	&	-0.630	&	0.033	&	-6.167	&	0.045	&	6264.1444	&	0.397	&	0.876	&	0.066	&	-0.716	&	0.053	&	-4.315	&	0.079	\\
6010.3662	&	0.614	&	0.802	&	0.032	&	-0.673	&	0.032	&	-5.981	&	0.045	&	6264.1553	&	0.406	&	0.678	&	0.021	&	-0.633	&	0.016	&	-4.742	&	0.024	\\
6010.3807	&	0.626	&	0.805	&	0.029	&	-0.637	&	0.030	&	-6.273	&	0.040	&	6264.1661	&	0.415	&	0.624	&	0.014	&	-0.596	&	0.011	&	-4.784	&	0.016	\\
6012.3223	&	0.223	&	0.839	&	0.015	&	-0.686	&	0.015	&	-4.789	&	0.021	&	6272.0819	&	0.927	&	-0.068	&	0.013	&	-1.123	&	0.010	&	-5.937	&	0.015	\\
6012.3369	&	0.235	&	0.890	&	0.015	&	-0.667	&	0.016	&	-4.687	&	0.022	&	6272.0928	&	0.936	&	-0.184	&	0.012	&	-1.130	&	0.010	&	-6.035	&	0.014	\\
6012.3514	&	0.247	&	0.921	&	0.015	&	-0.683	&	0.016	&	-4.695	&	0.022	&	6272.1037	&	0.945	&	-0.184	&	0.013	&	-1.119	&	0.010	&	-6.205	&	0.015	\\
6012.3660	&	0.259	&	0.953	&	0.016	&	-0.657	&	0.016	&	-4.470	&	0.022	&	6272.1146	&	0.954	&	-0.270	&	0.013	&	-1.142	&	0.010	&	-6.080	&	0.015	\\
\hline
\end{tabular}
\end{table*}

\begin{table*}
\tiny
\centering
\contcaption{}
\begin{tabular}{rrrrrrrrrrrrrrrr}
\hline
HJD & Rot. & H$\beta$ & $\sigma$ & He\,{\sc ii} & $\sigma$ & H$\alpha$ & $\sigma$ & HJD & Rot. & H$\beta$ & $\sigma$ & He\,{\sc ii} & $\sigma$ & H$\alpha$ & $\sigma$\\
(2450000+) & Phase & EW & EW & EW & EW & EW & EW & (2450000+) & Phase & EW & EW & EW & EW & EW & EW \\
\hline 
6272.1259	&	0.963	&	-0.239	&	0.012	&	-1.109	&	0.010	&	-6.133	&	0.014	&	6352.8845	&	0.403	&	0.720	&	0.016	&	-0.624	&	0.013	&	-5.283	&	0.019	\\
6272.1368	&	0.972	&	-0.266	&	0.012	&	-1.120	&	0.010	&	-6.145	&	0.014	&	6352.8898	&	0.408	&	0.793	&	0.016	&	-0.622	&	0.013	&	-5.245	&	0.020	\\
6272.1477	&	0.981	&	-0.254	&	0.012	&	-1.091	&	0.010	&	-6.143	&	0.014	&	6352.8950	&	0.412	&	0.756	&	0.016	&	-0.632	&	0.013	&	-5.412	&	0.020	\\
6272.1586	&	0.990	&	-0.330	&	0.013	&	-1.109	&	0.010	&	-6.315	&	0.015	&	6352.9002	&	0.416	&	0.646	&	0.017	&	-0.632	&	0.013	&	-5.604	&	0.020	\\
6282.8909	&	0.820	&	0.553	&	0.010	&	-0.944	&	0.008	&	-5.024	&	0.012	&	6352.9054	&	0.421	&	0.742	&	0.017	&	-0.635	&	0.014	&	-5.603	&	0.020	\\
6282.9019	&	0.829	&	0.498	&	0.011	&	-0.923	&	0.008	&	-5.151	&	0.013	&	6352.9106	&	0.425	&	0.714	&	0.018	&	-0.607	&	0.015	&	-5.726	&	0.021	\\
6282.9128	&	0.838	&	0.447	&	0.011	&	-0.970	&	0.009	&	-5.267	&	0.013	&	6352.9159	&	0.429	&	0.631	&	0.020	&	-0.630	&	0.016	&	-5.713	&	0.023	\\
6282.9236	&	0.847	&	0.347	&	0.011	&	-0.982	&	0.008	&	-5.515	&	0.013	&	6352.9213	&	0.434	&	0.597	&	0.019	&	-0.632	&	0.016	&	-5.811	&	0.023	\\
6282.9487	&	0.867	&	0.230	&	0.023	&	-1.082	&	0.019	&	-5.749	&	0.028	&	6352.9265	&	0.438	&	0.553	&	0.018	&	-0.631	&	0.015	&	-5.794	&	0.022	\\
6282.9596	&	0.876	&	0.108	&	0.016	&	-1.057	&	0.012	&	-6.071	&	0.018	&	6352.9318	&	0.442	&	0.515	&	0.019	&	-0.620	&	0.015	&	-5.953	&	0.022	\\
6282.9705	&	0.885	&	0.097	&	0.013	&	-1.089	&	0.010	&	-6.193	&	0.015	&	6352.9370	&	0.447	&	0.529	&	0.019	&	-0.616	&	0.015	&	-6.009	&	0.022	\\
6282.9814	&	0.894	&	0.076	&	0.015	&	-1.120	&	0.012	&	-6.345	&	0.018	&	6352.9422	&	0.451	&	0.539	&	0.019	&	-0.597	&	0.016	&	-5.899	&	0.023	\\
6284.0777	&	0.796	&	0.612	&	0.012	&	-0.967	&	0.010	&	-4.592	&	0.014	&	6352.9474	&	0.455	&	0.518	&	0.019	&	-0.632	&	0.015	&	-6.012	&	0.022	\\
6284.0886	&	0.805	&	0.562	&	0.012	&	-0.987	&	0.010	&	-4.691	&	0.014	&	6352.9527	&	0.459	&	0.499	&	0.020	&	-0.575	&	0.016	&	-6.079	&	0.023	\\
6284.0995	&	0.814	&	0.520	&	0.011	&	-1.003	&	0.009	&	-4.786	&	0.013	&	6352.9579	&	0.464	&	0.426	&	0.022	&	-0.605	&	0.018	&	-6.193	&	0.025	\\
6284.1104	&	0.823	&	0.495	&	0.012	&	-1.033	&	0.010	&	-4.964	&	0.014	&	6353.8716	&	0.215	&	0.626	&	0.020	&	-0.751	&	0.016	&	-6.390	&	0.023	\\
6284.1219	&	0.832	&	0.431	&	0.013	&	-1.037	&	0.010	&	-5.068	&	0.015	&	6353.8768	&	0.220	&	0.639	&	0.019	&	-0.730	&	0.016	&	-6.400	&	0.023	\\
6284.1328	&	0.841	&	0.311	&	0.013	&	-1.055	&	0.010	&	-5.293	&	0.014	&	6353.8820	&	0.224	&	0.633	&	0.020	&	-0.699	&	0.016	&	-6.248	&	0.023	\\
6284.1437	&	0.850	&	0.237	&	0.013	&	-1.043	&	0.010	&	-5.334	&	0.015	&	6353.8873	&	0.228	&	0.641	&	0.022	&	-0.698	&	0.018	&	-6.196	&	0.026	\\
6284.1546	&	0.859	&	0.178	&	0.013	&	-1.083	&	0.011	&	-5.498	&	0.015	&	6353.8925	&	0.233	&	0.715	&	0.021	&	-0.712	&	0.017	&	-6.193	&	0.025	\\
6288.0497	&	0.064	&	-0.352	&	0.038	&	-1.047	&	0.030	&	-6.018	&	0.044	&	6353.8977	&	0.237	&	0.735	&	0.021	&	-0.711	&	0.017	&	-6.107	&	0.025	\\
6288.0697	&	0.080	&	-0.160	&	0.039	&	-0.969	&	0.031	&	-6.050	&	0.046	&	6353.9029	&	0.241	&	0.755	&	0.022	&	-0.680	&	0.018	&	-6.054	&	0.026	\\
6288.0819	&	0.090	&	-0.131	&	0.013	&	-0.880	&	0.010	&	-5.876	&	0.015	&	6353.9081	&	0.246	&	0.775	&	0.025	&	-0.728	&	0.021	&	-5.937	&	0.030	\\
6288.0928	&	0.099	&	-0.113	&	0.014	&	-0.861	&	0.011	&	-5.808	&	0.016	&	6353.9138	&	0.250	&	0.820	&	0.023	&	-0.707	&	0.019	&	-6.119	&	0.028	\\
6288.1037	&	0.108	&	0.017	&	0.018	&	-0.826	&	0.014	&	-5.685	&	0.020	&	6353.9191	&	0.255	&	0.858	&	0.023	&	-0.719	&	0.019	&	-5.927	&	0.027	\\
6288.1146	&	0.117	&	-0.004	&	0.033	&	-0.784	&	0.027	&	-5.509	&	0.038	&	6353.9243	&	0.259	&	0.883	&	0.024	&	-0.641	&	0.019	&	-5.776	&	0.028	\\
6289.0665	&	0.900	&	-0.086	&	0.013	&	-1.094	&	0.010	&	-6.072	&	0.015	&	6353.9295	&	0.263	&	0.908	&	0.023	&	-0.689	&	0.019	&	-5.766	&	0.027	\\
6289.0774	&	0.909	&	-0.085	&	0.013	&	-1.103	&	0.010	&	-6.108	&	0.015	&	6353.9347	&	0.267	&	0.866	&	0.024	&	-0.659	&	0.019	&	-5.713	&	0.028	\\
6289.0883	&	0.918	&	-0.165	&	0.013	&	-1.133	&	0.010	&	-6.309	&	0.015	&	6353.9400	&	0.272	&	0.929	&	0.026	&	-0.739	&	0.022	&	-5.709	&	0.031	\\
6289.0992	&	0.927	&	-0.160	&	0.012	&	-1.136	&	0.010	&	-6.401	&	0.014	&	6353.9452	&	0.276	&	0.920	&	0.030	&	-0.644	&	0.024	&	-5.569	&	0.034	\\
6289.1102	&	0.936	&	-0.222	&	0.014	&	-1.157	&	0.011	&	-6.602	&	0.016	&	6353.9504	&	0.280	&	1.033	&	0.032	&	-0.650	&	0.026	&	-5.483	&	0.036	\\
6289.1211	&	0.945	&	-0.301	&	0.013	&	-1.136	&	0.010	&	-6.668	&	0.015	&	6354.7124	&	0.907	&	0.128	&	0.017	&	-1.176	&	0.013	&	-6.028	&	0.020	\\
6289.1320	&	0.954	&	-0.331	&	0.015	&	-1.167	&	0.012	&	-6.726	&	0.017	&	6354.7176	&	0.912	&	0.040	&	0.017	&	-1.182	&	0.014	&	-6.213	&	0.020	\\
6289.1429	&	0.963	&	-0.387	&	0.015	&	-1.173	&	0.012	&	-6.756	&	0.017	&	6354.7229	&	0.916	&	0.039	&	0.017	&	-1.206	&	0.014	&	-6.238	&	0.020	\\
6289.9441	&	0.622	&	0.676	&	0.011	&	-0.595	&	0.008	&	-4.921	&	0.013	&	6354.7281	&	0.920	&	0.057	&	0.017	&	-1.208	&	0.014	&	-6.296	&	0.020	\\
6289.9550	&	0.631	&	0.720	&	0.010	&	-0.583	&	0.008	&	-4.900	&	0.013	&	6354.7333	&	0.924	&	0.005	&	0.017	&	-1.223	&	0.013	&	-6.328	&	0.020	\\
6289.9659	&	0.640	&	0.786	&	0.010	&	-0.560	&	0.008	&	-4.828	&	0.013	&	6354.7385	&	0.929	&	-0.041	&	0.019	&	-1.244	&	0.015	&	-6.427	&	0.023	\\
6289.9768	&	0.649	&	0.878	&	0.010	&	-0.565	&	0.008	&	-4.628	&	0.012	&	6354.7438	&	0.933	&	-0.033	&	0.019	&	-1.250	&	0.015	&	-6.501	&	0.023	\\
6289.9877	&	0.658	&	0.847	&	0.010	&	-0.557	&	0.008	&	-4.610	&	0.012	&	6354.7490	&	0.937	&	-0.092	&	0.020	&	-1.271	&	0.016	&	-6.532	&	0.024	\\
6289.9986	&	0.667	&	0.920	&	0.010	&	-0.548	&	0.008	&	-4.442	&	0.012	&	6354.7551	&	0.942	&	-0.133	&	0.020	&	-1.285	&	0.016	&	-6.623	&	0.024	\\
6290.0094	&	0.676	&	0.947	&	0.011	&	-0.565	&	0.009	&	-4.474	&	0.013	&	6354.7603	&	0.947	&	-0.104	&	0.020	&	-1.284	&	0.016	&	-6.668	&	0.024	\\
6290.0203	&	0.685	&	0.926	&	0.011	&	-0.559	&	0.009	&	-4.337	&	0.013	&	6354.7655	&	0.951	&	-0.120	&	0.021	&	-1.286	&	0.016	&	-6.791	&	0.024	\\
6343.8391	&	0.962	&	-0.013	&	0.011	&	-1.132	&	0.009	&	-5.563	&	0.013	&	6354.7708	&	0.955	&	-0.133	&	0.021	&	-1.318	&	0.017	&	-6.716	&	0.024	\\
6343.8500	&	0.971	&	-0.065	&	0.012	&	-1.118	&	0.010	&	-5.606	&	0.014	&	6354.7760	&	0.960	&	-0.157	&	0.021	&	-1.305	&	0.017	&	-6.855	&	0.024	\\
6343.8609	&	0.980	&	-0.072	&	0.013	&	-1.122	&	0.010	&	-5.736	&	0.015	&	6354.7812	&	0.964	&	-0.189	&	0.021	&	-1.302	&	0.017	&	-6.963	&	0.025	\\
6343.8718	&	0.989	&	-0.070	&	0.014	&	-1.108	&	0.011	&	-5.751	&	0.016	&	6354.7864	&	0.968	&	-0.189	&	0.021	&	-1.292	&	0.017	&	-6.997	&	0.025	\\
6344.8483	&	0.792	&	0.849	&	0.010	&	-0.827	&	0.008	&	-3.541	&	0.012	&	6354.7917	&	0.972	&	-0.185	&	0.022	&	-1.334	&	0.017	&	-7.097	&	0.025	\\
6351.8353	&	0.540	&	0.536	&	0.017	&	-0.578	&	0.013	&	-5.354	&	0.020	&	6355.8575	&	0.849	&	0.563	&	0.018	&	-1.000	&	0.015	&	-4.650	&	0.022	\\
6351.8406	&	0.545	&	0.644	&	0.018	&	-0.567	&	0.014	&	-5.564	&	0.021	&	6355.8627	&	0.854	&	0.579	&	0.018	&	-1.031	&	0.015	&	-4.676	&	0.021	\\
6351.8458	&	0.549	&	0.550	&	0.017	&	-0.597	&	0.014	&	-5.514	&	0.021	&	6355.8680	&	0.858	&	0.525	&	0.021	&	-1.006	&	0.017	&	-4.680	&	0.024	\\
6351.8510	&	0.553	&	0.558	&	0.016	&	-0.570	&	0.013	&	-5.383	&	0.019	&	6355.8732	&	0.862	&	0.496	&	0.020	&	-1.027	&	0.016	&	-4.743	&	0.023	\\
6351.8562	&	0.557	&	0.604	&	0.016	&	-0.566	&	0.013	&	-5.355	&	0.020	&	6355.8784	&	0.867	&	0.475	&	0.020	&	-1.059	&	0.016	&	-4.738	&	0.023	\\
6351.8614	&	0.562	&	0.590	&	0.016	&	-0.566	&	0.013	&	-5.382	&	0.020	&	6355.8837	&	0.871	&	0.453	&	0.023	&	-1.079	&	0.019	&	-4.773	&	0.026	\\
6351.8667	&	0.566	&	0.584	&	0.016	&	-0.570	&	0.013	&	-5.389	&	0.019	&	6355.8889	&	0.875	&	0.416	&	0.022	&	-1.063	&	0.018	&	-5.007	&	0.025	\\
6351.8719	&	0.570	&	0.611	&	0.016	&	-0.537	&	0.013	&	-5.337	&	0.019	&	6355.8941	&	0.879	&	0.368	&	0.023	&	-1.074	&	0.019	&	-4.966	&	0.026	\\
6351.8779	&	0.575	&	0.607	&	0.016	&	-0.570	&	0.013	&	-5.251	&	0.019	&	6355.9007	&	0.885	&	0.341	&	0.021	&	-1.114	&	0.017	&	-5.073	&	0.024	\\
6351.8832	&	0.580	&	0.631	&	0.016	&	-0.542	&	0.013	&	-5.275	&	0.019	&	6355.9059	&	0.889	&	0.282	&	0.020	&	-1.083	&	0.017	&	-5.147	&	0.023	\\
6351.8884	&	0.584	&	0.685	&	0.016	&	-0.543	&	0.013	&	-5.217	&	0.019	&	6355.9111	&	0.893	&	0.277	&	0.020	&	-1.086	&	0.016	&	-5.059	&	0.023	\\
6351.8936	&	0.588	&	0.700	&	0.016	&	-0.556	&	0.013	&	-5.272	&	0.019	&	6355.9164	&	0.898	&	0.215	&	0.021	&	-1.117	&	0.017	&	-5.295	&	0.024	\\
6351.8988	&	0.592	&	0.679	&	0.016	&	-0.523	&	0.013	&	-5.184	&	0.019	&	6355.9216	&	0.902	&	0.226	&	0.023	&	-1.122	&	0.019	&	-5.396	&	0.026	\\
6351.9041	&	0.597	&	0.704	&	0.016	&	-0.525	&	0.013	&	-5.197	&	0.020	&	6355.9268	&	0.906	&	0.234	&	0.024	&	-1.181	&	0.019	&	-5.311	&	0.027	\\
6351.9093	&	0.601	&	0.758	&	0.016	&	-0.522	&	0.013	&	-5.122	&	0.019	&	6355.9320	&	0.911	&	0.258	&	0.024	&	-1.149	&	0.020	&	-5.540	&	0.028	\\
6351.9145	&	0.605	&	0.802	&	0.016	&	-0.534	&	0.013	&	-5.175	&	0.019	&	6355.9372	&	0.915	&	0.189	&	0.028	&	-1.195	&	0.023	&	-5.626	&	0.031	\\
6352.8793	&	0.399	&	0.771	&	0.016	&	-0.615	&	0.013	&	-5.252	&	0.019	\\
\hline
\end{tabular}
\end{table*}

\bsp	
\label{lastpage}
\end{document}